\documentclass[fleqn,usenatbib]{mnras}

\usepackage{newtxtext,newtxmath}

\usepackage[T1]{fontenc}

\DeclareRobustCommand{\VAN}[3]{#2}
\let\VANthebibliography\thebibliography
\def\thebibliography{\DeclareRobustCommand{\VAN}[3]{##3}\VANthebibliography}

\usepackage{graphicx}	
\usepackage{amsmath}	
\usepackage{verbatim}

\newcommand{\xmm}{\hbox{\textit{XMM-Newton}}}
\newcommand{\xray}{\hbox{X-ray}}
\newcommand{\chandra}{\hbox{\textit{Chandra}}}

\newcommand{\lum}{{erg~s$^{-1}$}}

\def\bharm{$\overline{{\rm BHAR}/M_\star}$}
\def\bhar{$\rm \overline{BHAR}$}

\def\mstar{$M_\star$}

\def\sigmaone{$\Sigma_{1}$}

\def\lx{$L_{\rm X}$}
\def\lxlimit{$L_{\rm X, limit}$}
\def\dn{$\rm D_{n}4000$}

\title[AGNs in the lifecycle of galaxies]{The incidence of AGN in galaxies with different stellar population ages}

\author[]{Q. Ni,$^{1,2}$\thanks{E-mail: qingling1001@gmail.com}
J. Aird,$^{1}$
A. Merloni,$^{2}$
K. L. Birchall,$^{3}$
J. Buchner,$^{2}$
M. Salvato,$^{2}$
G. Yang$^{4,5}$
\\
$^{1}$Institute for Astronomy, University of Edinburgh, Royal Observatory, Edinburgh EH9 3HJ, UK\\
$^{2}$Max-Planck-Institut f\"{u}r extraterrestrische Physik (MPE), Gie{\ss}enbachstra{\ss}e 1, D-85748 Garching bei M\"unchen, Germany\\
$^{3}$School of Physics \& Astronomy, University of Leicester, University Road, Leicester LE1 7RH, UK\\
$^{4}$Kapteyn Astronomical Institute, University of Groningen, P.O. Box 800, 9700 AV Groningen, The Netherlands\\
$^{5}$SRON Netherlands Institute for Space Research, Postbus 800, 9700 AV Groningen, The Netherlands
}

\date{Accepted XXX. Received YYY; in original form ZZZ}

\pubyear{2015}

\begin{document}
\label{firstpage}
\pagerange{\pageref{firstpage}--\pageref{lastpage}}
\maketitle

\begin{abstract}
It has been argued that recycled gas from stellar mass loss in galaxies might serve as an important fuelling source for black holes (BHs) in their centers.
Utilizing spectroscopic samples of galaxies from the Sloan Digital Sky Survey (SDSS) at $z = 0$--0.35 and the Large Early Galaxy Astrophysics Census (LEGA-C) survey at $z = 0.6$--1 that have X-ray coverage from \xmm\ or \chandra, we test this stellar mass loss fuelling scenario by investigating how AGN activity and BH growth vary with the break strength at 4000~\AA, \dn\ (which is closely related to the age of stellar populations), as younger galaxies are considered to have higher stellar mass loss rates. We found that when controlling for host-galaxy properties, the fraction of log~\lx/\mstar\ $> 32$ (which roughly corresponds to Eddington ratios $\gtrsim 1$\%) AGN  and sample-averaged black hole accretion rate (\bhar) decrease with \dn\ among \dn\ $\lesssim$ 1.9 galaxies, suggesting a higher level of AGN activity among younger galaxies, which supports the stellar mass loss fuelling scenario.
For the oldest and most massive galaxies at $z = 0$--0.35, this decreasing trend is not present anymore. We found that, among these most massive galaxies at low redshift, the fraction of low specific-accretion-rate (31 $<$ log \lx/\mstar\ $<$ 32) AGNs increases with \dn, which may be associated with additional fuelling from hot halo gas and/or enhanced accretion capability.
\end{abstract}




\section{Introduction}


In the past decades, the understanding of how galaxies evolve over cosmic history has progressed rapidly as a result of accumulating data from various imaging and spectroscopic surveys \citep[e.g.][]{Kauffmann2003, Faber2007, Madau2014, vdw2014, Barro2017, Wu2018}. Galaxies appear to follow a range of evolutionary pathways, whereby their stellar populations and gas properties change over time, although the physical processes that drive these transformations are still unclear. While supermassive black holes (BHs) only occupy a small space in the galaxy centers, they are thought to play an important role in galaxy evolution.
As the mass of BHs is found to be tightly correlated with the mass of their host bulges \citep[e.g.][]{Magorrian1998, MH2003, MM2013, KH2013}, BHs appear to coevolve with their host galaxies.
While we know that central supermassive BHs grow primarily by accreting gas and can be seen as Active Galactic Nuclei (AGNs), the exact feeding mechanism of BHs remains unclear. 
It is important to investigate what drives the growth of BHs, which will ultimately reveal the physical mechanisms behind the potential coevolution scenario.

It has been found that BH accretion rate tracks the star formation rate over cosmic history, suggesting cold gas supply as a common fuel for both the galaxy and the BH \citep[e.g.][]{Aird2010, KH2013}.
However, the noticeable fraction of AGNs in quiescent galaxies (given their low level of star formation activity and thus inferred low level of global cold gas content) indicates that additional mechanisms may fuel the growth of central BHs after quenching \citep[e.g.][]{Kocevski2017,Wang2017,Aird2019,Aird2022}.
Among galaxies where the cold gas supply is not sufficient, stellar mass loss may provide an important, additional source of fuel for accretion onto the BH \citep[e.g.][]{Hopkins2006, Ciotti2007, Kauffmann2009}.
In this scenario, as galaxies with younger stellar populations have stronger stellar winds and higher mass loss rates that provide more fuel for the central BHs, BH growth is expected to vary among galaxies with different ages of stellar populations.
Through characterizing the AGN activity across the lifecycle of galaxies, we can investigate the role of stellar mass loss in fuelling BHs.

As spectra can provide direct information about the age of stellar populations through features such as the break strength at 4000~\AA, \dn, large samples of galaxies and AGNs with spectroscopic coverage are needed to investigate the potential fuelling through stellar mass loss.
In the local universe, the Sloan Digital Sky Survey (SDSS) provides a large legacy sample of galaxies \citep[e.g.][]{Strauss2002}, and the X-ray information provided by serendipitous catalogs from accumulating X-ray observations over the entire sky \citep[e.g.][]{Evans2010, Webb2020} provides opportunities to effectively identify AGNs among this large galaxy sample \citep[e.g.][]{Brandt2015, Brandt2021}. 
More recently, deep spectroscopic surveys \citep[e.g.][]{vdw2016, McLure2018} have started to help build representative samples of galaxies at intermediate reshifts. When deep X-ray coverage is available, these surveys can also be effectively utilized to study the incidence of AGNs in galaxies with different stellar population ages \citep[e.g.][]{Mountrichas2022, Georgantopoulos2023}.

In this paper, we use samples of galaxies from SDSS and the Large Early Galaxy Census (LEGA-C; e.g. \citealt{vdw2016}) survey to investigate how AGN activity and BH growth vary with \dn\ across the galaxy lifecycle at $z = 0-0.35$ and $z = 0.6-1$.
We also test whether AGN activity and BH growth vary with \dn\ when other galaxy properties are controlled, thus revealing whether the age of stellar populations has a fundamental influence on AGN activity/BH growth.

The paper is structured as follows. In Section~\ref{s-sample}, we describe the sample construction process. 
In Section~\ref{s-ar}, we detail the analysis results and discuss what they imply in Section~\ref{s-dc}. The conclusions are presented in Section~\ref{s-cf}.
Throughout this paper, stellar masses ($M_\star$) are given in units of $M_\odot$; star formation rates (SFRs) and sample-averaged black hole accretion rates (\bhar) are given in units of $M_\odot$~yr$^{-1}$.
$L_X$ represents X-ray luminosity at rest-frame 2--10 keV in units of erg s$^{-1}$. Quoted uncertainties are at the $1\sigma$\ (68\%) confidence level. 
A cosmology with $H_0=70$~km~s$^{-1}$~Mpc$^{-1}$, $\Omega_M=0.3$, and $\Omega_{\Lambda}=0.7$ is assumed.

\section{DATA \& SAMPLE SELECTION} \label{s-sample}

Two samples of X-ray AGNs with spectroscopic coverage are utilized in this study: one sample includes SDSS galaxies with \xmm\ coverage (see Section~\ref{4xmmsample}); one sample includes LEGA-C galaxies with \chandra\ coverage in the COSMOS field (see Section~\ref{cosmossample}). In Section~\ref{ss-agnf}, we discussed how the AGN fraction and \bhar\ are estimated. 

\subsection{Constructing a sample of SDSS galaxies with \xmm\ coverage} \label{4xmmsample}
\subsubsection{Selecting galaxies in MPA-JHU catalog with \xmm\ coverage}
The SDSS galaxy/AGN sample utilized in the study is built upon the MPA-JHU value-added catalog.\footnote{\url{http://www.mpa-garching.mpg.de/SDSS/DR7/}}
The MPA-JHU catalog provides galaxy property measurements for SDSS DR8 spectra classified by the pipeline as a galaxy (SPECTROTYPE = GALAXY), including \dn, \mstar, and SFR used in this study. We only consider objects observed as the prime targets in the SDSS Legacy Survey Main Galaxy Sample in this study.\footnote{Galaxies in the SDSS Legacy Survey are observed with 3$''$-diameter fibers.}

\dn\ in the MPA-JHU catalog is measured according to the \citet{Balogh1999} definition, as the ratio of the flux in the 4000--4100 \AA\ continuum to that in the 3850--3950 \AA\ continuum. 
\mstar\ values are measured following model grids based on SDSS photometry, as described in \citet{Kauffmann2003}.
SFR values are measured by combining emission line measurements from \citet{Brinchmann2004} with aperture corrections as described in \citet{Salim2007}.
\mstar\ and SFR values in the MPA-JHU catalog are largely consistent with those derived from Galex-SDSS-WISE SED fitting \citep[e.g.][]{Salim2016}, and have been widely adopted.

We use the RapidXMM database \citep{Ruiz2022} to select galaxies in MPA-JHU catalog within \xmm\ coverage.
RapidXMM provides \xmm\ aperture photometry as well as upper limits within HEALPix cells of size $\approx$ 3 arcsec.
As the 4XMM-DR11 catalog \citep{Webb2020} is utilized to match galaxies with their \xmm\ counterparts for X-ray AGN identification (see Section~\ref{ss-4xmmmatching}), we limit our sample to SDSS galaxies with \xmm\ point observations made before 2020 December 17.

\subsubsection{Matching selected galaxies with \xmm\ counterparts} \label{ss-4xmmmatching}

The 4XMM-DR11 catalog used in this study to provide \xray\ counterparts to SDSS galaxies contains sources drawn from a total of 12210 \xmm\ observations made between 2000 February 1 and 2020 December 17, including $\sim$ 600,000 unique X-ray sources over $\approx$ 1239 deg$^2$.

As SDSS galaxies in the MPA-JHU catalog are a subset of SDSS objects, we first match \xray\ sources in the 4XMM DR11 catalog with SDSS DR8 \citep{AM2011} and catWISE \citep{Marocco2021} catalogs, and use the obtained optical/NIR counterparts positions to further match with MPA-JHU galaxies.
When we perform the source matching, we only consider 4XMM sources that are within 30 arcmin of MPA-JHU sources, thus restricting to 4XMM sources that lie within the footprint of the MPA-JHU catalog. 
The input SDSS and WISE catalogs only include objects within 1 arcmin of 4XMM sources, to save computation time.
Following the method in \citet{Ni2021b}, we use {\sc NWAY} \citep{Salvato2018} to perform the source matching, with priors obtained from sources reported in the \chandra\ Source Catalog (CSC) 2.0 \citep{Evans2010} that are matched to the 4XMM sources within the 95\% uncertainties (as \chandra\ detections have better positional accuracy than \xmm\ detections). 
Utilizing the SDSS and WISE sources that are matched to these \chandra\ sources within 5$''$, we obtain priors in the WISE W1$+$W2 vs. W1$-$W2 color space and in the $i$-band magnitude space.
With these priors, we match 4XMM sources to their optical/NIR counterparts.
Matched primary counterparts with $p_{\rm any} > 0.1$ and $p_{\rm i} > 0.1$ are adopted as reliable matches.
We then match MPA-JHU galaxies to these obtained optical/NIR positions of 4XMM sources with a 1$''$ matching radius.
To assess the matching accuracy, we use the 4XMM sources that have \chandra\ counterparts, and compare the level of agreement between matched MPA-JHU sources when either 4XMM or CSC position is used.
The matched MPA-JHU counterparts of 4XMM sources have a $\approx$96\% agreement with those of CSC sources.

\subsection{LEGA-C galaxies in the COSMOS field} \label{cosmossample}

To extend the redshift range probed in this study, we also include galaxies/AGNs in the LEGA-C Data Release 3 \citep{vdW2021} which have \chandra\ \xray\ coverage from the COSMOS-Legacy survey \citep{Civano2016}.\footnote{Galaxies in the LEGA-C survey are observed with 1$''$ $\times$ $8''$ slits.}
We only utilize sources in the LEGA-C catalog with PRIMARY = 1, FLAG\_MORPH = 0, and FLAG\_SPEC $<$ 2;  \dn\ values are available for 2129 of these sources, measured according to the \citet{Balogh1999} definition as well.
We also examine the spectra visually and remove spectra with broad emission lines from the sample.
We cross-match these sources to the photometric sample of galaxies in the COSMOS field in \citet{Ni2021}, which provides \mstar\ and SFR measurements for sources within both the COSMOS and UltraVISTA regions utilizing photometric data in 38 bands (including 24 broad bands) from NUV to FIR \citep{Laigle2016} and the SED-fitting code \texttt{CIGALE} \citep[e.g.][]{Boquien2019, Yang2020}. X-ray counterparts from \chandra\ are also matched for this photometric sample of galaxies in \citet{Ni2021}. 
The detailed process to obtain \mstar\ and SFR measurements can be seen in section 2.1 of \citet{Ni2021}. An AGN component is included in the SED fitting, in addition to the galaxy component. As discussed in \citet{Ni2021}, adding the X-ray information or not during the SED fitting does not significantly affect the Bayesian \mstar\ and SFR measurements. Uncertainties of the \mstar\ and SFR values obtained are also discussed in section 2.1 and appendix A of \citet{Ni2021}. 
We note that \mstar\ measurements are generally robust when comparing results from different SED fitting methods, with a scatter of $\approx$ 0.1 dex; the scatter of SFR measurements can be up to $\approx$ 0.4 dex.
This finally provides a sample of 1792 galaxies in the COSMOS field.

\subsection{Sample properties} \label{ss-sp}

\begin{table}
 \begin{center}
 \caption{Summary of sample properties. (1) Name of the sample. (2) Redshift range of the sample. (3) \mstar\ range of the sample. (4) Number of galaxies in the sample. (5) Number of log \lx/\mstar\ $>$ 32 AGNs.}
  \begin{tabular}{ccccccccccc}
  \hline\hline
Sample &  Redshift  &  Mass   & Number of   & Number of \\
Name   &   Range    &  Range  &  Galaxies   &   AGNs \\
(1)    &  (2)       &  (3)    &  (4)        & (5)                \\
\hline
4XMM &  0--0.35    & log\mstar\ $>$ 9  & 22576 & 89 \\ \vspace{0.1 cm}
COSMOS   &  0.6--1.0   & log\mstar\ $>$ 10 & 1496   & 38  \\ 
\hline
  \end{tabular}
  \end{center}
  \label{ts}
\end{table}

\begin{figure*}
\begin{center}
\includegraphics[scale=0.53]{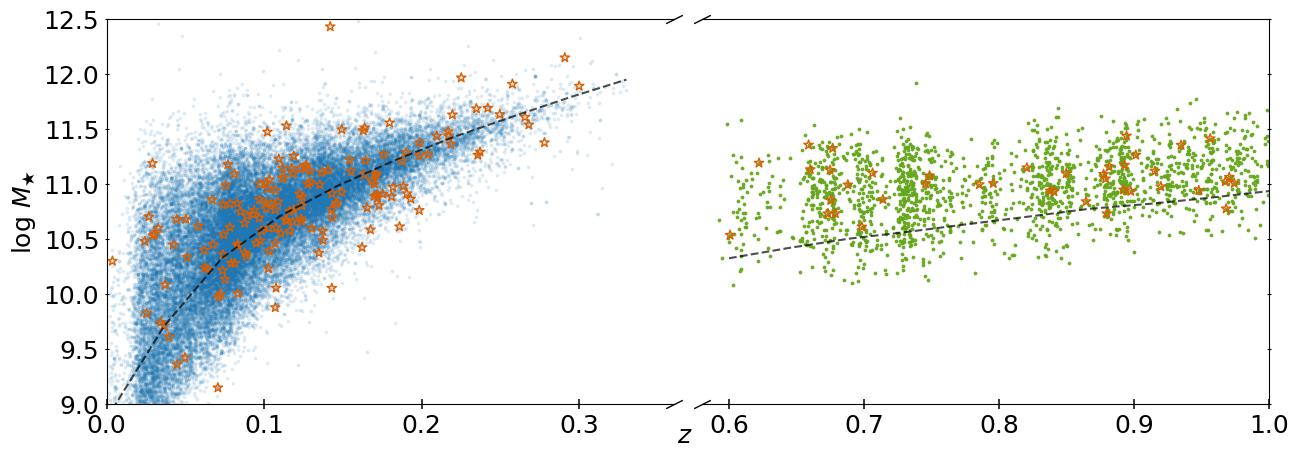}
\caption{\mstar\ as a function of $z$ for SDSS main sample galaxies in the MPA-JHU catalog with \xmm\ coverage (blue dots) and galaxies included the LEGA-C survey (green dots). X-ray AGNs with log \lx/\mstar\ $>$ 32 are marked as orange stars. The dashed curve indicates the 90\% \mstar\ completeness limit as a function of redshift.
}
\label{mz}
\end{center}
\end{figure*}

\begin{figure*}
\begin{center}
\includegraphics[scale=0.55]{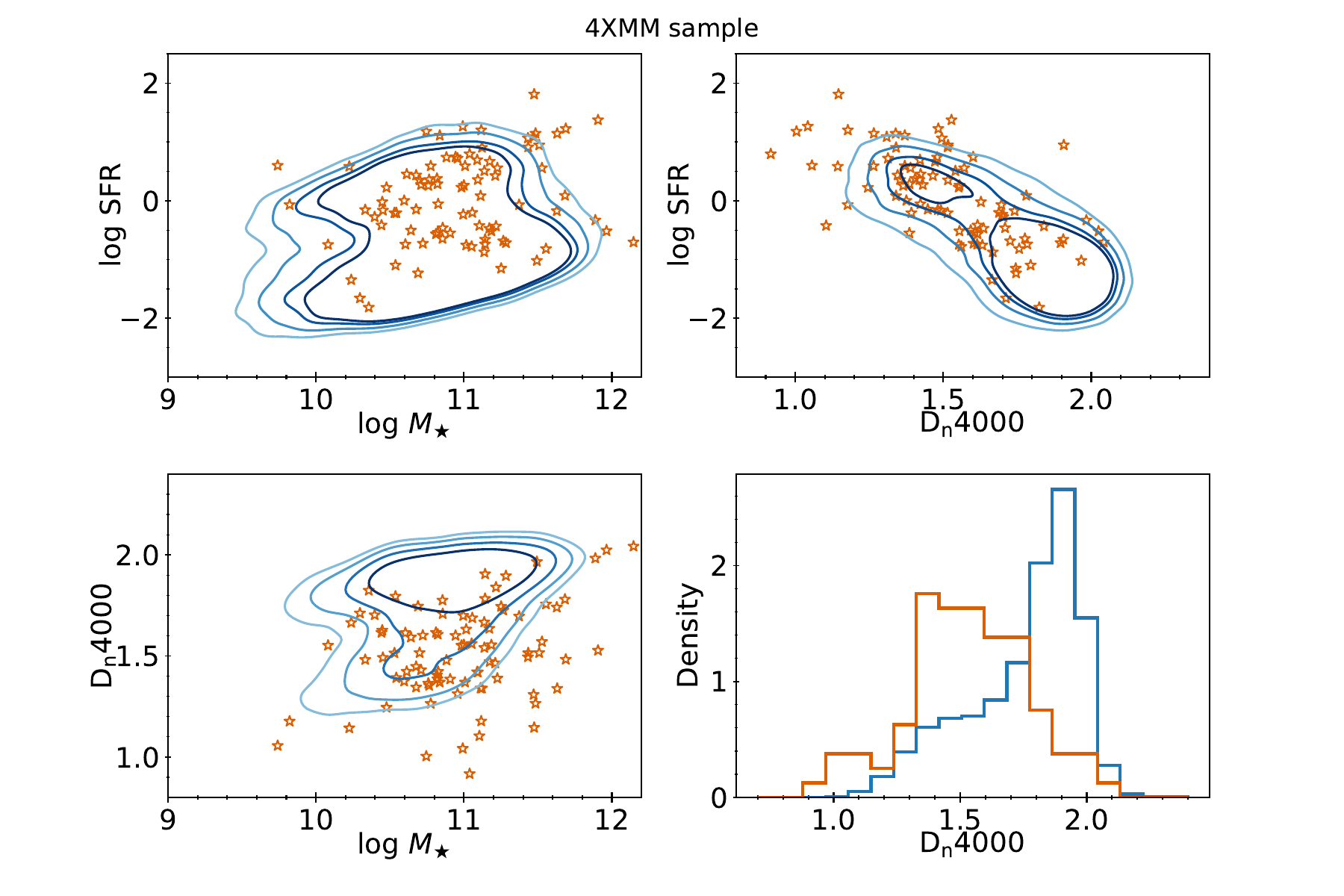}
\caption{Galaxies in the 4XMM sample in the SFR vs. \mstar\ plane (upper-left), SFR vs. \dn\ plane (upper-right), and \dn\ vs. \mstar\ plane (bottom-left). 
The contours encircle 68 per cent, 80 per cent, 90 per cent, and 95 per cent of galaxies.
Log \lx/\mstar\ $>$ 32 AGNs are represented by the orange stars.
In the bottom-right panel, the \dn\ distribution of galaxies is represented by the blue histogram; the \dn\ distribution of log \lx/\mstar\ $>$ 32 AGNs is represented by the orange histogram.
}
\label{samplep}
\end{center}
\end{figure*}

\begin{figure*}
\begin{center}
\includegraphics[scale=0.55]{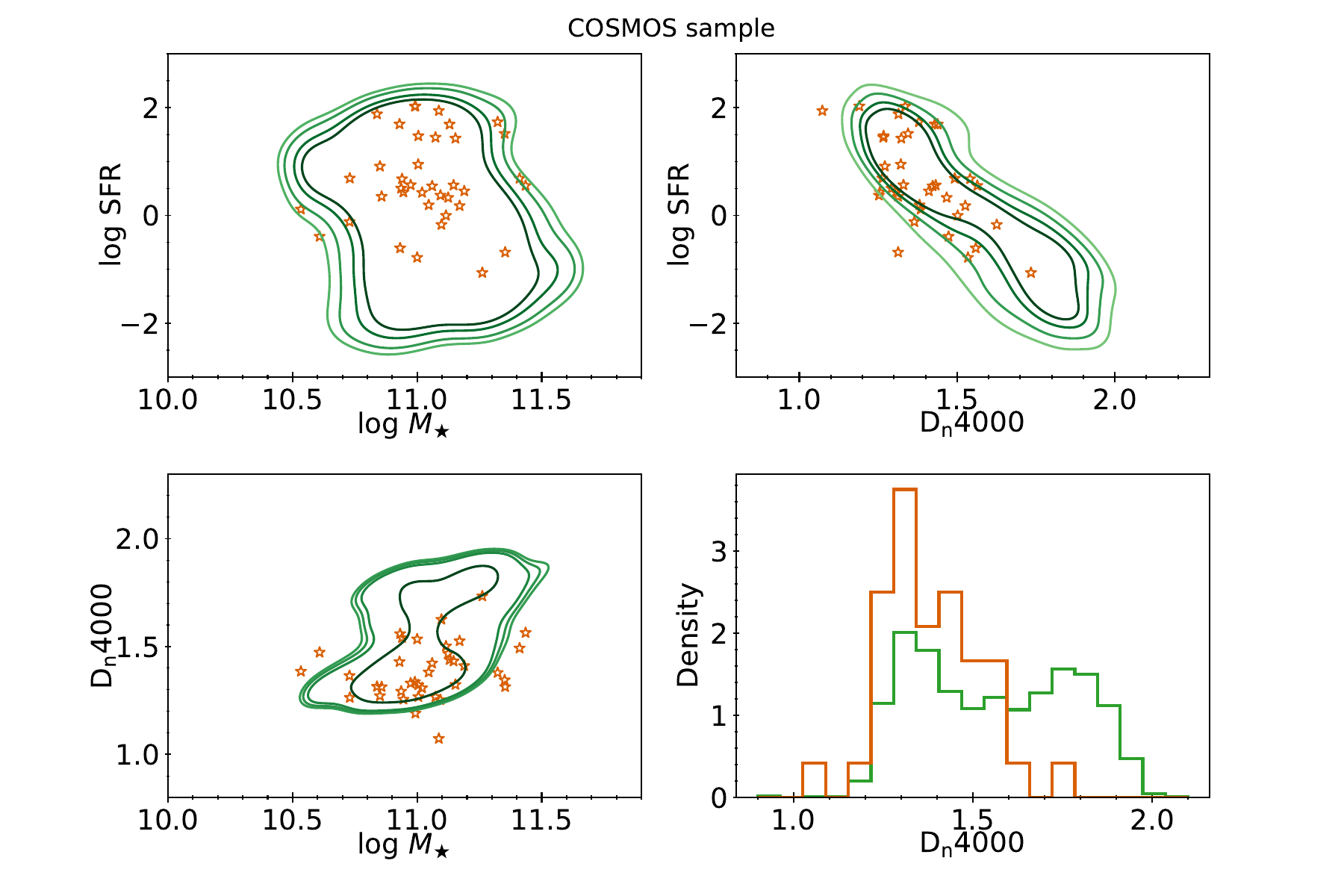}
\caption{Galaxies in the COSMOS sample in the SFR vs. \mstar\ plane (upper-left), SFR vs. \dn\ plane (upper-right), and \dn\ vs. \mstar\ plane (bottom-left). 
The contours encircle 68 per cent, 80 per cent, 90 per cent, and 95 per cent of galaxies.
Log \lx/\mstar\ $>$ 32 AGNs are represented by the orange stars.
In the bottom-right panel, the \dn\ distribution of galaxies is represented by the green histogram; the \dn\ distribution of log \lx/\mstar\ $>$ 32 AGNs is represented by the orange histogram.
}
\label{samplep_legac}
\end{center}
\end{figure*}

In Figure~\ref{mz}, we present the \mstar\ vs. $z$ distributions for the selected SDSS galaxies and LEGA-C galaxies.
For SDSS galaxies with \xmm\ coverage, we plot the 90\%\ \mstar\ completeness curve of galaxies in the SDSS main galaxy sample. The limiting \mstar\ is derived following section 3.2 of \citet{Ilbert2013} given the $r$-band Petrosian magnitude limit of 17.77 of the SDSS main galaxy sample \citep{Strauss2002}.
In our study, we only utilize SDSS galaxies above this mass-completeness curve, and we refer to this galaxy sample as the 4XMM sample throughout the remainder of this work. 
The LEGA-C primary targets are drawn from a $K_s$-band selected parent sample; these targets are representative of the parent sample when taking into account the selection correction factor \texttt{SCOR} provided in the LEGA-C catalog (see Appendix~A of \citealt{vdW2021} for details). This selection correction factor is utilized in our analyses to weight each galaxy, although we find that excluding this factor does not impact our results materially (see Section~\ref{ss-agn-dn} for details).
The 90\% \mstar\ completeness curve for the parent sample (which has a redshift-dependent $K_s$-band limiting magnitude of $20.7 - 7.5 \times {\rm log}((1 + z)/1.8)$) is shown on the plot.
In our study, we only utilize LEGA-C galaxies above this mass-completeness curve of the parent sample, and we refer to this galaxy sample as the COSMOS sample throughout the remainder of this work. 
We carry out our studies with mass-complete samples to avoid any potential bias associated with an incomplete characterization of galaxy populations.
In Table~1, we present the properties of the 4XMM sample and the COSMOS sample.
In Figures~\ref{samplep} and \ref{samplep_legac}, we present how galaxies/AGNs in our samples distribute on the SFR vs. \mstar\ plane, SFR vs. \dn\ plane, and \dn\ vs. \mstar\ plane; we also present the \dn\ distribution of galaxies/AGNs.
In Appendix~\ref{a-dncomtamination}, we show that the contamination from AGN emission to \dn\ measurements is small for AGNs in our sample, and will not materially affect our results.

\subsection{Obtaining AGN fraction and sample-averaged black-hole accretion rate} \label{ss-agnf}

\begin{figure*}
\begin{center}
\includegraphics[scale=0.5]{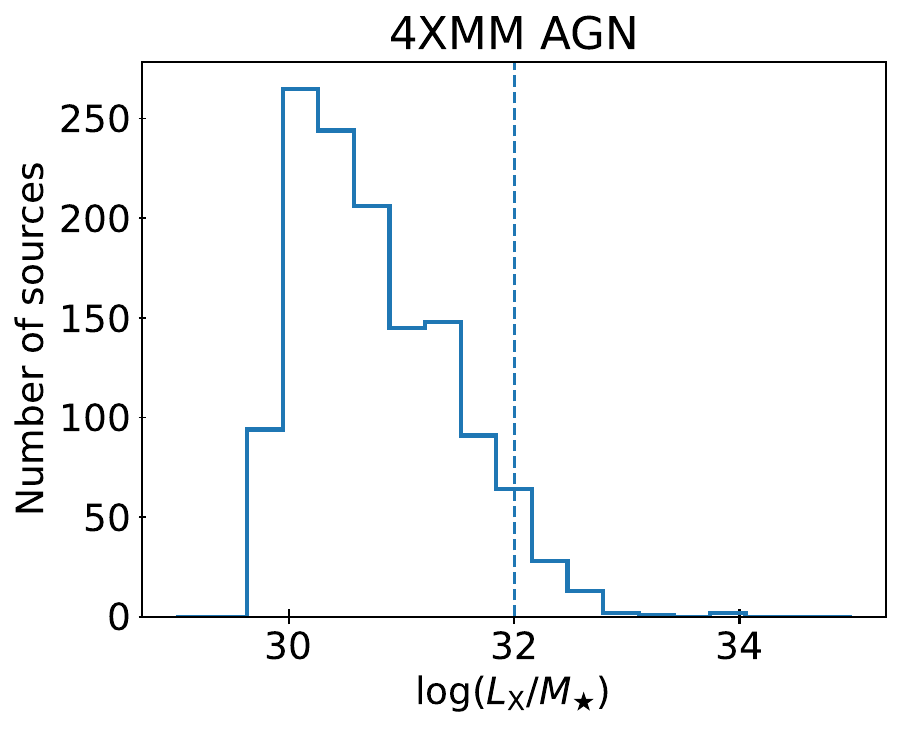}
\includegraphics[scale=0.5]{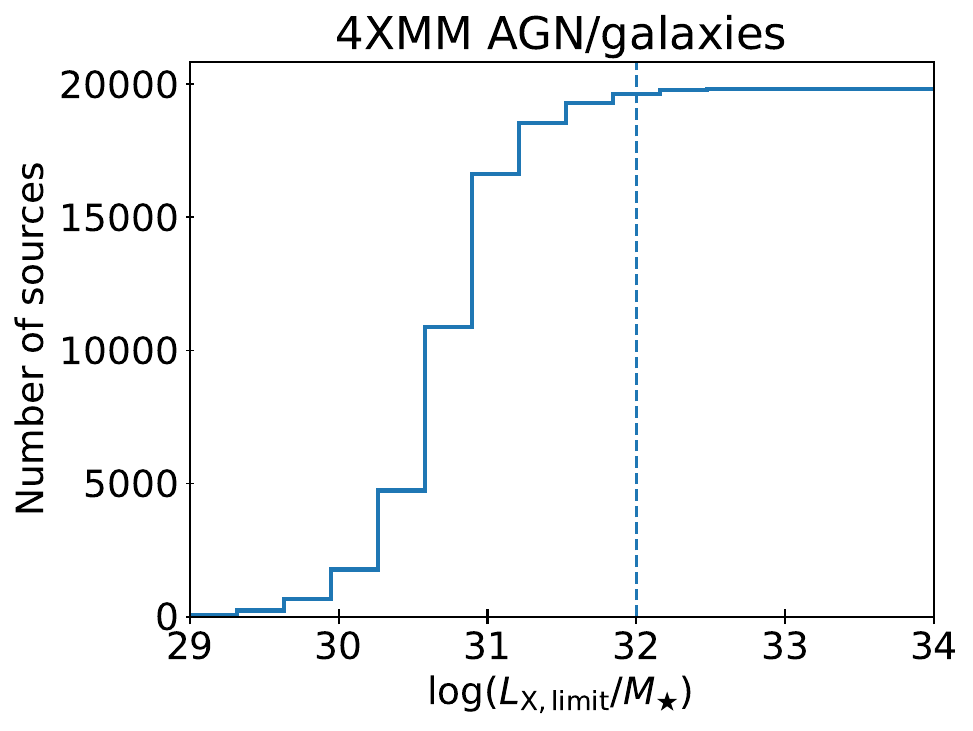}
\includegraphics[scale=0.5]{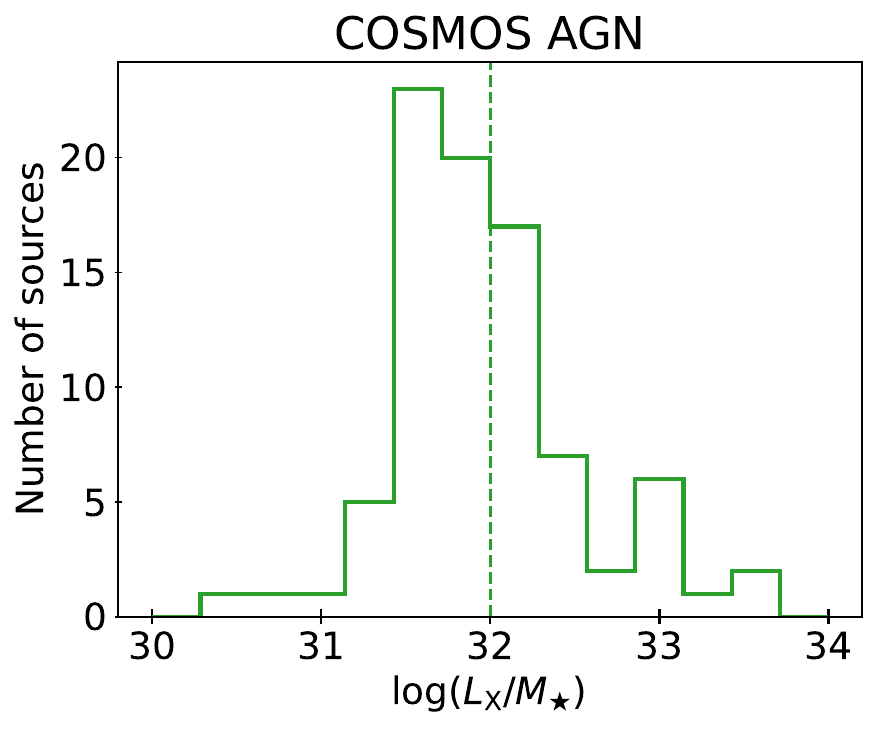}
\includegraphics[scale=0.5]{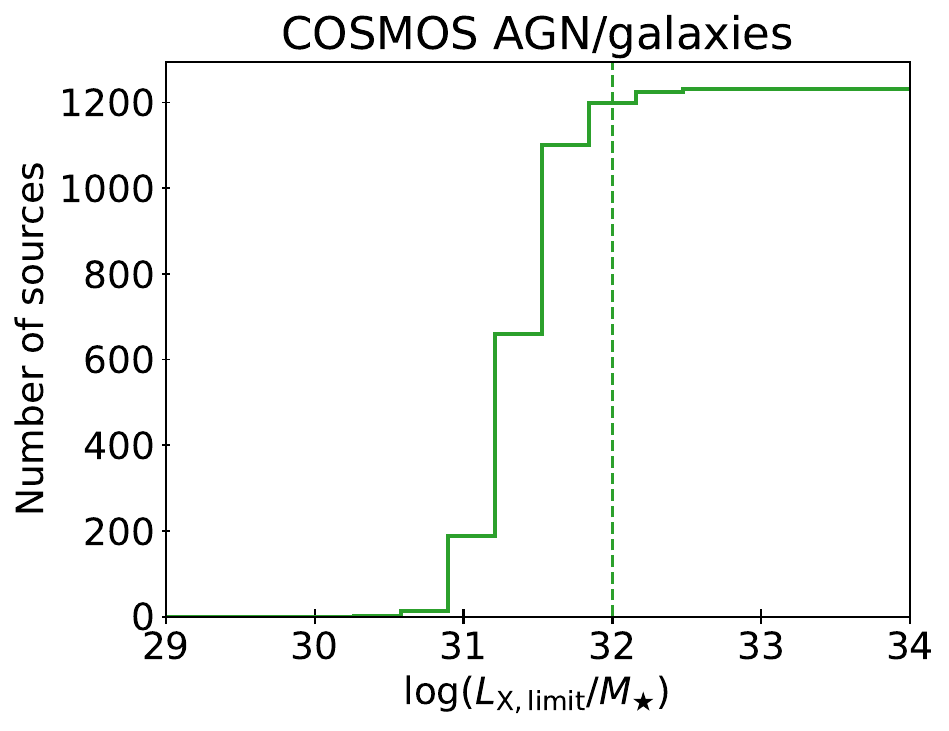}
\caption{\textit{Left panels:} The observed distribution of log \lx/\mstar\ for X-ray detected sources in the 4XMM (top left) and COSMOS (bottom left) samples. \textit{Right panels:} The cumulative distribution of the detection limit, log \lxlimit/\mstar\, above which an AGN could be detected for all galaxies in the 4XMM (top right) and COSMOS (bottom right) samples. The vertical lines represent the \lx/\mstar\ limit adopted in our work when calculating AGN fraction. While a significant number of AGNs detected have log \lx/\mstar\ $< 32$, we note that a large fraction of galaxies in both samples have log \lxlimit/\mstar\ $\leqslant 32$ to enable an unbiased characterization of AGN fraction.
}
\label{lxm_distribution}
\end{center}
\end{figure*}

The AGN fraction is defined in terms of \lx/\mstar\ \citep[as advocated by][]{Bongiorno2016, Aird2018, Aird2019, Aird2022, Birchall2022}, which measures the rate of black hole growth relative to the stellar mass of the host galaxy (i.e.~the ``specific black hole accretion rate'') and thus accounts for the overall stellar-mass-selection bias whereby weakly accreting AGN in more massive galaxies have a higher \lx\ and are thus easier to detect \citep[see][]{Aird2012}. 
AGN fractions throughout this work, unless otherwise stated, refer to objects with log~\lx/\mstar\ $\geqslant$ 32, which roughly corresponds to Eddington ratios of $\gtrsim 1\%$ following the conversion factors from Equation 2 of \citet{Aird2018}.
For the 4XMM sample, we adopt the X-ray fluxes of detected \xmm\ sources from the 4XMM DR11 catalog \citep{Webb2020}. 
We convert the X-ray fluxes to \lx\ assuming a power-law model with Galactic absorption and $\Gamma = 1.7$ following the preference order of 4.5--12 keV band, and 0.2--12 keV band, thus minimizing the effects of X-ray obscuration. About $90\%$ of the X-ray sources in our 4XMM sample have 4.5--12 keV band flux measurements available.
For the COSMOS sample, we adopt \lx\ calculated from \citet{Ni2021}, which is converted from X-ray fluxes following the 2--7 keV, 0.5--7 keV, and 0.5--2 keV order, also assuming the $\Gamma = 1.7$ power-law model.
For X-ray sources in the COSMOS sample, $\approx 70\%$ of them have 2--7 keV band flux measurements available, and 0.5--7 keV/0.5--2 keV band flux measurements are used for $\approx 27\%$/$3\%$ of them.
In the left panels of Figure~\ref{lxm_distribution}, we present the \lx/\mstar\ distribution of X-ray detected sources which have \lx\ values greater than the contributions from X-ray binaries (XRBs) in our samples.
The XRB luminosity ($L_{\rm X,XRB}$) is estimated through a redshift-dependent function of $M_\star$ and SFR (model 269, \citealt{Fragos2013}), which is derived utilizing observations in \citet{Lehmer2016}.

When deriving the AGN fraction, we correct \lx\ to account for the modest systematic effect from obscuration with correction factors detailed below. Utilizing X-ray sources in \chandra\ Deep Field-South \citep{Luo2017} that have similar X-ray flux level as \chandra\ COSMOS sources but with more counts, \citet{Yang2018} compared the intrinsic \lx\ from spectral modeling with \lx\ calculated following the scheme mentioned above, and found that the overall underestimation of X-ray emission due to obscuration is $\approx 20\%$.
We apply this correction factor throughout this work for all the \lx\ values of X-ray detected sources in the COSMOS sample when calculating the AGN fraction. 
The obscuration correction factor is obtained in a similar manner for the 4XMM sample, utilizing the XMMFITCAT-Z spectral fit catalag \citep{Ruiz2021} that provides intrinsic \lx\ measurements for 3XMM-DR6 sources. We match X-ray sources in our sample with sources in the XMMFITCAT-Z catalog that have spec-$z$ available. With this matched sample, we derive the obscuration correction factor needed by obtaining the average value of the intrinsic \lx\ reported divided by \lx\ calculated in this work; \lx-dependent weights are applied to the matched sample to recover the \lx\ distribution of the X-ray AGNs in the whole 4XMM sample. We found that the overall underestimation of X-ray emission due to obscuration is $\approx 10\%$.
We apply this correction factor throughout this work for all the \lx\ values of X-ray detected sources in the 4XMM sample when calculating the AGN fraction.\footnote{We note that the change in AGN fraction  associated with applying the obscuration correction factor is generally much smaller than the statistical uncertainty of AGN fraction. Thus, X-ray absorption should not bias our results materially.}

When deriving the AGN fraction, we have also taken into account the varying sensitivity of \xray\ observations that provide \xray\ coverage to our samples.
For the 4XMM sample, we derive the sensitivity upper limit of the relevant RapidXMM HealPix from the background level reported at the position of each MPA-JHU galaxy.
Following Equations 3 and 4 of \citet{Chen2018}, we derive the minimum number of counts required for a source to be detected in the 0.2--12 keV band given the background level, and derive the corresponding flux sensitivity with the corresponding energy conversion factor (ECF) which are derived assuming a power-law spectrum with $\Gamma = 1.7$.\footnote{\url{https://www.cosmos.esa.int/web/xmm-newton/epic-upper-limits}}
Since the 4XMM catalog uses {\sc DET\_ML} $= 6$ in the 0.2--12 keV band as the source detection criterion, we set the probability of the detected source being a random Poisson fluctuation due to the background as $2.5 \times 10^{-3}$ when utilizing the equations.
For the COSMOS sample, we derive the sensitivity map following the method in \citet{Aird2017}.

Combining with the redshift information, the lower limit of the \lx\ of a source in order to be detected, \lxlimit, can be obtained for every galaxy in both the 4XMM sample and the COSMOS sample.
We note that a power-law model with Galactic absorption and $\Gamma = 1.7$ is assumed through the whole conversion process.
In the right panels of Figure~\ref{lxm_distribution}, we present the \lxlimit/\mstar\ distribution of galaxies in the 4XMM and COSMOS samples.
We only derive AGN fraction utilizing galaxies with log \lxlimit/\mstar\ $\leqslant$ 32, i.e. these where we have the sensitivities to detect an AGN with log \lx/\mstar\ $>$ 32, if it exists in the given galaxy:
\begin{align}
f_{{\rm AGN, log}~L_{\rm X}/M_\bigstar > 32} = \frac{N_{{\rm det, log}~L_{\rm X}/M_\bigstar > 32}}{N_{{\rm galaxy, log}~L_{\rm X, limit}/M_\bigstar \leqslant 32}}.
\end{align}
The uncertainty of the AGN fraction is obtained via bootstrapping the sample (i.e. randomly drawing the same number of objects from the sample with replacement) 1000 times. For each bootstrapped sample, the AGN fraction is calculated, and the 16th and 84th percentiles of the obtained AGN fraction distribution give the estimation of the 1$\sigma$ uncertainty. When no AGN is detected in a sample, we report the $1\sigma$ confidence upper limits derived following \citet{Cameron2011}.

We also estimate the \textit{long-term average BH growth} from \bhar\ of a given sample of galaxies sharing similar properties, following the method described in \citet{Ni2019, Ni2021} that includes contributions from both X-ray detected sources and X-ray undetected sources.
We apply the obscuration correction factor mentioned earlier (1.1 for the 4XMM sample and 1.2 for the COSMOS sample) when taking into account the X-ray emission from X-ray detected sources.\footnote{We note that the change in \bhar\ associated with applying the obscuration correction factor is generally smaller than the statistical uncertainty of \bhar. Thus, X-ray absorption should not bias our results materially.}
The \xray\ emission of a group of X-ray undetected sources is taken into account via X-ray stacking techniques in the 0.2--12 keV band for the 4XMM sample and the 0.5--7 keV band for the COSMOS sample.   
With the source counts rate and background counts rate reported in RapidXMM, the net 0.2--12~keV count rate at each galaxy position in the 4XMM sample can be obtained, which is then converted to the 0.2--12 keV flux with the corresponding energy conversion factor derived assuming a power-law spectrum with $\Gamma = 1.7$.
For the COSMOS sample, the stacking process is described in \citet{Ni2021}, which gives the 0.5--7 keV net count rate/flux at each galaxy position.
We derive the average X-ray luminosity $\overline{L_{\rm X,stack}}$ from the average flux and the average redshift of the stacked sample.

With \lx\ for individual X-ray detected sources and $\overline{L_{\rm X,stack}}$ for X-ray undetected sources, we can obtain sample-averaged AGN bolometric luminosity following Equation 3 of \cite{Ni2021} assuming the \lx-dependent bolometric correction from \citet{Hopkins2007}:
\begin{equation}\label{equ:lx}
\overline{L_{\rm bol}} = \frac{ \bigg[{{\sum\limits_{n=0}^{N_{\rm det}}}}  (L_{\rm X} -L_{\rm X,XRB}) k_{\rm bol}\bigg]+ (\overline{L_{\rm X, stack}} -  \overline{L_{\rm X, XRB}} )N_{\rm non}  \overline{k_{\rm bol}}}{N_{\rm det}+N_{\rm non}}
\end{equation}
We also subtract the contributions from \xray\ binaries (XRBs) from \lx\ and $\overline{L_{\rm X,stack}}$ before applying the bolometric correction. The contributions from XRBs are generally small compared to the overall X-ray luminosity.
Then, sample-averaged AGN bolometric luminosity can be converted to \bhar\ adopting a constant radiative efficiency of 0.1 following Equation 4 in \citet{Ni2021}.
The uncertainty of \bhar\ is obtained via bootstrapping the sample 1000 times. For each bootstrapped sample, \bhar\ is calculated, and the 16th and 84th percentiles of the obtained \bhar\ distribution give the estimation of the 1$\sigma$ uncertainty associated with \bhar\ of the sample.

\section{Analysis Results} \label{s-ar}

\subsection{AGN fraction and \bhar\ as a function of \dn} \label{ss-agn-dn}

\begin{figure*}
\begin{center}
\includegraphics[scale=0.44]{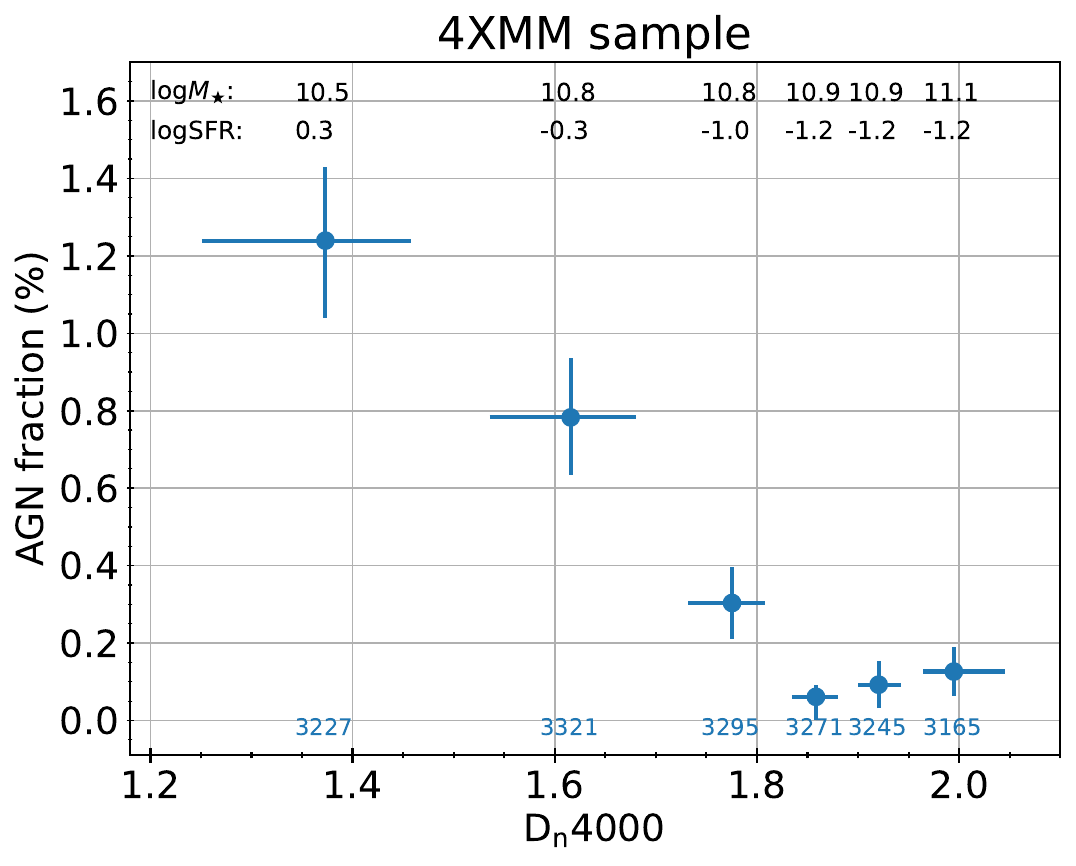}~~~~~~
\includegraphics[scale=0.44]{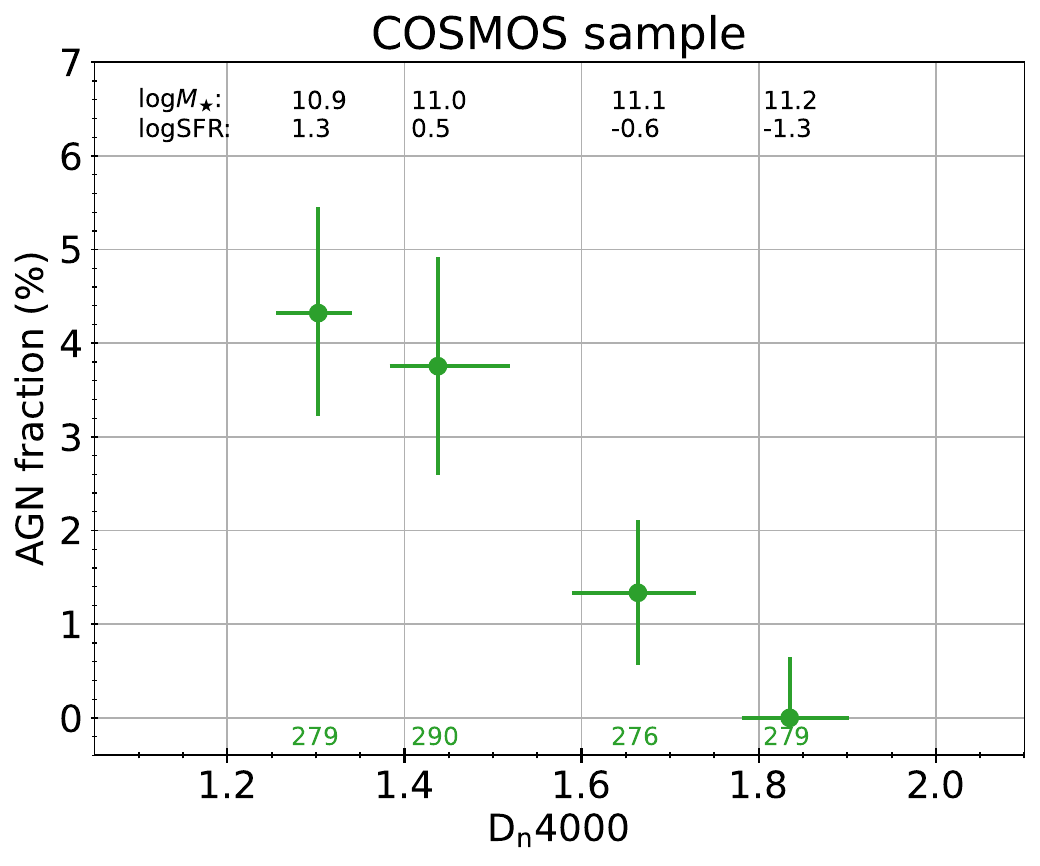}\\
(a)~~~~~~~~~~~~~~~~~~~~~~~~~~~~~~~~~~~~~~~~~~~~~~~~~~~~~~~~~~~~~~~~~~~~~~~~~~~~~~~~~~~~~~~~~~~~~~~~~~~~~~~~~~~~~~~~~~~(d)\\
\includegraphics[scale=0.44]{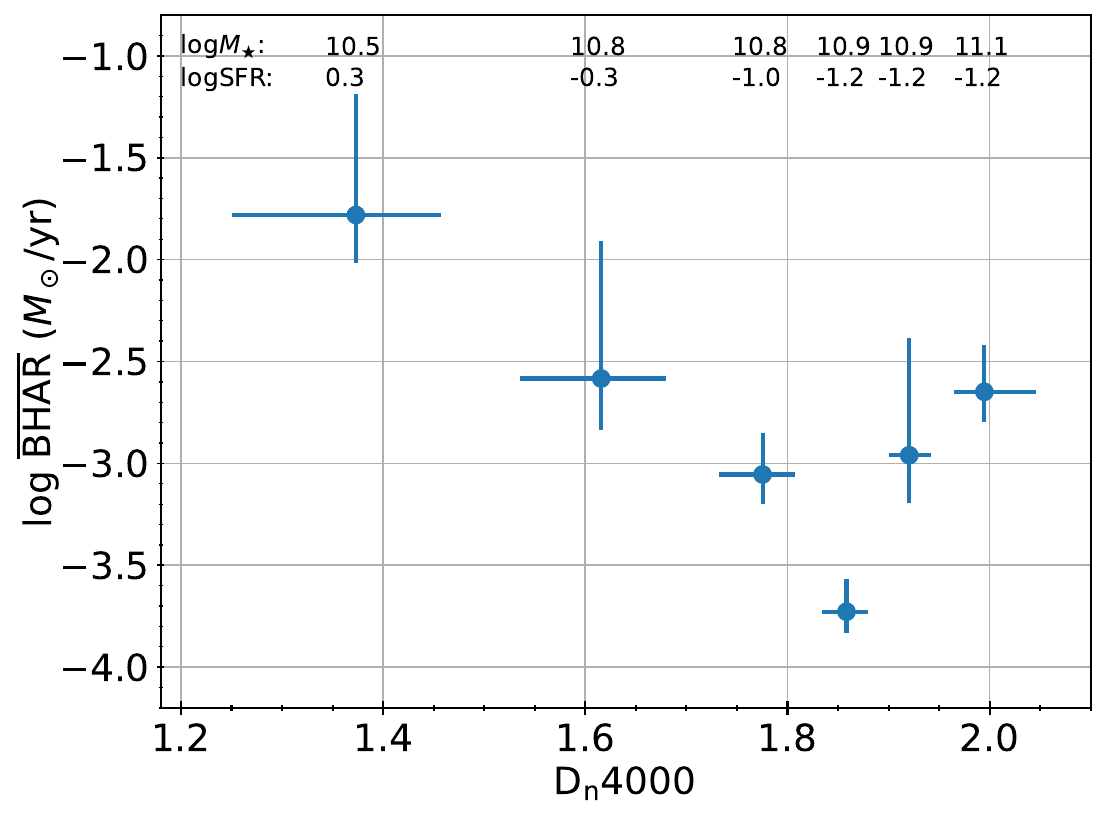}
\includegraphics[scale=0.44]{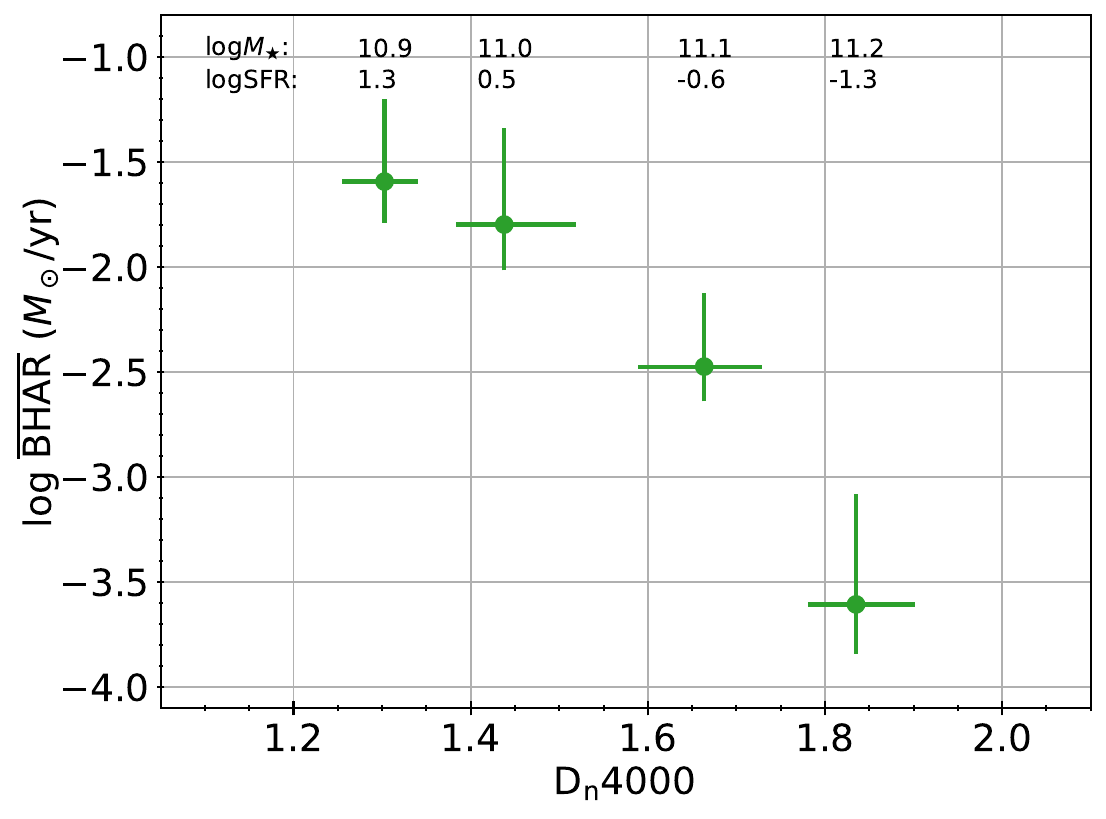}\\
(b)~~~~~~~~~~~~~~~~~~~~~~~~~~~~~~~~~~~~~~~~~~~~~~~~~~~~~~~~~~~~~~~~~~~~~~~~~~~~~~~~~~~~~~~~~~~~~~~~~~~~~~~~~~~~~~~~~~~(e)\\
\includegraphics[scale=0.44]{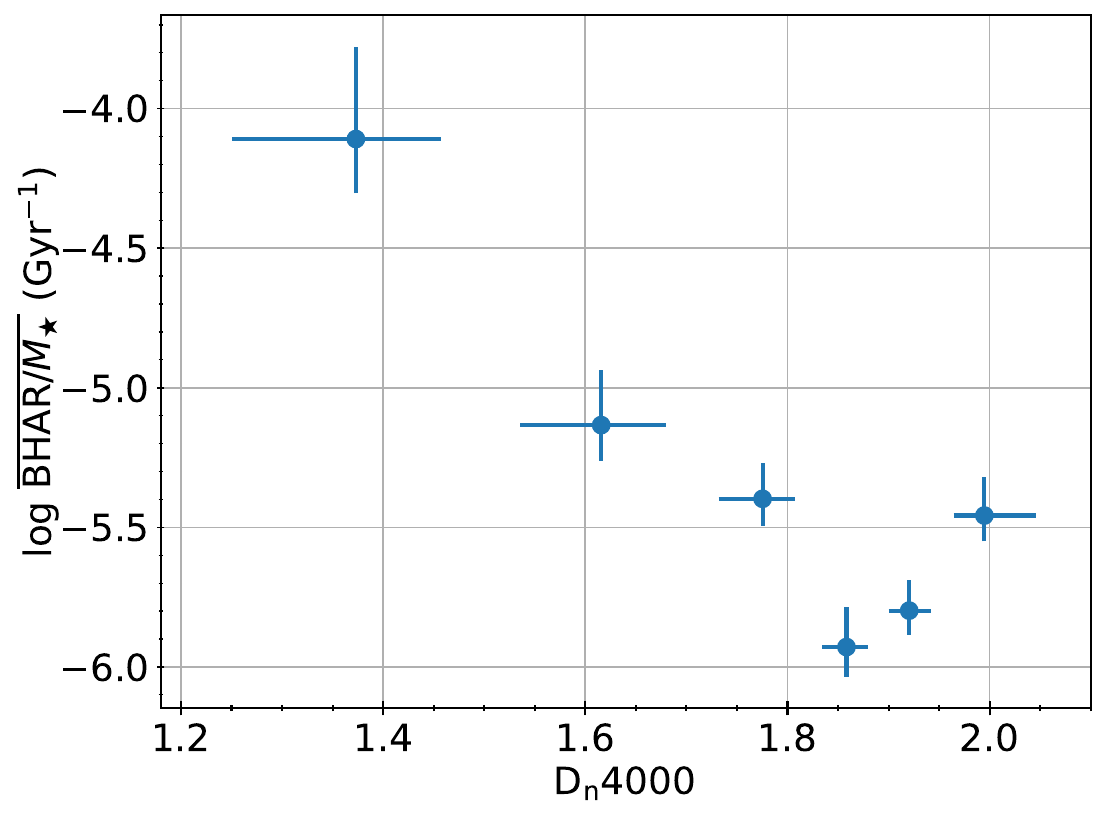}
\includegraphics[scale=0.44]{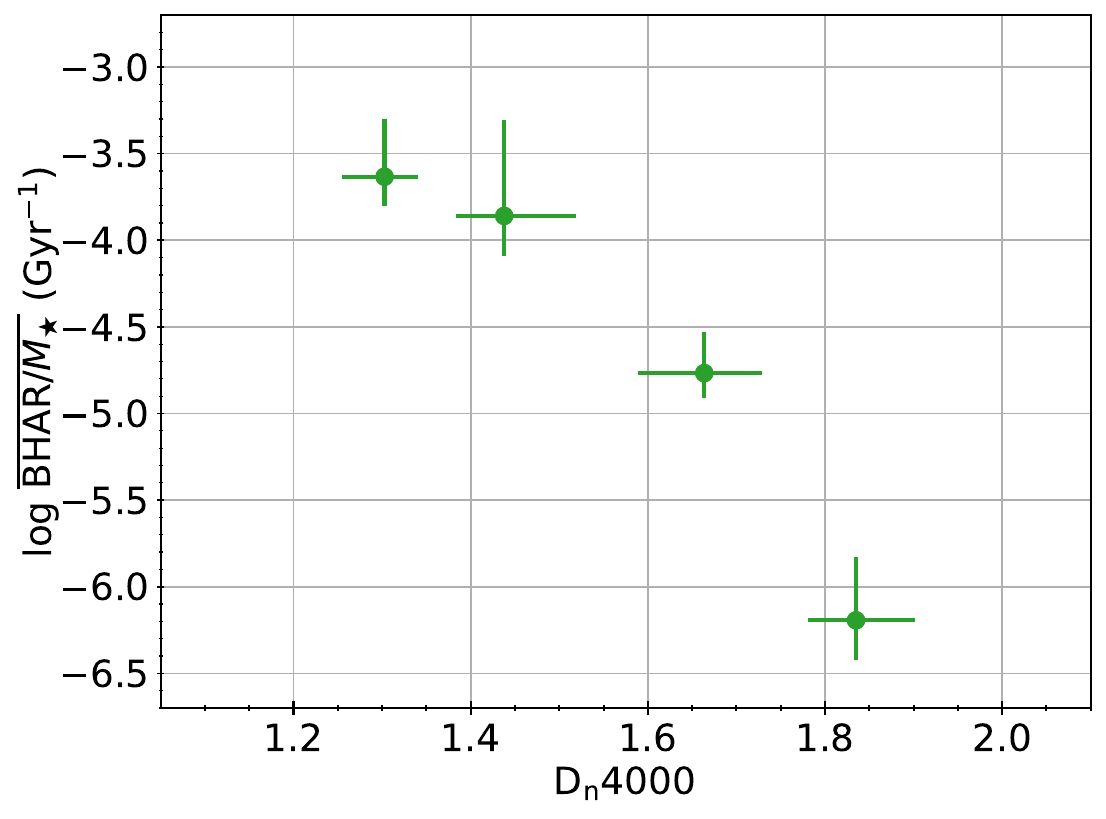}\\
(c)~~~~~~~~~~~~~~~~~~~~~~~~~~~~~~~~~~~~~~~~~~~~~~~~~~~~~~~~~~~~~~~~~~~~~~~~~~~~~~~~~~~~~~~~~~~~~~~~~~~~~~~~~~~~~~~~~~~(f)\\
\caption{\textit{(a):} AGN fraction as a function of \dn\ among galaxies in the 4XMM sample.
The horizontal position of each data point represents the median \dn\ of the sources in each bin, with $x$-axis error bars demonstrating the 16th and 84th percentiles of the \dn\ values in each bin.
The $y$-axis error bars represent the 1$\sigma$ confidence interval of AGN fraction from bootstrapping. 
We also list the median log \mstar\ and log SFR of each bin on the top of the plot.
The numbers in the bottom of the plot represent the number of galaxies in each bin used to derive the AGN fraction (with log \lxlimit/\mstar\ $\leqslant 32$).
\textit{(b):} \bhar\ as a function of \dn\ among galaxies in the 4XMM sample.
The horizontal position of each data point represents the median \dn\ of the sources in each bin, with $x$-axis error bars demonstrating the 16th and 84th percentiles of the \dn\ values in each bin.
The $y$-axis error bars represent the 1$\sigma$ confidence interval of \bhar\ from bootstrapping. 
We also list the median log \mstar\ and log SFR of each bin on the top of the plot.
\textit{(c):} \bharm\ a function of \dn\ among galaxies in the 4XMM sample. 
The horizontal position of each data point represents the median \dn\ of the sources in each bin, with $x$-axis error bars demonstrating the 16th and 84th percentiles of the \dn\ values in each bin.
The $y$-axis error bars represent the 1$\sigma$ confidence interval of \bharm\ from bootstrapping.
\textit{(d):} Similar to panel~(a), but for AGN fraction as a function of \dn\ among galaxies in the COSMOS sample.
\textit{(e):} Similar to panel~(b), but for \bhar\ as a function of \dn\ among galaxies in the COSMOS sample.
\textit{(f):} Similar to panel~(c), but for \bharm\ as a function of \dn\ among galaxies in the COSMOS sample.
}
\label{agnf_age}
\end{center}
\end{figure*}

For objects in the 4XMM sample, we bin them into 6 bins with equal number of objects per bin according to their \dn\ values.
Figure~\ref{agnf_age}a shows that the AGN fraction presents a clear decreasing trend with \dn\ at \dn\ $\lesssim 1.85$, and slightly increases at \dn\ $\gtrsim 1.85$.
We note that this result will not be materially affected by the \lx/\mstar\ threshold we adopt (see Appendix~\ref{a-lxpdf} for details).
We further plot \bhar\ as a function of \dn\ for the 4XMM sample (see Figure~\ref{agnf_age}b).
We can see that \bhar\ also decreases as \dn\ increases from $\sim1.25$ to $\sim1.85$. 
At \dn\ $\gtrsim$~1.85, the trend appears to reverse, and \bhar\ significantly increases with increasing \dn.
To account for differences in the average \mstar\ for galaxies across our \dn\ bins, we also plot \bharm\ as a function of \dn\ (see Figure~\ref{agnf_age}c), which shows an overall similar trend as \bhar.

Similarly, we bin objects in the COSMOS sample into 4 bins according to their \dn\ values, with equal number of objects per bin.
For the COSMOS sample, when we calculate AGN fraction as well as \bhar\ for a given subsample in this study, we weight the contribution from each object by the \texttt{SCOR} parameter in the LEGA-C catalog, which accounts for the selection effects from the parent sample; we also verified that the analysis results do not vary materially when we do not weight each object by \texttt{SCOR}.
Figure~\ref{agnf_age}d shows that, at the \dn\ range we probe (\dn\ $\approx$ 1.25--1.9), AGN fraction decreases with \dn\ in general.
In terms of \bhar\ as well as \bharm\ (see Figure~\ref{agnf_age}e and Figure~\ref{agnf_age}f), similarly, a decreasing trend in general is observed with increasing \dn.
We note that, at \dn\ $\lesssim$ 1.5, this decreasing trend of AGN fraction, \bhar, and \bharm\ is not very significant -- the large error bars as a result of the limited sample size prohibits us from drawing any significant conclusion.

\begin{figure*}
\begin{center}
\includegraphics[scale=0.45]{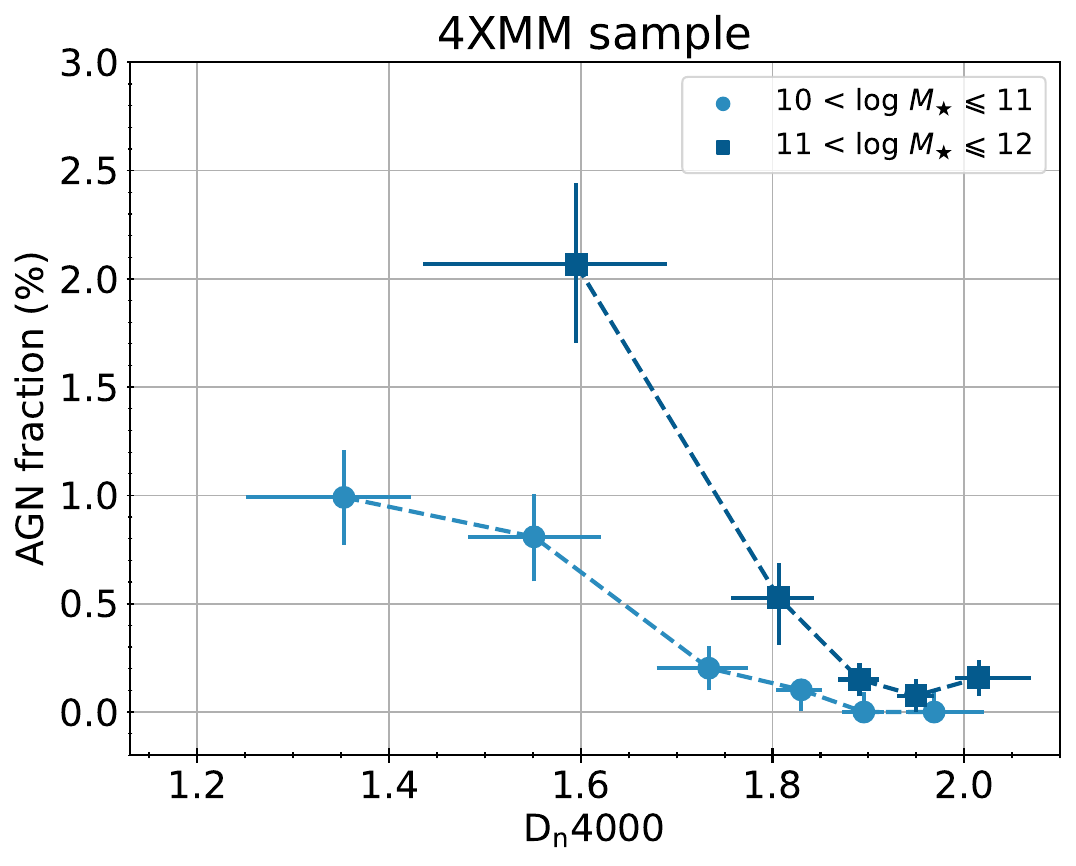}~~~~
\includegraphics[scale=0.45]{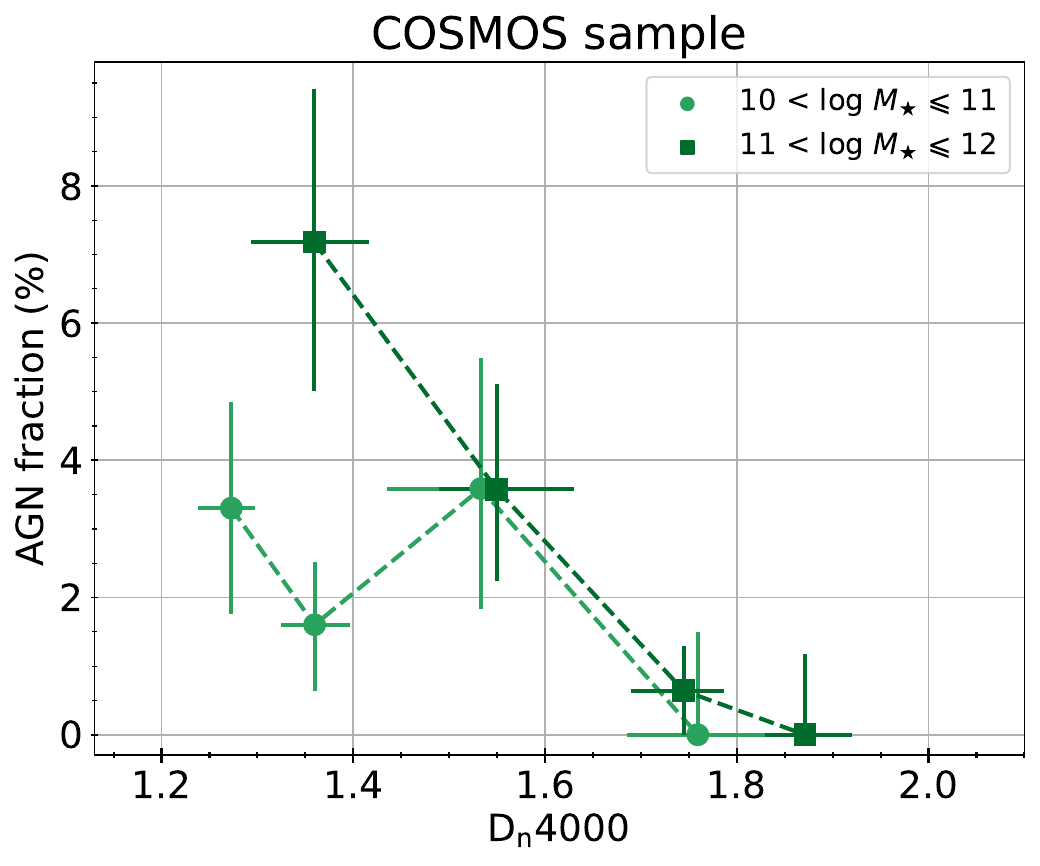}\\
(a)~~~~~~~~~~~~~~~~~~~~~~~~~~~~~~~~~~~~~~~~~~~~~~~~~~~~~~~~~~~~~~~~~~~~~~~~~~~~~~~~~~~~~~~~~~~~~~~~~~~~~~~~~~~~~~~~~~~(d)\\
\includegraphics[scale=0.45]{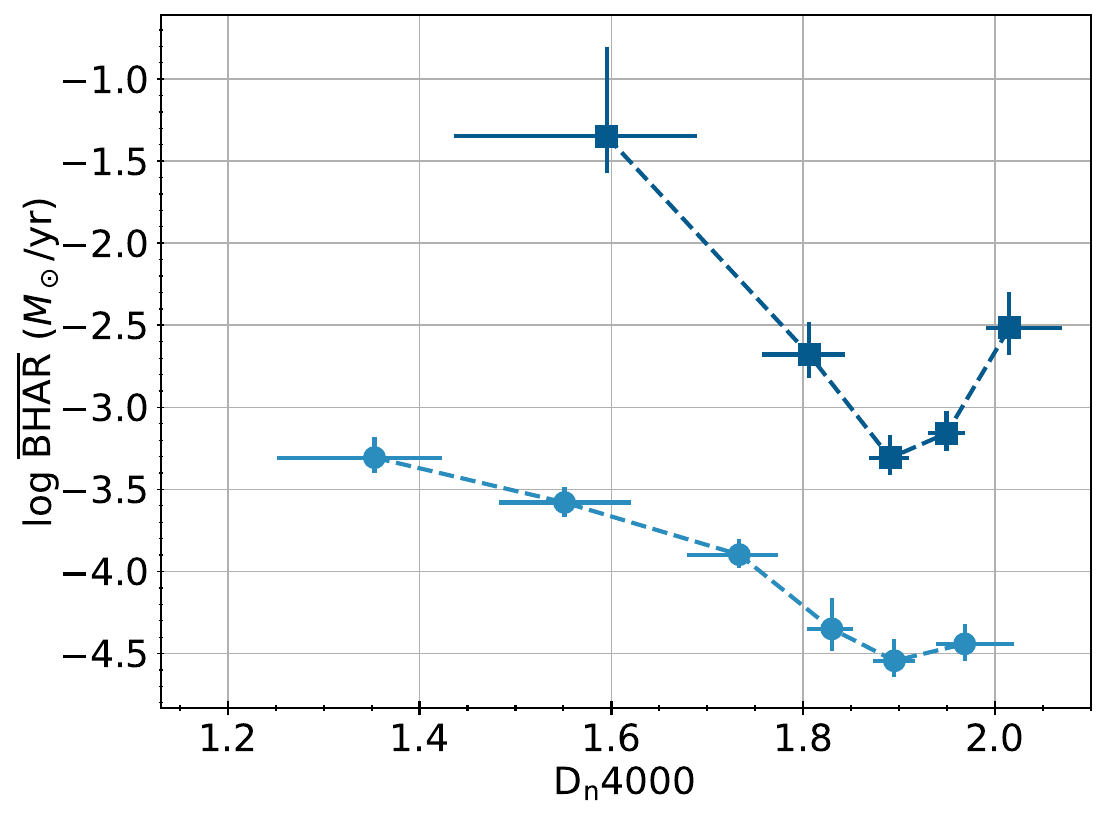}
\includegraphics[scale=0.45]{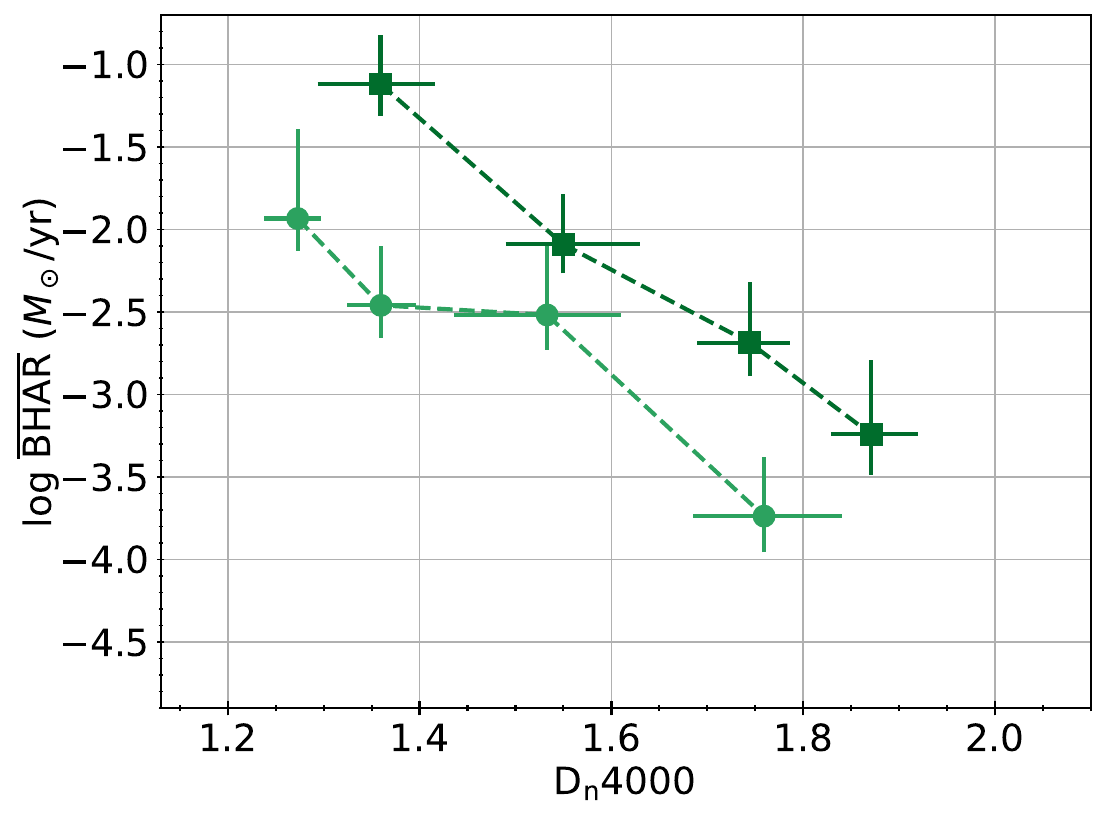}\\
(b)~~~~~~~~~~~~~~~~~~~~~~~~~~~~~~~~~~~~~~~~~~~~~~~~~~~~~~~~~~~~~~~~~~~~~~~~~~~~~~~~~~~~~~~~~~~~~~~~~~~~~~~~~~~~~~~~~~~(e)\\
\includegraphics[scale=0.45]{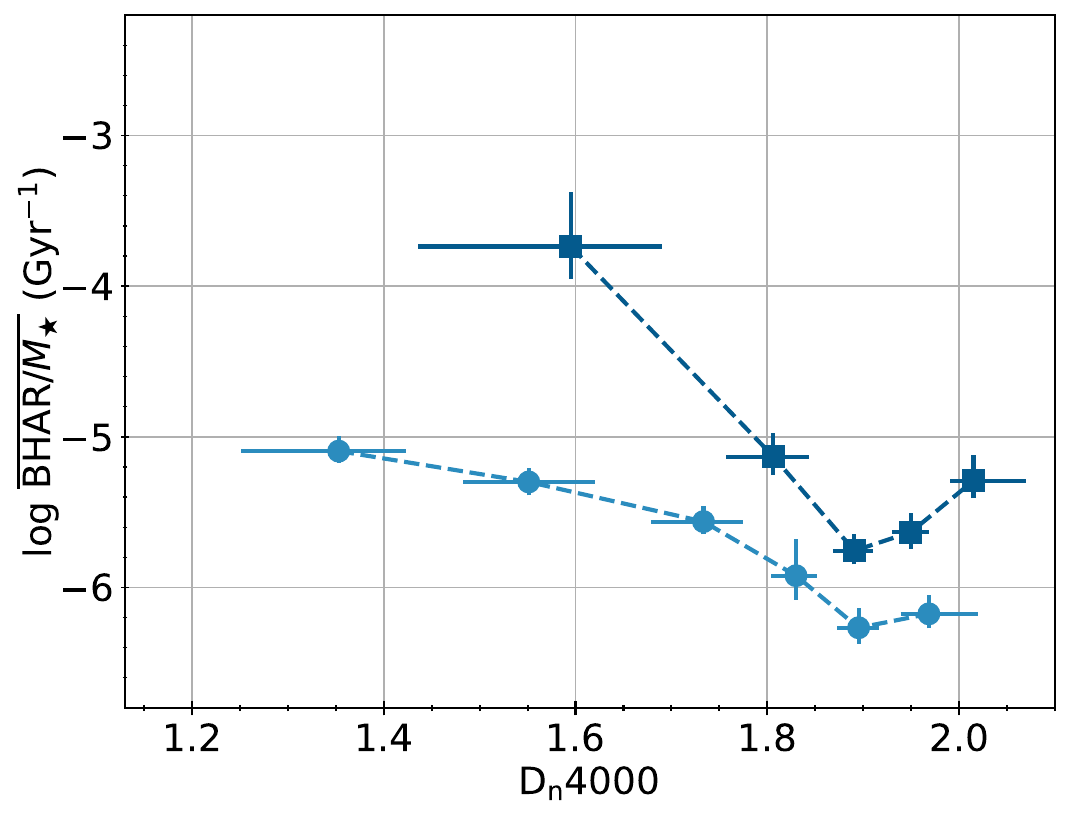}
\includegraphics[scale=0.45]{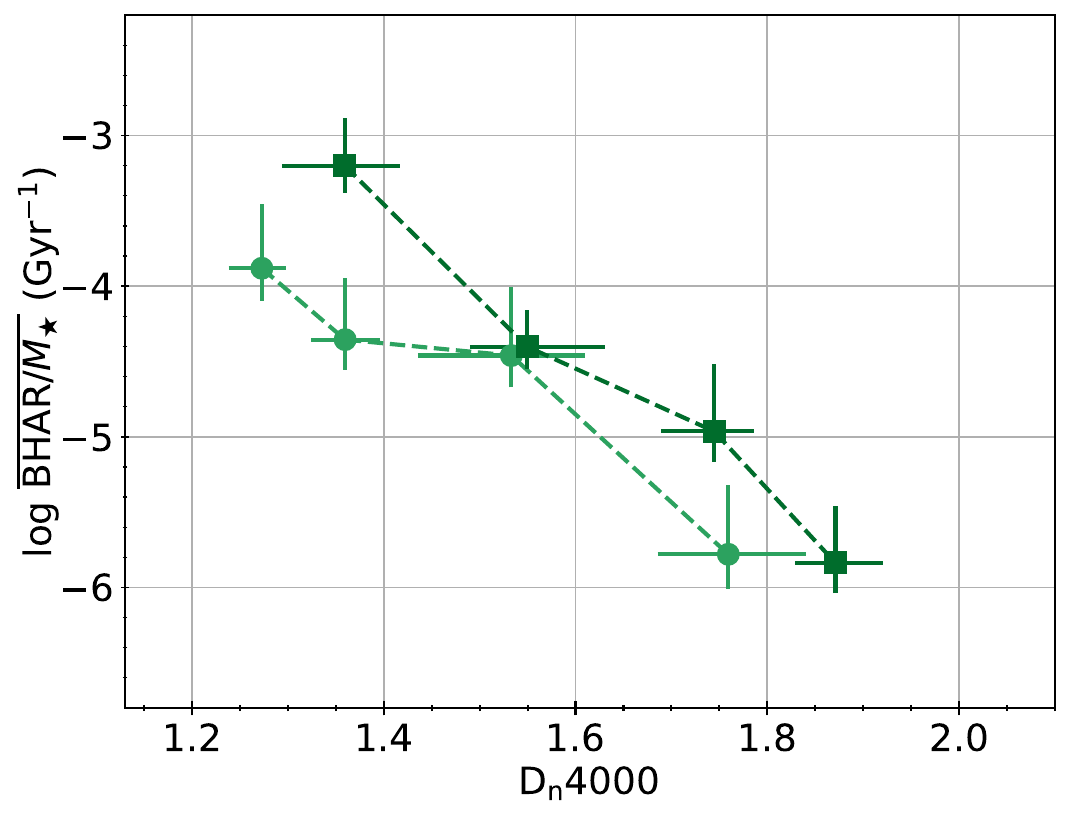}\\
(c)~~~~~~~~~~~~~~~~~~~~~~~~~~~~~~~~~~~~~~~~~~~~~~~~~~~~~~~~~~~~~~~~~~~~~~~~~~~~~~~~~~~~~~~~~~~~~~~~~~~~~~~~~~~~~~~~~~~(f)\\
\caption{\textit{(a):} AGN fraction as a function of \dn\ among galaxies in two \mstar\ bins (represented by different colors and symbols) in the 4XMM sample.
The horizontal position of each data point represents the median \dn\ of the sources in each bin, with $x$-axis error bars demonstrating the 16th and 84th percentiles of the \dn\ values in each bin.
The $y$-axis error bars represent the 1$\sigma$ confidence interval of AGN fraction from bootstrapping. 
\textit{(b):} \bhar\ as a function of \dn\ among galaxies in the 4XMM sample.
The horizontal position of each data point represents the median \dn\ of the sources in each bin, with $x$-axis error bars demonstrating the 16th and 84th percentiles of the \dn\ values in each bin.
The $y$-axis error bars represent the 1$\sigma$ confidence interval of \bhar\ from bootstrapping. 
\textit{(c):} \bharm\ as a function of \dn\ among galaxies in the 4XMM sample. 
The horizontal position of each data point represents the median \dn\ of the sources in each bin, with $x$-axis error bars demonstrating the 16th and 84th percentiles of the \dn\ values in each bin.
The $y$-axis error bars represent the 1$\sigma$ confidence interval of \bharm\ from bootstrapping.
\textit{(d):} Similar to panel~(a), but for AGN fraction as a function of \dn\ among galaxies in two \mstar\ bins in the COSMOS sample.
\textit{(e):} Similar to panel~(b), but for \bhar\ as a function of \dn\ among galaxies in two \mstar\ bins in the COSMOS sample.
\textit{(f):} Similar to panel~(c), but for \bharm\ as a function of \dn\ among galaxies in two \mstar\ bins in the COSMOS sample.
}
\label{agnf_age_mbins}
\end{center}
\end{figure*}

We further investigate AGN fraction and \bhar\ in different \mstar\ ranges for both the 4XMM sample and the COSMOS sample; the results can be seen in Figure~\ref{agnf_age_mbins}.
We can see that for different \mstar\ ranges in the 4XMM sample, AGN fraction, \bhar, and \bharm\ also decrease with \dn\ at \dn\ $\lesssim$ 1.9.
The increasing trend of \bhar\ at larger \dn\ is more prominent among objects with $11 <$ log~\mstar\ $\leqslant 12$.
For different \mstar\ ranges in the COSMOS sample, we also observe decreasing trends of AGN fraction, \bhar, and \bharm\  with \dn, though we note that for \dn\ $\lesssim$ 1.5 objects with $10 <$ log~\mstar\ $\leqslant 11$, the decreasing trend is not very significant, which might be caused by the limited sample size.

\subsection{AGN fraction and \bhar\ as a function of \dn\ when controlling for other host-galaxy parameters} \label{ss-agn-dn-ms}

We note that, while in Section~\ref{ss-agn-dn} we characterized the incidence of X-ray AGNs among galaxies with different stellar ages (as \dn\ is closely associated with the age of the stellar populations), we can still not quantify how the difference in \dn\ (which indicates the difference in the mean stellar population age) directly affects AGN activity and BH growth, as host-galaxy properties such as \mstar\ and SFR vary across different \dn\ bins, which are known to closely related with AGN activity and BH growth.

\begin{figure}
\begin{center}
\includegraphics[scale=0.45]{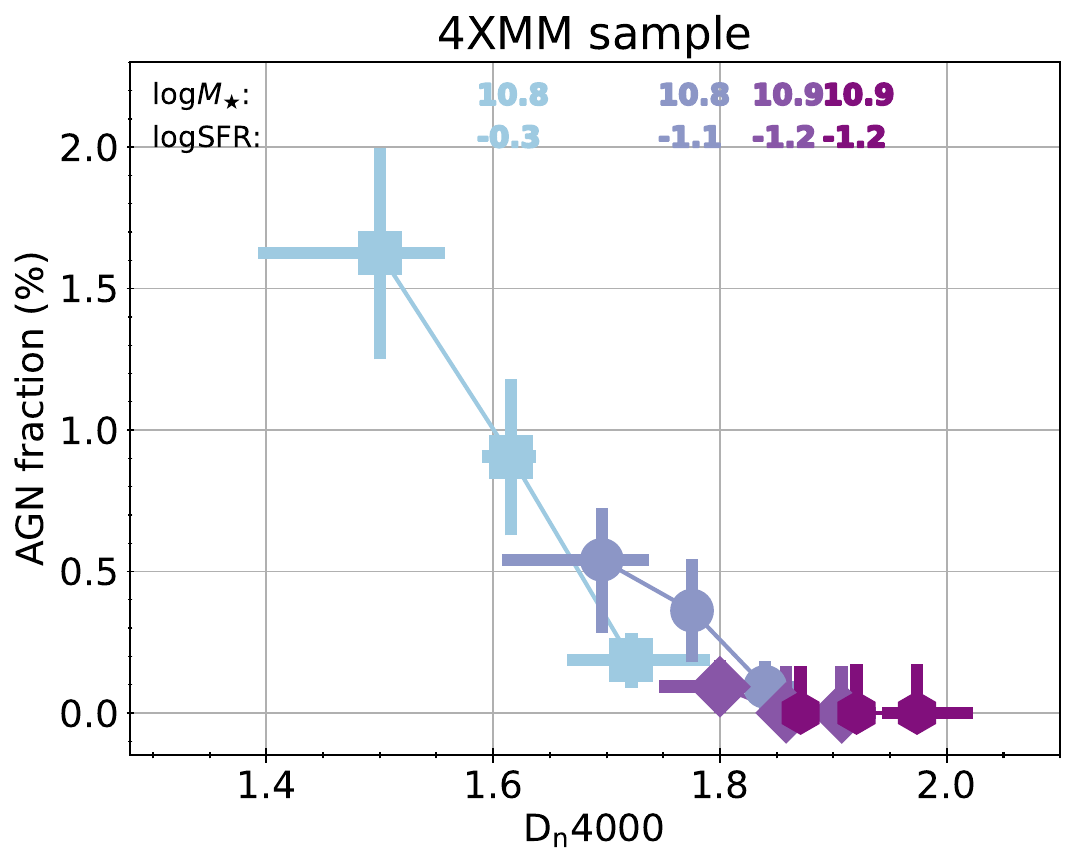}
\includegraphics[scale=0.44]{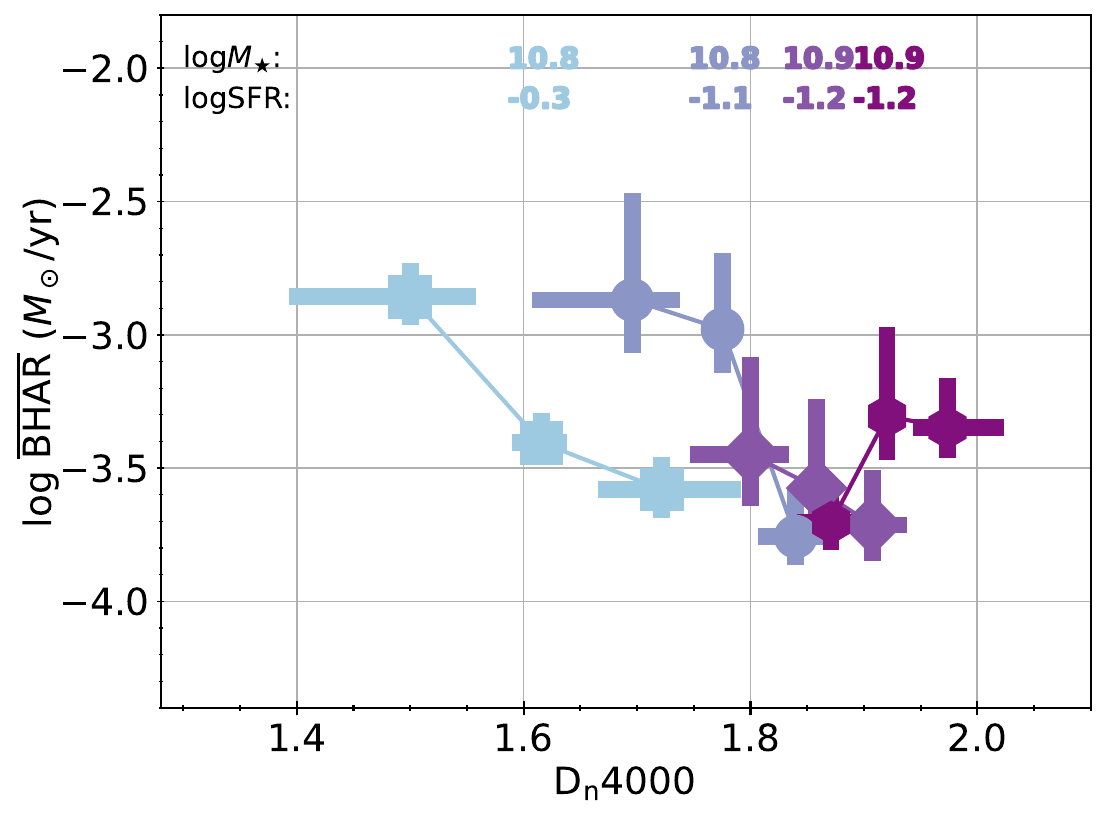}
\includegraphics[scale=0.44]{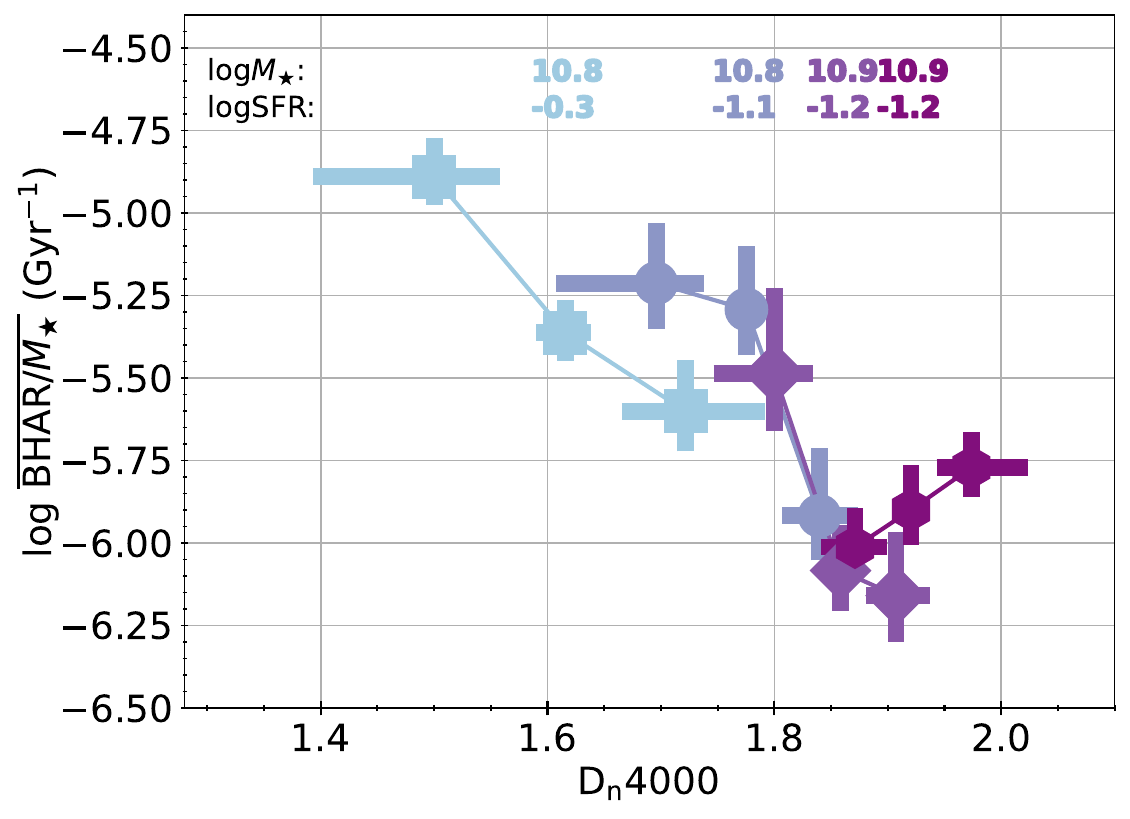}
\caption{\textit{Top}: AGN fraction as a function of \dn\ among galaxies in the 4XMM sample when controlling for \mstar, SFR, and $z$.
Different symbols and colors represent a set of subsamples with similar \mstar, SFR, and $z$ values (as listed on top of the panel with the same color). The horizontal position of each data point represents the median \dn\ of the sources in each sample, with $x$-axis error bars demonstrating the 16th and 84th percentiles of the \dn\ values.
The $y$-axis error bars represent the 1$\sigma$ confidence interval of AGN fraction from bootstrapping. 
\textit{Middle:} Similar to the top panel, but for \bhar\ as a function of \dn\ among galaxies in the 4XMM sample when controlling for \mstar, SFR, and $z$.
\textit{Bottom:} Similar to the top panel, but for \bharm\ as a function of \dn\ among galaxies in the 4XMM sample when controlling for \mstar, SFR, and $z$.
}
\label{agnf_age_rw}
\end{center}
\end{figure}

We thus would like to study AGN fraction and \bhar\ as a function of \dn\ when controlling for other host-galaxy parameters.
To achieve this, for each bin (except the first and last bins) of galaxies in the 4XMM sample in Figure~\ref{agnf_age}, we sort this subsample with their \dn\ values, and keep the central 1/3 of the objects to form a reference bin.\footnote{While choosing this relatively small subsample size reduces the statistical power compared to the larger bin size used in Section 3.1, these narrower bins are necessary to ensure sufficient sources with similar \mstar, SFR and $z$ values in the adjacent bins that are used to create the comparison samples.} The first 1/3 of the objects are merged with the bin on the left, and the last 1/3 of the objects are merged with the bin on the right. We then select the nearest neighbour of objects in the reference bin among objects in its left/right bin in the \mstar, SFR, and $z$ space utilizing the \texttt{NearestNeighbors} algorithm in the \texttt{scikit-learn} python package, to constitute two comparison samples with similar \mstar, SFR, and $z$ properties, but one with smaller \dn\ and one with larger \dn. In Figure~\ref{agnf_age_rw}, we show AGN fractions of all these subsamples, with each set of subsamples sharing similar \mstar, SFR, and $z$ values represented by different colors and symbols. Comparing within each subsample set reveals how AGN fraction varies with \dn\ when controlling for other host-galaxy parameters, and a significant decreasing trend is observed within each subsample set at \dn\ $\lesssim$ 1.9.\footnote{We note that while AGN fraction varies with \dn, it also varies when other host-galaxy properties change, so that at a fixed \dn, the AGN fraction from different subsamples differs due to the differences in \mstar\ and SFR.}
We also demonstrate that our results hold when defining the AGN fraction by an \lx\ limit of $10^{42}$ \lum\ in the Appendix~\ref{a-lx}.
We show \bhar\ and \bharm\ of all these subsamples as well in Figure~\ref{agnf_age_rw},  with each set of subsamples sharing similar \mstar, SFR, and $z$ values represented by different colors and symbols.
Comparing within each subsample set shows that \bhar\ and \bharm\ also decrease with \dn\ at \dn\ $\lesssim 1.9$ when controlling for \mstar, SFR, and $z$.\footnote{While X-ray emission from normal star-forming galaxies has potential dependence on stellar ages \citep[e.g.][]{Gilbertson2022}, the XRB contribution is little among the objects we investigated ($\sim$ 5--15\%), so should not bias our results.} At \dn\ $\gtrsim 1.9$, \bhar\ and \bharm\ increase with \dn\ when controlling for \mstar, SFR, and $z$.

\begin{figure}
\begin{center}
\includegraphics[scale=0.45]{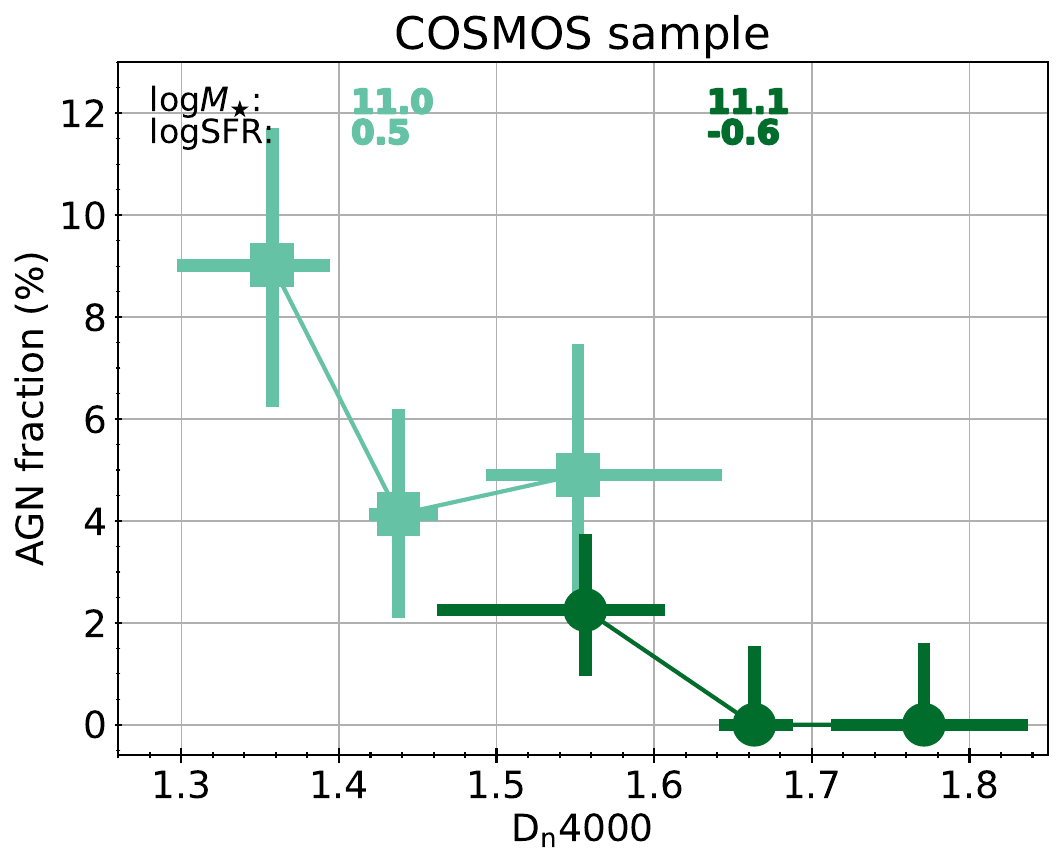}
\includegraphics[scale=0.44]{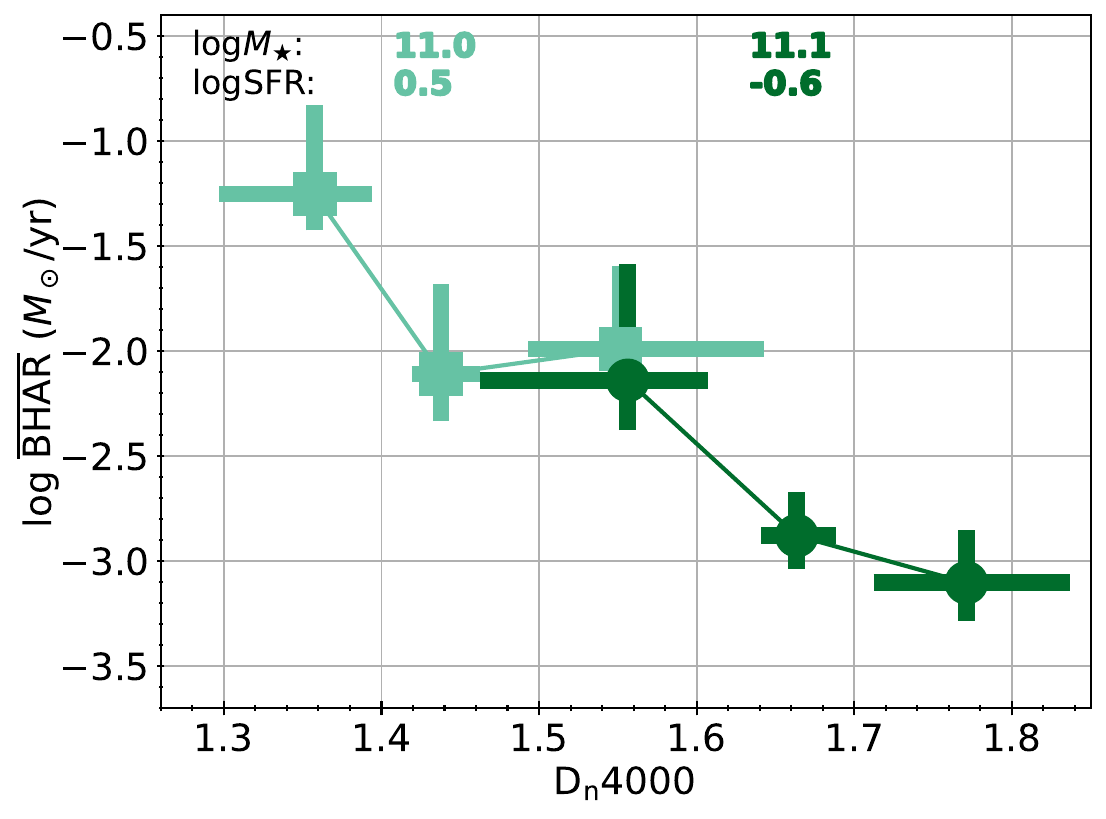}
\includegraphics[scale=0.44]{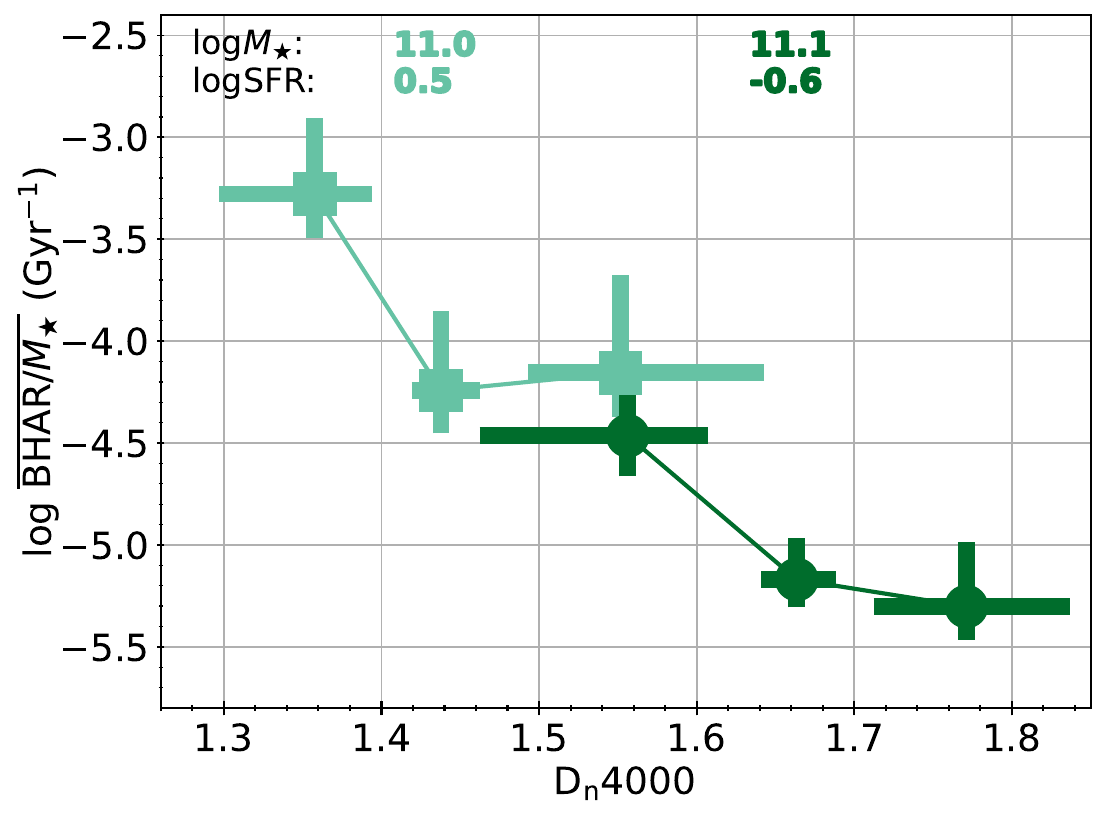}
\caption{\textit{Top}: Similar to the top panel of~Figure~\ref{agnf_age_rw}, but for AGN fraction as a function of \dn\ when controlling for \mstar, SFR, and $z$ among galaxies in the COSMOS sample.
\textit{Middle:} Similar to the middle panel of~Figure~\ref{agnf_age_rw}, but for \bhar\ as a function of \dn\ when controlling for \mstar, SFR, and $z$ among galaxies in the COSMOS sample.
\textit{Bottom:} Similar to the middle panel of~Figure~\ref{agnf_age_rw}, but for \bharm\ as a function of \dn\ when controlling for \mstar, SFR, and $z$ among galaxies in the COSMOS sample.
}
\label{legac_agnf_age_rw}
\end{center}
\end{figure}

Similarly, we study how AGN fraction and BH growth vary with \dn\ within subsample sets when controlling for other host-galaxy parameters for the COSMOS sample.
The results are shown in Figure~\ref{legac_agnf_age_rw}.
We also observe a trend of decreasing AGN fraction, \bhar\ and \bharm\ with \dn, similar to that seen at lower redshifts in our 4XMM sample, although at  \dn\ $\lesssim 1.5$ this trend has a relatively low significance level.

We also show that host morphological properties do not affect the results in this subsection materially in Appendix~\ref{a-morph}.

\subsection{The incidence of low-accretion-rate AGN as a function of \dn\ compared with log \lx/\mstar\ $>$ 32 AGN} \label{ss-loweddagn}

\begin{figure}
\begin{center}
\includegraphics[scale=0.4]{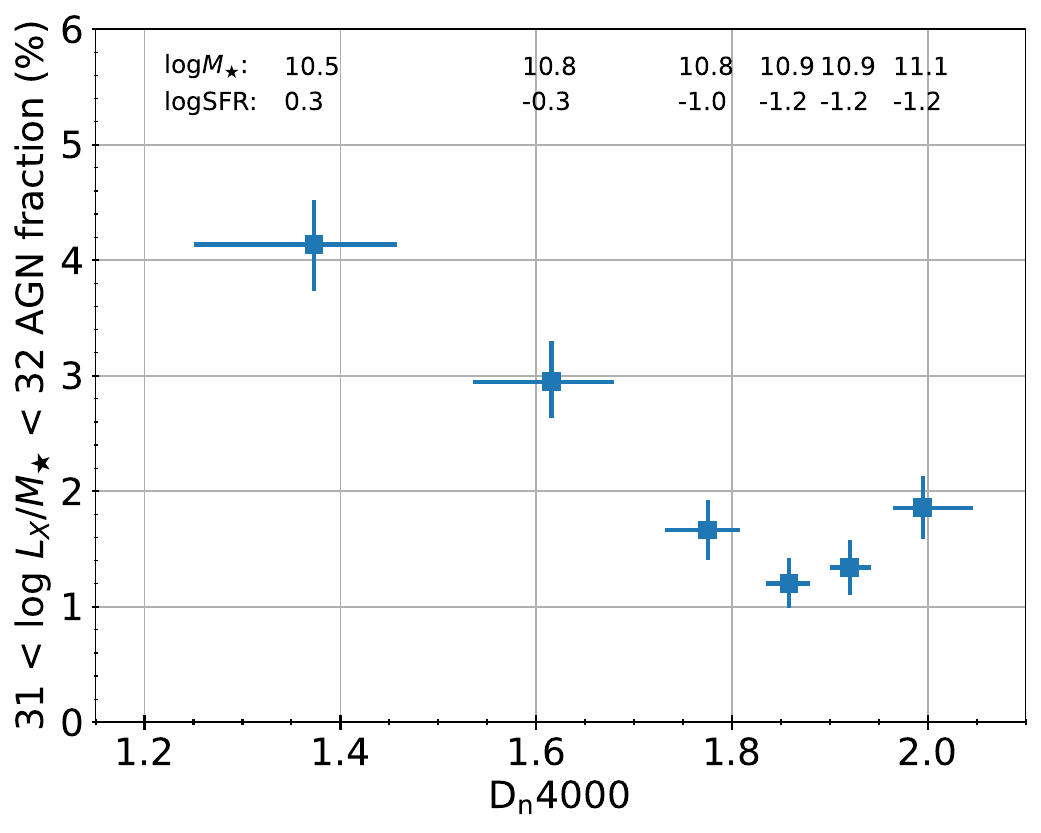}
\caption{31 $<$ log \lx/\mstar\ $<$ 32 AGN fraction as a function of \dn\ among galaxies in the 4XMM sample.
The horizontal position of each data point represents the median \dn\ of the sources in each bin, with $x$-axis error bars demonstrating the 16th and 84th percentiles of the \dn\ values in each bin.
The $y$-axis error bars represent the 1$\sigma$ confidence interval of AGN fraction from bootstrapping. 
We also list the median log \mstar\ and log SFR value of each bin on the top of the plot.
}
\label{lowedd_dn}
\end{center}
\end{figure}

In Section~\ref{ss-agn-dn}, we find that for objects in the 4XMM sample, while \bhar\ and \bharm\ significantly increase with \dn\ at \dn\ $\gtrsim 1.85$, the AGN fraction does not increase significantly, which might be caused by our definition for AGNs, as we only look at log \lx/\mstar\ $>$ 32 AGNs in Section~\ref{ss-agn-dn} and do not take low-accretion-rate AGNs into account when calculating the AGN fraction. This inspired us to examine the incidence of low-accretion-rate AGN as a function of \dn. As stated in Section~\ref{ss-agnf}, we only consider an X-ray detected source as AGN when \lx\ is greater than the contribution from XRBs. As can be seen in Figure~\ref{lxm_distribution}, we have a considerable number of AGNs with log \lx/\mstar\ $\leqslant$ 32 detected in the 4XMM sample, and there are also a large number of galaxies in the 4XMM sample with log \lxlimit/\mstar\ $\leqslant$ 32. We thus use these objects to study how the fraction of AGNs with 31 $<$ log \lx/\mstar\ $<$ 32 varies as a function of \dn, and the result is shown in Figure~\ref{lowedd_dn}. We can see that the fraction increases at \dn\ $\gtrsim$ 1.85 obviously. 

\begin{figure}
\begin{center}
\includegraphics[scale=0.4]{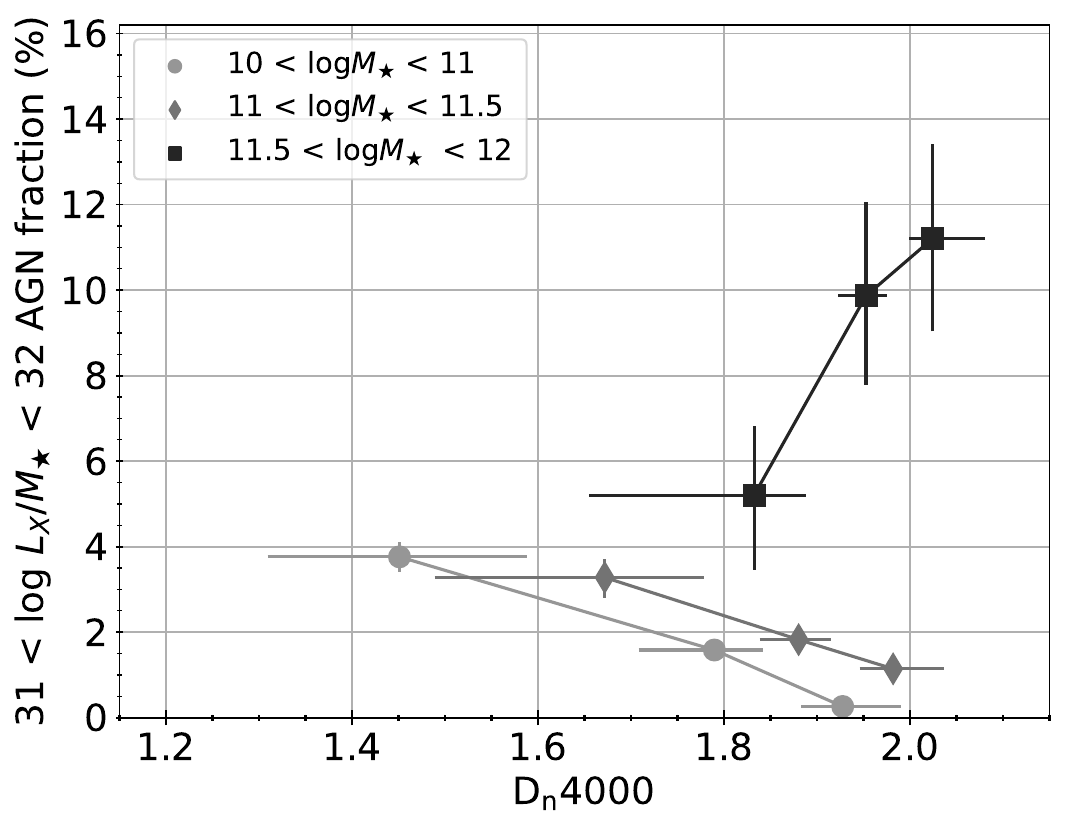}
\caption{31 $<$ log \lx/\mstar\ $<$ 32 AGN fraction as a function of \dn\ among galaxies with different \mstar\ ranges in the 4XMM sample; different symbols and colors represent subsamples of galaxies with different \mstar\ ranges.
The horizontal position of each data point represents the median \dn\ of the sources in each bin, with $x$-axis error bars demonstrating the 16th and 84th percentiles of the \dn\ values in each bin.
The $y$-axis error bars represent the 1$\sigma$ confidence interval of AGN fraction from bootstrapping. 
}
\label{lowedd}
\end{center}
\end{figure}

We further probe if this increase of low-accretion-rate AGN fraction with \dn\ is linked with \mstar, by studying the low-accretion-rate AGN fraction as a function of \dn\ in different \mstar\ bins.
As we can see in Figure~\ref{lowedd}, for galaxies/AGN with log~\mstar\ $<$ 11.5, AGNs tend to live among younger galaxies. In contrast, for galaxies/AGNs with log \mstar\ $>$ 11.5, the fraction of AGNs with 31 $<$  log \lx/\mstar\ $<$ 32 increases with \dn. 
When we perform the same analyses for log \lx/\mstar\ $>$ 32 AGNs, we found that the fractions of log \lx/\mstar\ $>$ 32 AGNs at different \mstar\ ranges all drop with \dn.

\subsection{Comparing with the incidence of AGNs selected at different wavelength bands} \label{ss-radiomir}

We note that our results in previous subsections are based on X-ray-selected AGNs. While X-ray selection is known to be able to provide the most complete and unbiased sample of AGNs, we would like to test how different AGN selection methods could potentially affect our results.
In this subsection, we select AGNs based purely on detection in a waveband in a given catalog, in contrast to the careful measurement of AGN fraction to specific black hole accretion rate limits (corrected for any incompleteness) adopted in our X-ray-based analyses above. Extending this more robust approach to MIR- and radio-selected samples is deferred to a future work. 

\subsubsection{Comparing with the incidence of AGNs selected at MIR}
\begin{figure}
\begin{center}
\includegraphics[scale=0.4]{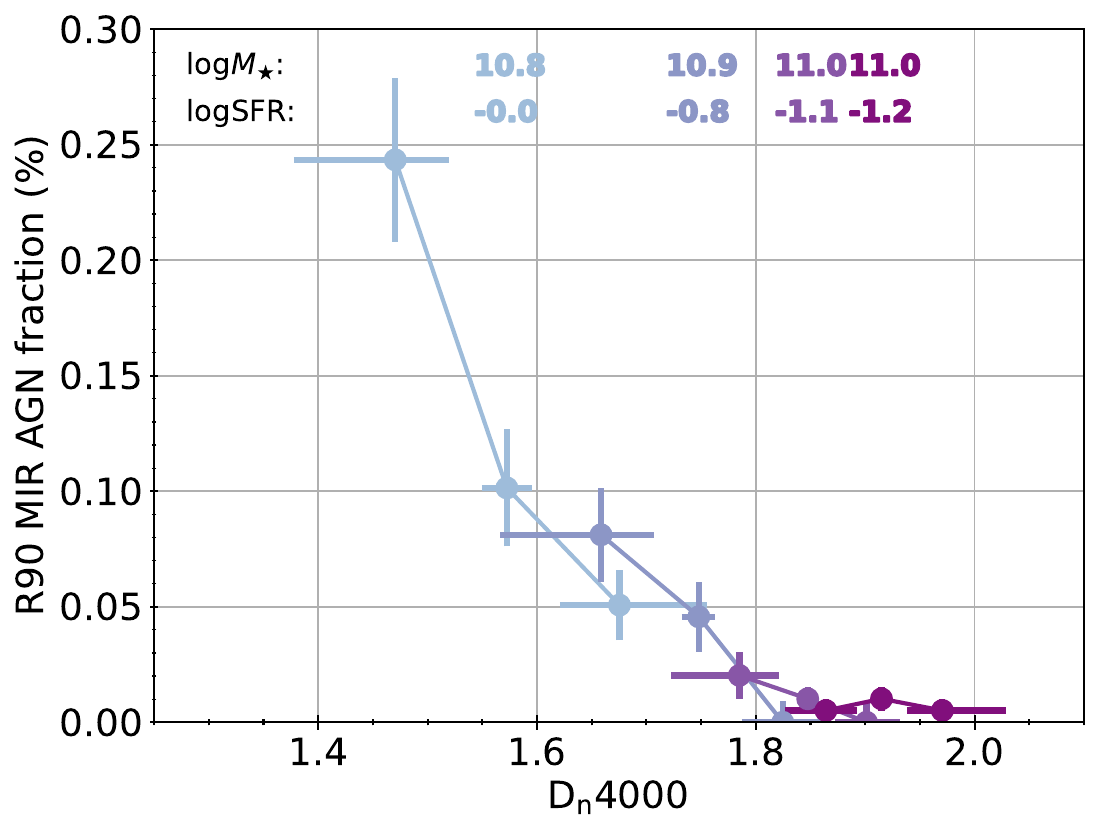}
\caption{MIR AGN fraction as a function of \dn\ among SDSS galaxies when controlling for \mstar, SFR, and $z$.
Different colors represent a set of subsamples with similar \mstar, SFR, and $z$ values (as listed on top of the panel with the same color). The horizontal position of each data point represents the median \dn\ of the sources in each sample, with $x$-axis error bars demonstrating the 16th and 84th percentiles of the \dn\ values.
The $y$-axis error bars represent the 1$\sigma$ confidence interval of AGN fraction from bootstrapping. 
}
\label{miragn}
\end{center}
\end{figure}

We obtain a MIR-selected AGN sample among MPA-JHU galaxies in the main SDSS galaxy sample from the R90 catalog of \citet{Assef2018}, which consists of AGN candidates with 90\% reliability.
We perform the same analyses as those in Section~\ref{ss-agn-dn-ms} to check how the fraction of MIR-selected AGNs varies with \dn\ when controlling for host-galaxy properties, and whether the observed trend is consistent with what we observed in the X-ray.
The analysis results are shown in Figure~\ref{miragn}.
We observe a similar trend as these in Section~\ref{ss-agn-dn-ms}: when controlling for host-galaxy properties, the fraction of MIR-selected AGNs decreases with \dn\ at \dn\ $\lesssim 1.9$.
As MIR-selected AGNs are generally biased against AGNs in most massive galaxies, the MIR-selected AGN fraction at high \dn\ in Figure~\ref{miragn} is too small to observe any trend.

\subsubsection{Comparing with the incidence of AGNs selected at radio wavelength}
\begin{figure*}
\begin{center}
\includegraphics[scale=0.3]{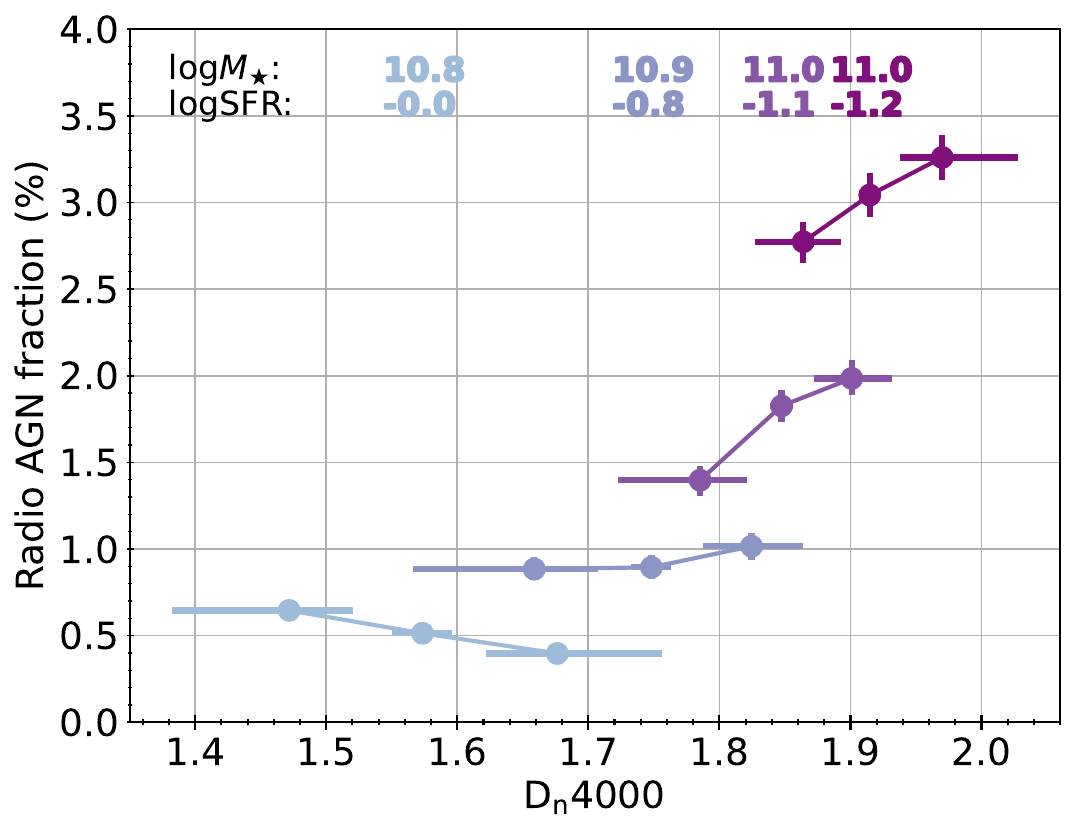}
\includegraphics[scale=0.3]{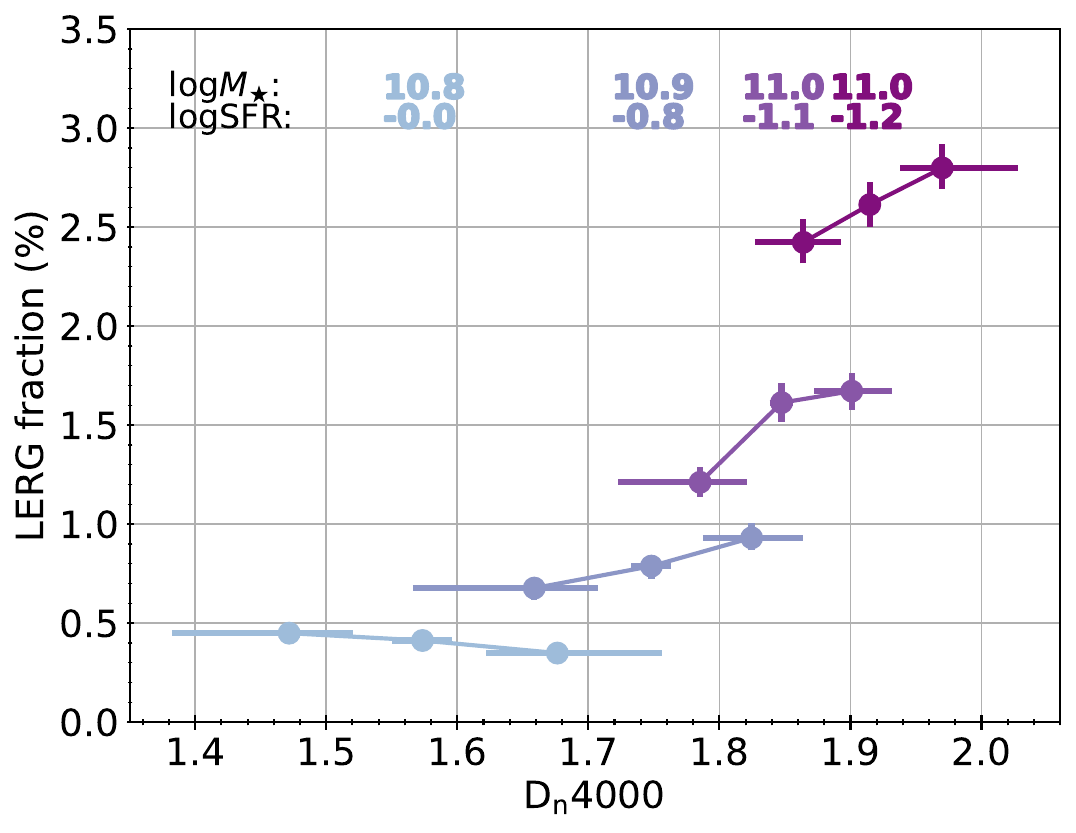}
\includegraphics[scale=0.3]{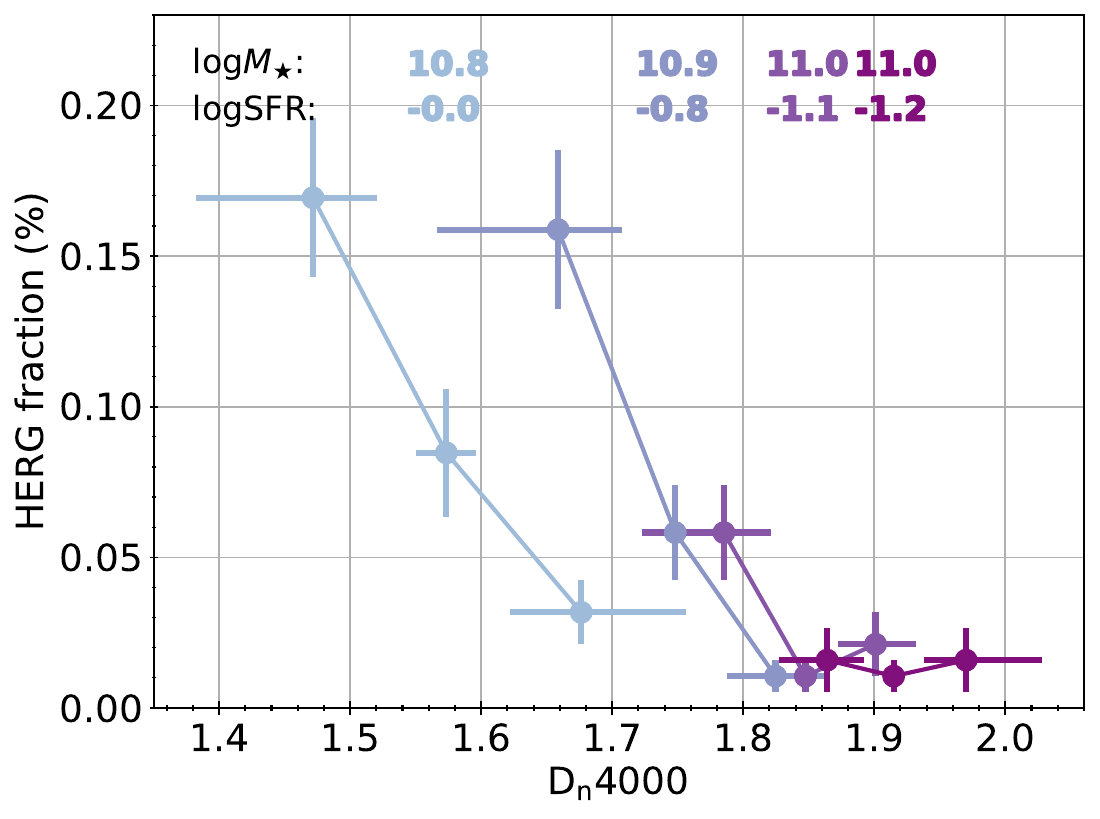}
\caption{\textit{Left panel:} Radio AGN fraction as a function of \dn\ among SDSS galaxies when controlling for \mstar, SFR, and $z$.
Different colors represent a set of subsamples with similar \mstar, SFR, and $z$ values (as listed on top of the panel with the same color). The horizontal position of each data point represents the median \dn\ of the sources in each sample, with $x$-axis error bars demonstrating the 16th and 84th percentiles of the \dn\ values.
The $y$-axis error bars represent the 1$\sigma$ confidence interval of AGN fraction from bootstrapping. 
\textit{Middle panel:} Similar to the left panel, but for LERGs.
\textit{Right panel:} Similar to the left panel, but for HERGs.
}
\label{radioagn}
\end{center}
\end{figure*}

To study how the fraction of radio-selected AGNs varies with \dn, we utilize the radio AGN sample from \citet{BH2012}, constructed by combining SDSS data with the NRAO (National Radio Astronomy Observatory) VLA (Very Large Array) Sky Survey (NVSS) and the Faint Images of the Radio Sky at Twenty centimetres (FIRST) survey.
The classification of radio AGNs into high-excitation radio galaxies (HERGs) and low-excitation radio galaxies (LERGs) is available for this sample.
This radio AGN sample mainly consists of LERGs (that have small Eddington ratios; $<$ 1\%), with a small fraction of HERGs reported.
We match these radio AGNs to MPA-JHU galaxies in the main SDSS galaxy sample, and perform the same analyses as those in Section~\ref{ss-agn-dn-ms} to check how the fraction of radio-selected AGNs in general, HERGs, and LERGs, vary with \dn\ when controlling for host-galaxy properties.
The analysis results are shown in Figure~\ref{radioagn}.

Unlike X-ray-selected AGNs and MIR-selected AGNs, radio-selected AGNs in this sample (dominated by the LERG population) are more likely to be found among old galaxies with large \dn\ values. This has been known for decades, as the radio AGN fraction is strongly linked with \mstar\ \citep[e.g.][]{Best2005}.
As can be seen in the right panel of Figure~\ref{radioagn}, HERG fraction always decreases with \dn\ when controlling for host-galaxy properties at \dn\ $\lesssim$ 1.9, similar to what we found for the X-ray-selected AGN fraction.
In the middle panel of Figure~\ref{radioagn}, we can see that at \dn\ $\lesssim$ 1.7, the LERG fraction also decreases with \dn\ when controlling for \mstar, SFR, and $z$. In contrast, at high \dn\ values (\dn\ $\gtrsim$ 1.7), the LERG fraction increases with \dn\ when controlling for other parameters, similar to what we found among low-accretion-rate X-ray-selected AGNs among massive (log \mstar\ $\gtrsim$ 11.5) galaxies. 

\section{Discussions} \label{s-dc}

In Section~\ref{ss-agn-dn}, we characterize how the log \lx/\mstar\ $>$ 32 AGN fraction, \bhar, and \bharm\ vary with \dn\ at two different redshift ranges; in Section~\ref{ss-agn-dn-ms}, we found that when controlling for host-galaxy properties (\mstar, SFR, and $z$), the fraction of log \lx/\mstar\ $>$ 32 AGNs and \bhar\ decrease with \dn\ among galaxies with \dn\ $\lesssim$ 1.9. We discuss the potential reason for these findings in Section~\ref{ss-smloss}.
In Section~\ref{ss-loweddagn}, we found that among the most massive galaxies at low redshift, the fraction of 31 $<$ log \lx/\mstar\ $<$ 32 AGNs increases with \dn, and we discuss the potential reason for this as well as the increase of \bharm\ among the oldest galaxies at low redshift in Section~\ref{ss-loweddr}.

\subsection{Stellar mass loss as a potential fuel for BH growth} \label{ss-smloss}

\begin{figure*}
\begin{center}
\includegraphics[scale=0.45]{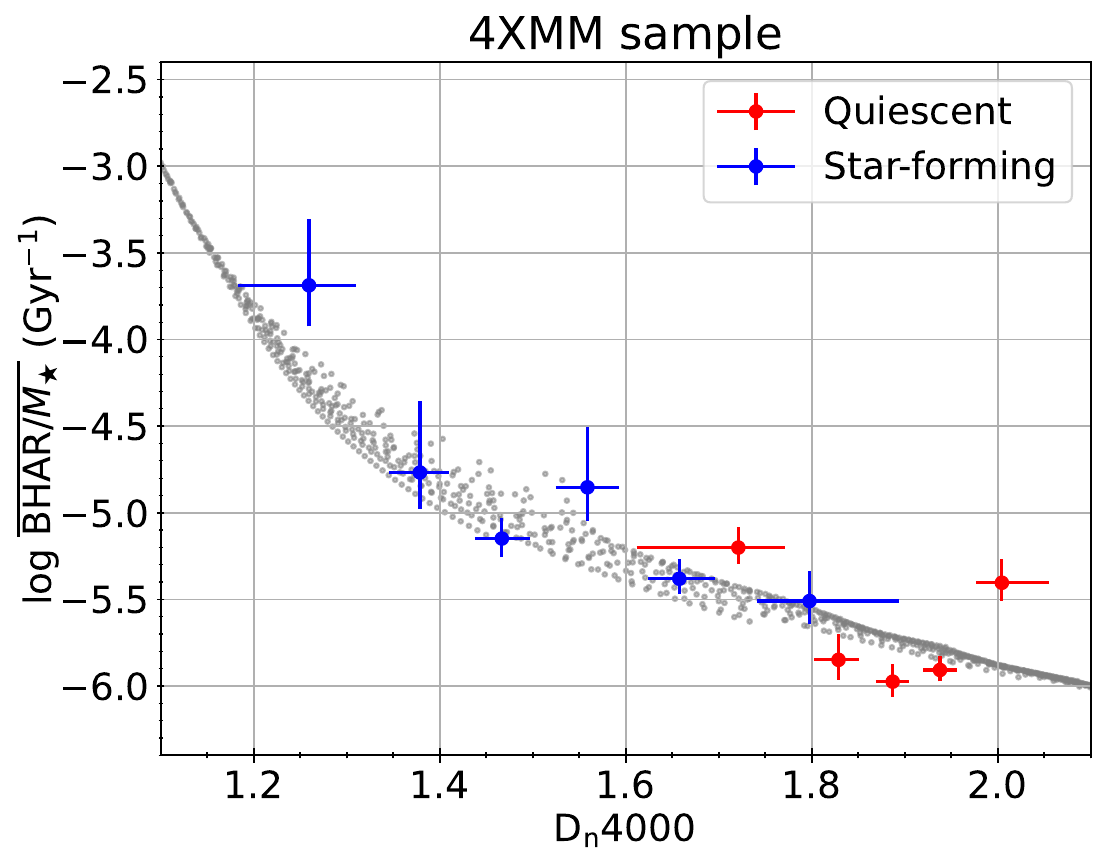}
\includegraphics[scale=0.45]{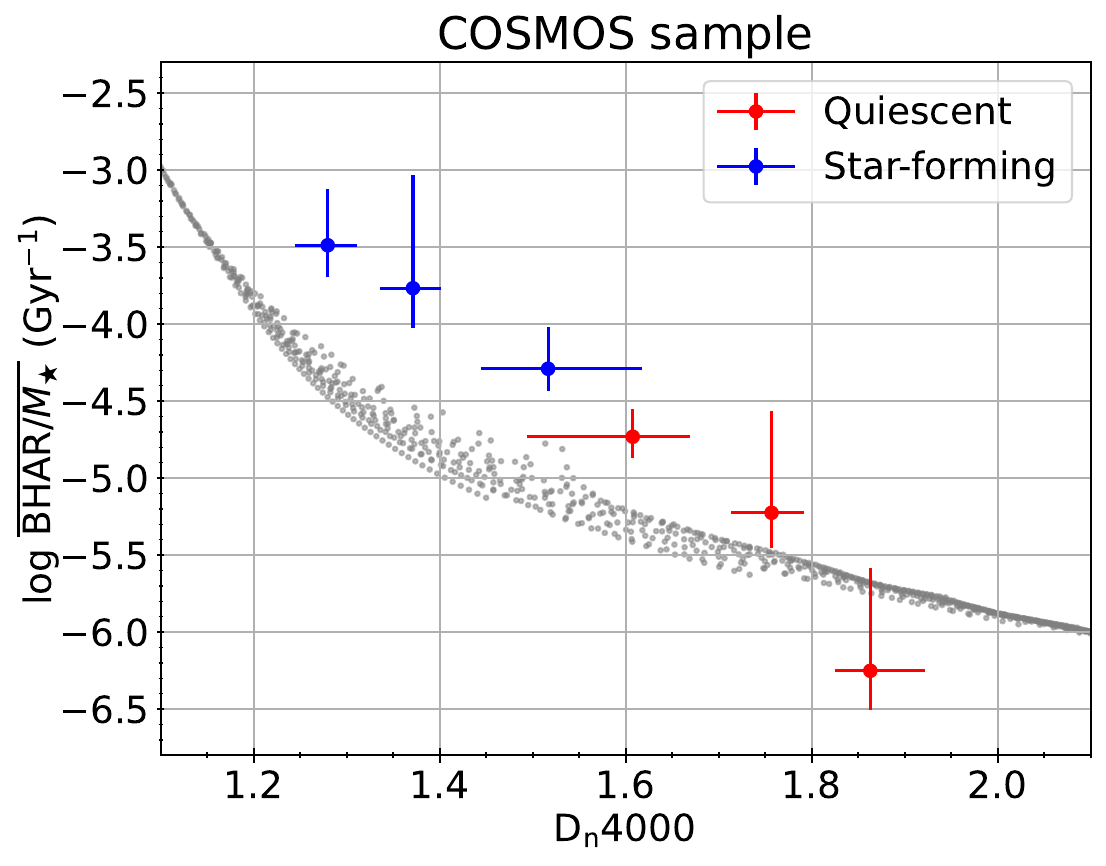}
\caption{\textit{Left panel:} \bharm\ as a function of \dn\ among star-forming/quiescent galaxies in the 4XMM sample, represented by the blue/red symbols. 
The horizontal position of each data point represents the median \dn\ of the sources in each bin, with $x$-axis error bars demonstrating the 16th and 84th percentiles of the \dn\ values in each bin.
The $y$-axis error bars represent the 1$\sigma$ confidence interval of \bharm\ from bootstrapping.
In the background, stellar mass loss rate per stellar mass unit per Gyr scaled by a factor of 1/5000 is shown as grey dots.
\textit{Right panel:} Similar to the left panel, but for \bharm\ as a function of \dn\ among star-forming/quiescent galaxies in the COSMOS sample.
}
\label{bharpermstar_qs}
\end{center}
\end{figure*}

We showed in Section~\ref{ss-agn-dn} that galaxies in both the 4XMM sample and COSMOS sample display a decrease of AGN fraction, \bhar, or \bharm\ with \dn\ at \dn\ $\lesssim$ 1.9, suggesting a higher level of AGN activity and BH growth among younger galaxies.
It is plausible that among younger galaxies, higher amounts of fuels are available for the central BHs.
\citet{Kauffmann2009} argue that stellar mass loss may serve as an important source of fuel for BHs when the cold gas supply is not plentiful. As younger galaxies have higher stellar mass loss rates, a higher level of AGN activity and BH growth among younger galaxies is expected if recycled gas from stellar mass loss serves as an important fuelling source, consistent with our finding.
We further plot \bharm\ among star-forming galaxies and quiescent galaxies in both the 4XMM sample and the COSMOS sample as a function of \dn\ separately in Figure~\ref{bharpermstar_qs}, and compare with stellar mass loss rate expected at a given \dn.\footnote{The separation between star-forming galaxies and quiescent galaxies is performed following the criterion in Section 2.4 of \citet{Ni2021}, which utilizes the star formation main sequence.}
We use {\sc python-FSPS} \citep{Conroy2009,Conroy2010} to predict how stellar mass loss rates vary as a function of \dn.
We adopt the universal initial mass function as parametrized by \citet{Chabrier2003}. Star-forming histories are generated using a range of formation times, with exponential decline time-scales or delayed-exponential time-scales ranging from 0.1 Gyr to 3 Gyr.
In the background of Figure~\ref{bharpermstar_qs}, we show the predicted stellar mass loss rate (in units of the fraction of the stellar mass returned per Gyr) scaled by a factor of 1/5000 as a function of \dn\ for galaxies with different star-forming histories. 
We could see that \bhar\ per solar mass roughly traces the stellar mass loss per solar mass predicted by the FSPS model divided by a factor of $\sim 5000$ among almost all galaxies in the 4XMM sample (except for the \dn\ $\approx 2$ bin), indicating that stellar mass loss might be the fuelling source for these galaxies: not only for the quiescent galaxies, but also for the star-forming galaxies.
For the whole 4XMM sample, the log \bhar/${\overline {\rm SFR}}$ value is $\approx -2.5$, and the scaling factor of 5000 is consistent with this value assuming that stellar mass loss also serves as the major fuel for star formation and the fraction of gas that turns into stars is $\sim 0.03-0.4$ \citep[e.g.][]{Ciotti2007}.
As the galaxies in our 4XMM sample and COSMOS sample have similar median/mean \mstar\ and velocity dispersion, we assume that the fraction of stellar mass loss that can be captured by the central BH is similar for these two samples. If we compare the \bhar\ per solar mass in the COSMOS sample with the stellar mass loss rate scaled by a factor of 1/5000, we could see that \bharm\ only traces stellar mass loss rate well among the oldest galaxies in the COSMOS sample. 
Among most galaxies in the COSMOS sample, it is likely that the cold gas in the galaxy with origins other than the stellar mass loss is capable of serving as the major fuel for both the BH and the star formation process.

The decreasing trend of \xray\ AGN fraction and \bhar\ with \dn\ when controlling for other host-galaxy properties among most galaxies in the 4XMM sample and COSMOS sample observed in Section~\ref{ss-agn-dn-ms} further supports this scenario.
A similar trend is also observed in the case of MIR-selected AGN fraction and HERG fraction at low redshift (see Figures~\ref{miragn} and \ref{radioagn}).
As stellar population synthesis models predict that stellar mass loss rates decline as a function of mean stellar age (e.g. see the background dots in Figure~\ref{bharpermstar_qs}), we could see that the difference in X-ray AGN fraction (or \bhar) when controlling for \mstar, SFR, and $z$ is smaller at larger \dn. 
From the bottom panel of Figure~\ref{agnf_age_rw} (4XMM sample), we observe a $\sim$ 1$\times 10^{-5}$/5$\times 10^{-6}$/2$\times 10^{-6}$ Gyr$^{-1}$ drop in \bharm\ associated with a $\sim$ 3$\times 10^{-2}$/8$\times 10^{-3}$/4$\times 10^{-3}$ Gyr$^{-1}$ drop in stellar mass loss rate (inferred from the difference in \dn) in the first/second/third \dn\ subsample set.
In the fourth \dn\ subsample set, we observe an increase in \bhar\ when \dn\ increases.
From the bottom panel of Figure~\ref{legac_agnf_age_rw} (COSMOS sample), we observe a $\sim$ 5$\times 10^{-4}$/3$\times 10^{-5}$ Gyr$^{-1}$ drop in \bharm\ associated with a $\sim$ 9$\times 10^{-2}$/2$\times 10^{-2}$ Gyr$^{-1}$ drop in stellar mass loss rate in the first/second \dn\ subsample set.
Generally, higher levels of BH growth are associated with more stellar mass loss. At the same time, the relation between the difference in \bharm\ associated with the difference in stellar mass loss rate is not exactly linear, i.e. the predicted change in the stellar mass loss rate based on the change in \dn\ does not result in a consistent change in \bharm\ across all the subsamples used in Figures~\ref{agnf_age_rw} and \ref{legac_agnf_age_rw}. While stellar mass loss could be one potential fuelling mechanism, it is not always the dominant fuelling mechanism.
Also, the fraction of stellar mass loss that could be accreted by the central BH is likely to vary among galaxies, depending on other galaxy properties such as their total stellar masses (which is clearly indicated in Figure~\ref{agnf_age_mbins}: more massive galaxies tend to accrete stellar mass loss more efficiently, which is expected as they have larger potential wells and larger black hole masses), morphologies, and star formation histories.

\begin{figure*}
\begin{center}
\includegraphics[scale=0.45]{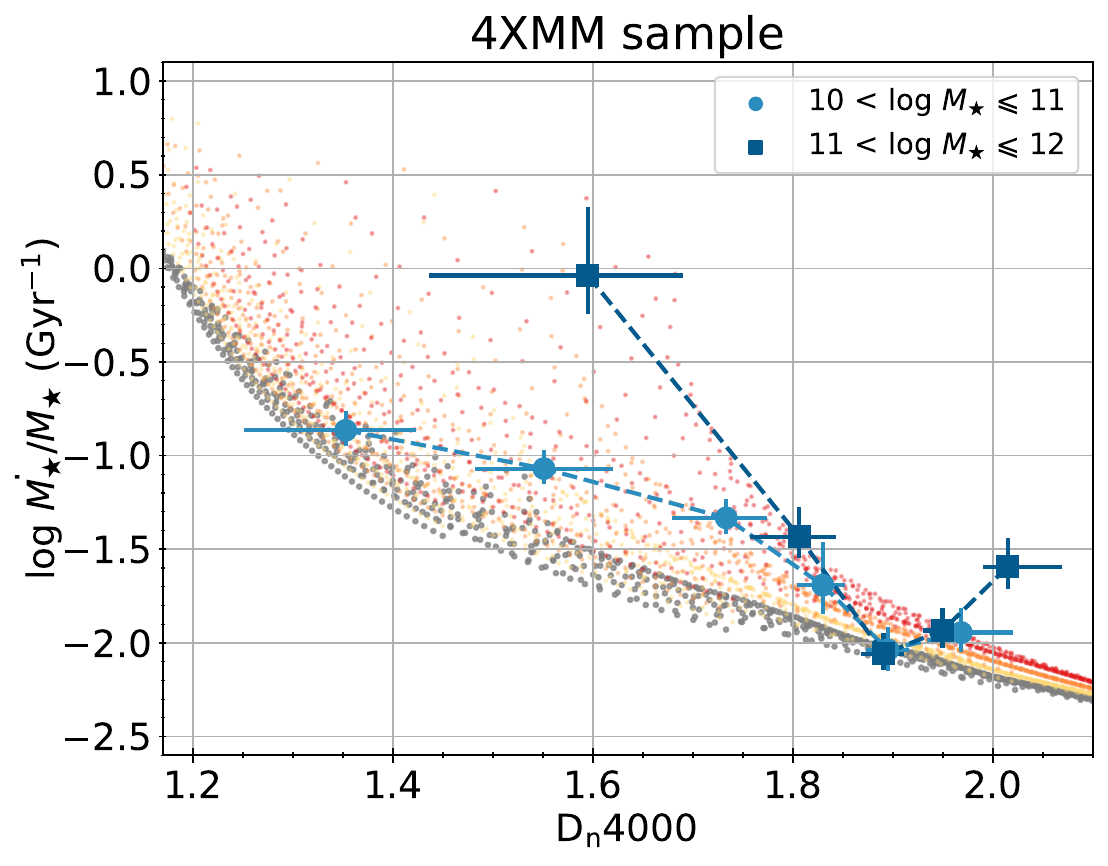}
\includegraphics[scale=0.45]{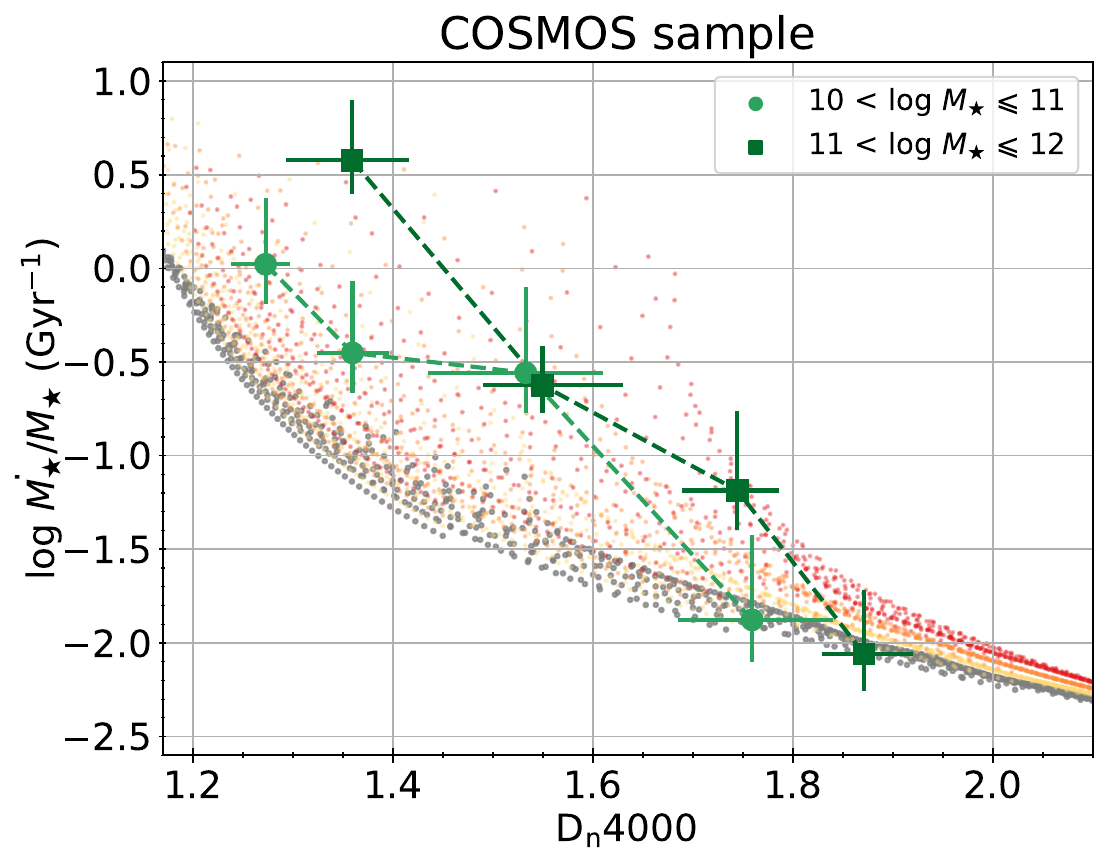}
\caption{\textit{Left panel:} Scaled \bharm\ as a function of \dn\ among galaxies in 4XMM sample in different \mstar\ ranges (the scaling factor is selected to align the \dn\ bin with the lowest \bharm\ value with stellar mass loss rate), plotted against stellar mass loss rate per stellar mass unit per Gyr (grey dots), as well as the average stellar mass loss rate over the past 1 Gyr (yellow dots), 2 Gyrs (orange dots), and 3 Gyrs (red dots).
The horizontal position of each data point represents the median \dn\ of the sources in each bin, with $x$-axis error bars demonstrating the 16th and 84th percentiles of the \dn\ values in each bin.
The $y$-axis error bars represent the 1$\sigma$ confidence interval of \bharm\ from bootstrapping.
\textit{Right panel:} Similar to the left panel, but for scaled \bharm\ as a function of \dn\ among galaxies in the COSMOS sample in different \mstar\ ranges.
}
\label{bharpermstar_mstarbins}
\end{center}
\end{figure*}

In Figure~\ref{bharpermstar_mstarbins}, we plot the stellar mass loss per stellar mass unit per Gyr as a function of \dn, as well as scaled \bharm\ of galaxies in subsamples with different \mstar\ ranges (the scaling factor is chosen so that the lowest \bharm\ align with stellar mass loss rate) for both the 4XMM sample and COSMOS sample.\footnote{We note that it is also plausible that for the bins we used for scaling calibration with \dn\ $\approx 1.9$, the \bharm\ is below the prediction from stellar mass loss rate due to AGN feedback from jets.}
For the 4XMM sample, \bharm\ does not track stellar mass loss well among log~\mstar~$>$ 11 galaxies. This might be due to the fact that the accretion efficiency from recycled gas is closely related to other factors in addition to \mstar, or there are other gas sources for the most massive galaxies at low redshift. It is also plausible that the onset of the cooling of recycled gas takes several Gyrs to happen among these most massive galaxies due to, e.g., AGN feedback (more massive galaxies tend to host more luminous AGNs that drive stronger outflows and exhibit more powerful jets), so that \bharm\ does not track instantaneous stellar mass loss rate at the scale of $\lesssim 100$ Myrs, but stellar mass loss accumulated in the past several Gyrs. In the background of Figure~\ref{bharpermstar_mstarbins}, we also plot the average stellar mass loss rate of stellar populations with different star formation histories over the past 1 Gyr, 2 Gyrs, and 3 Gyrs, represented by the yellow, orange, and red dots. We can see that if the cooling and accretion of the accumulated hot recycled gas is not a process that happens in a less than 1 Gyr timescale, it could explain the high \bharm\ among the relatively young massive galaxies in the 4XMM sample. If this is the case, it is also plausible that recycled gas contributes significantly to the fuel of BHs among star-forming galaxies in the COSMOS sample (see the right panel of Figure~\ref{bharpermstar_mstarbins}).

We note that it is also plausible that metallicity plays a role here, as the higher the \dn, the higher the metallicity \citep[e.g.][]{Gallazzi2005}, and it has been argued that BH growth might be more efficient in the low-metallicity regime \citep[e.g.][]{Toyouchi2019}. Disentangling the effects of age and metallicity would need a large and complete high-signal-to-noise spectroscopic sample from future surveys \citep[e.g.][]{DESI2016, 4most2019}. Since \dn\ is primarily utilized as an age-sensitive parameter \citep[e.g.][]{Kauffmann2003, Gallazzi2005, Kauffmann2009, Wu2018} and we do not observe similarly significant trends when utilizing more metallicity-sensitive parameters (e.g. [MgFe]’, [Mg$_2$Fe]; see \citealt{Gallazzi2005} and references therein), we interpret the link between AGN activity/BH growth with \dn\ mainly a result of the variation of AGN activity/BH growth among galaxies with different stellar population ages.

We also caution that we do not observe a solid decreasing AGN fraction or \bhar\ trend with \dn\ at \dn\ $\lesssim 1.5$ for the COSMOS sample.
The lack of a clear trend might be due to the limited sample size; it might be due to gas with origins other than stellar mass loss serving as the dominating fuel which ``washes out'' the role of stellar mass loss; it might also be attributed to a scenario where the recycled gas from stellar mass loss can not easily cool down or reach the BH in the galaxy center that does not significantly affect our lower redshift (4XMM) sample. One plausible reason for this scenario could be supernova (SN) feedback. SN can create a rarefied and hot environment, and SN winds can expel recycled gas from the galaxy.
The core-collapse SN rate should directly trace the SFR (and type Ia SN rate is small compared to the core-collapse SN; e.g. \citealt{Dekel2019}). The average SFR among star-forming galaxies in the COSMOS sample is much higher than that among the 4XMM sample (consistent with the fact that galaxies in the COSMOS sample have higher redshift), which is associated with stronger SN feedback. We also note that for 10 $<$ log \mstar\ $\leqslant$ 11 galaxies in the 4XMM sample, the decline of \bharm\ with \dn\ among relatively young galaxies is not as steep as that among 11 $<$ log \mstar\ $\leqslant$ 12 galaxies, which might also be attributed to SN winds that are more effective at expulsion when the central surface mass density is low \citep[e.g.][]{Hopkins2022}.

\subsection{Additional fuelling mechanism and/or enhanced accretion capability among old, massive galaxies in the local universe} \label{ss-loweddr}

We have shown in Figure~\ref{bharpermstar_qs} that \bhar\ among the oldest galaxies in the 4XMM sample also does not track the stellar mass loss rate, and appears to be higher than expected from the stellar mass loss fuelling. 
Among these most massive systems at \dn\ $\sim 2$, additional fuelling mechanisms (such as the fuelling from hot gas in the halo) may take place, leading to the high \bharm.\footnote{We note that hot gas among giant ellipticals also shines in the X-ray \citep[e.g.][]{Boroson2011}, but the contribution from the diffuse hot gas alone can not account for the excess amount of \bhar\ observed in the largest \dn\ bin in the 4XMM sample. For ellipticals with log \mstar\ $\sim 11$, we expect diffuse hot emission with log \lx\ $\sim 40$. Converting it to \bhar, it is at the level of log \bhar\ $\sim -5$ and log \bharm\ $\sim -7$ Gyr$^{-1}$, far below the values we observed.}
The additional fuelling mechanism may also be particularly effective in triggering low-accretion-rate AGNs among old, massive galaxies at low redshift, as can be seen in Figure~\ref{lowedd}. 
Hot halo gas fuelling has long been introduced as a source for powering radio AGNs, particularly LERGs, which tend to have old and massive hosts \citep[e.g.][]{BH2012}.
In Figure~\ref{radioagn}, we can see that the fraction of LERG increases with \dn. Also, the LERG population among old galaxies tends to increase with \dn\ when controlling for \mstar, SFR, $z$ (in contrast, the HERG population and the LERG population among young galaxies tend to decrease with \dn\ when controlling for \mstar, SFR, $z$).
It is plausible that the fuelling from the hot halo gas could explain both the increasing fraction of low-accretion-rate X-ray AGN and the increasing fraction of LERG among old galaxies. With this additional fuelling mechanism that is more effective among old galaxies, we also do not observe any sign of a decreasing trend of BH growth associated with \dn\ among the oldest galaxies in the 4XMM sample in Figure~\ref{agnf_age_rw}.

It is also plausible that the relatively high \bhar\ among old galaxies at low redshift is linked with enhanced capability of BH accretion among these galaxies \citep[e.g.][]{Gaspari2015, McDonald2021}. 
Using hydrodynamic simulations, \citet{Gaspari2015} suggest that among massive galaxies, chaotic cold accretion of condensed hot gas is less efficient when a rotating disk is present.
Thus, BHs in dispersion-dominated systems might accrete a larger fraction of gas supply than in rotation-dominated systems. As the fraction of elliptical galaxies increases with \dn, this might also explain the increasing number of low-accretion-rate AGNs as well as the increasing \bhar\ with \dn\ among old, massive galaxies in the 4XMM sample.

\section{Summary and Conclusions} \label{s-cf}

Utilizing spectroscopic samples of galaxies with \xray\ data coverage, we studied the incidence of AGNs among galaxies with different mean stellar population ages in this work.
The main points from this paper are the following:

\begin{enumerate}

\item We built two samples of galaxies/AGNs with both spectroscopic coverage and \xray\ coverage. One sample (4XMM sample) includes SDSS galaxies with \xmm\ coverage at $z= 0$--0.35; our other sample (COSMOS sample) includes LEGA-C galaxies with \chandra\ coverage at $z= 0.6$--1.0. \dn\ measurements from spectra are adopted as a tracer of the mean stellar population age of the galaxy. X-ray observations are utilized to estimate AGN fraction and \bhar\ for samples of galaxies (see Section 2). 

\item In Section~\ref{ss-agn-dn}, we characterized how the AGN fraction, \bhar\, as well as \bharm\ vary with \dn\ among galaxies in the 4XMM sample and COSMOS sample. In Section~\ref{ss-smloss}, we show that \bharm\ as a function of \dn\ roughly traces the scaled stellar mass loss rate predicted by \dn\ among galaxies in the 4XMM sample (except for the oldest galaxies) as well as old/quiescent galaxies in the COSMOS sample, indicating stellar mass loss as a potentially important (and possibly dominant) fuelling source.

\item In Section~\ref{ss-agn-dn-ms}, we found that when controlling for host-galaxy properties (\mstar, SFR, and $z$), the fraction of log \lx/\mstar\ $>$~32 AGNs and \bhar\ decrease with \dn\ among galaxies in the 4XMM sample (except for the oldest/most massive galaxies) and COSMOS sample, suggesting higher numbers of AGNs and higher levels of BH growth among younger galaxies.
We also observed similar trends in terms of the MIR-selected AGN fraction and the HERG fraction among SDSS galaxies in Section~\ref{ss-radiomir}.
These results further support the scenario of stellar mass loss as a potential fuelling source for AGN (see Section~\ref{ss-smloss}).

\item In Section~\ref{ss-agn-dn}, we observed a slight increase of \bhar\ and \bharm\ among the oldest galaxies in the local universe; in Section~\ref{ss-loweddagn}, we found that among the most massive galaxies in the local universe, the fraction of low specific-accretion-rate AGNs (31 $<$ log \lx/\mstar\ $<$ 32) increases significantly with \dn. The LERG fraction in the local universe also increases with \dn\ among the old, massive galaxies (see Section~\ref{ss-radiomir}).
Additional fuelling from the hot halo gas and potentially enhanced accretion capability among old, massive galaxies may explain these trends (see Section~\ref{ss-loweddr}).

\end{enumerate}

Our work shows that stellar mass loss may be an important fuelling source to trigger AGN activity and thus drive ongoing BH growth, not only among old quiescent galaxies, but also among young star-forming galaxies at low redshift. 

\section*{Acknowledgements}

We thank the anonymous referee for constructive feedback.
We thank Johan Comparat for the helpful discussion.
QN and JA acknowledge support from a UKRI Future Leaders Fellowship (grant code: MR/T020989/1).
KLB acknowledges funding from a Horizon 2020 grant (XMM2Athena). 
For the purpose of open access, the authors have applied a Creative Commons Attribution (CC BY) licence to any Author Accepted Manuscript version arising from this submission.

\section*{Data Availability}
The data underlying this article were accessed from the \xmm\ Science Archive, SDSS data releases, \chandra\ data archive, and ESO data portal.
The derived data generated in this research will be shared upon reasonable request to the corresponding author.

\bibliographystyle{mnras}
\bibliography{bib_agn_age} 

\appendix

\section{Assessing the reliability of \dn\ measurements for AGNs} \label{a-dncomtamination}
As we limit our sample to objects with galaxy-like spectra in this study (i.e. quasar-like sources with prominent broad emission lines are excluded), X-ray AGNs in our sample are type 2 AGNs with obscured disk emission. Thus, the AGN power-law continuum should have little contribution to the optical spectra.
However, it is possible for a galaxy to look younger in the spectrum when an AGN component (even if the contribution is small) is present, as quasars have a bluer continuum compared to galaxies.
In this appendix, we assess whether the AGN disk emission affects the reliability of \dn\ measurements for AGNs in our sample.

\begin{figure*}
\begin{center}
\includegraphics[scale=0.6]{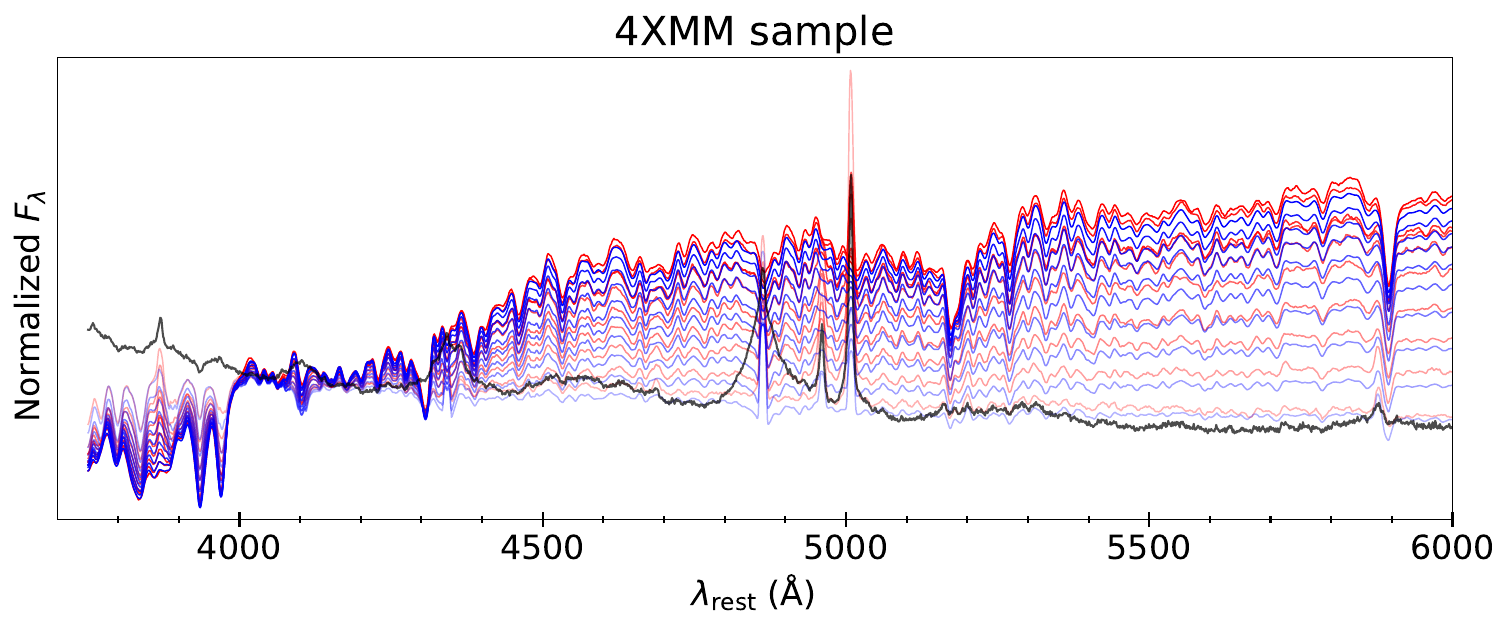}
\caption{Normalized composite spectra of galaxies or AGNs in different \dn\ bins (\dn\ $=$ 1.1--2.0 with a step of 0.1 and a bin size of 0.1) for the 4XMM sample, represented by the blue or red lines; the deeper the color, the larger the \dn\ value. The SDSS quasar composite spectrum from \citet{VB2001} is shown as the black line for comparison. 
}
\label{sdssspec}
\end{center}
\end{figure*}

\begin{figure*}
\begin{center}
\includegraphics[scale=0.6]{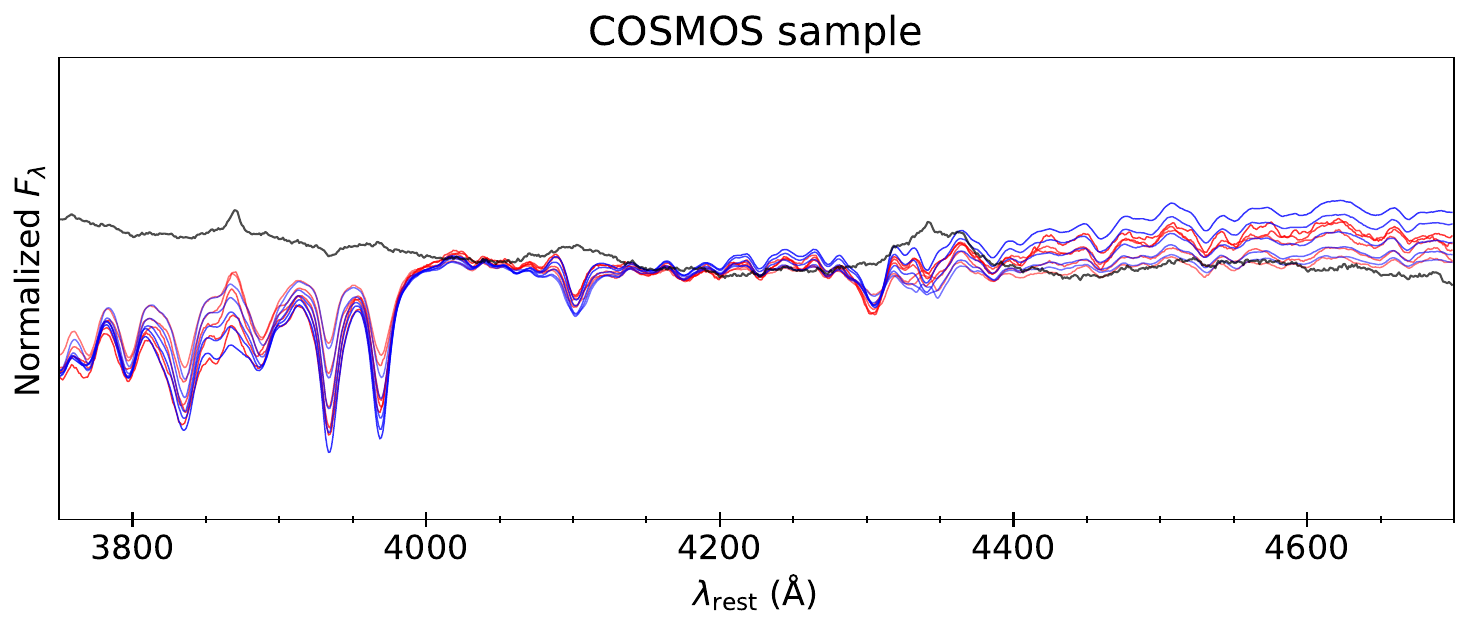}
\caption{Similar to Figure~\ref{sdssspec}, but for galaxies or AGNs in different \dn\ bins (\dn\ $=$ 1.3--1.7 with a step of 0.1 and a bin size of 0.1) in the COSMOS sample.
}
\label{legacspec}
\end{center}
\end{figure*}

\begin{figure*}
\begin{center}
\includegraphics[scale=0.45]{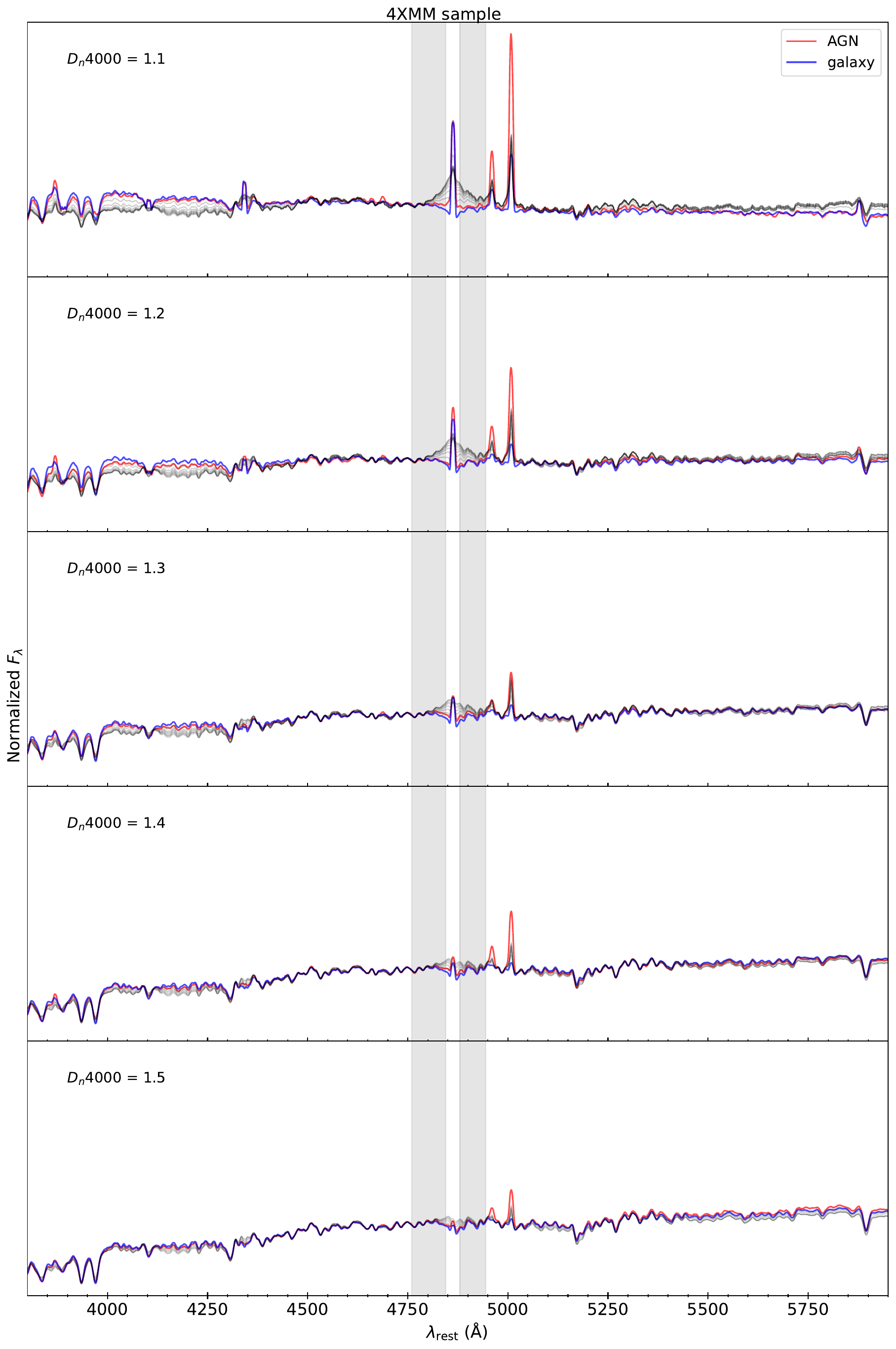}
\caption{In each panel, normalized composite spectra of galaxies or AGNs in the 4XMM sample at a given \dn\ bin (with a median \dn\ value as labeled) are represented by the blue or red lines; 
the black lines represent synthetic spectra created by combining galaxy composite spectra that have stronger \dn\ with the \citet{VB2001} composite quasar spectrum in a proportion that mimics the \emph{observed} \dn\ of the presented AGN composite spectrum. 
These synthetic spectra show excess emission around the broad H$\beta$ wings (indicated by the grey regions) compared to galaxy composite spectra.
We estimate the possible bias in the \dn\ measurements for AGNs in our sample by choosing the synthetic combination that best matches the observed excess flux of the AGN composite spectrum over the galaxy composite spectrum in the grey regions. 
All the spectra presented are normalized at 4750 \AA.
}
\label{sdss_dn}
\end{center}
\end{figure*}

\begin{figure*}
\begin{center}
\includegraphics[scale=0.45]{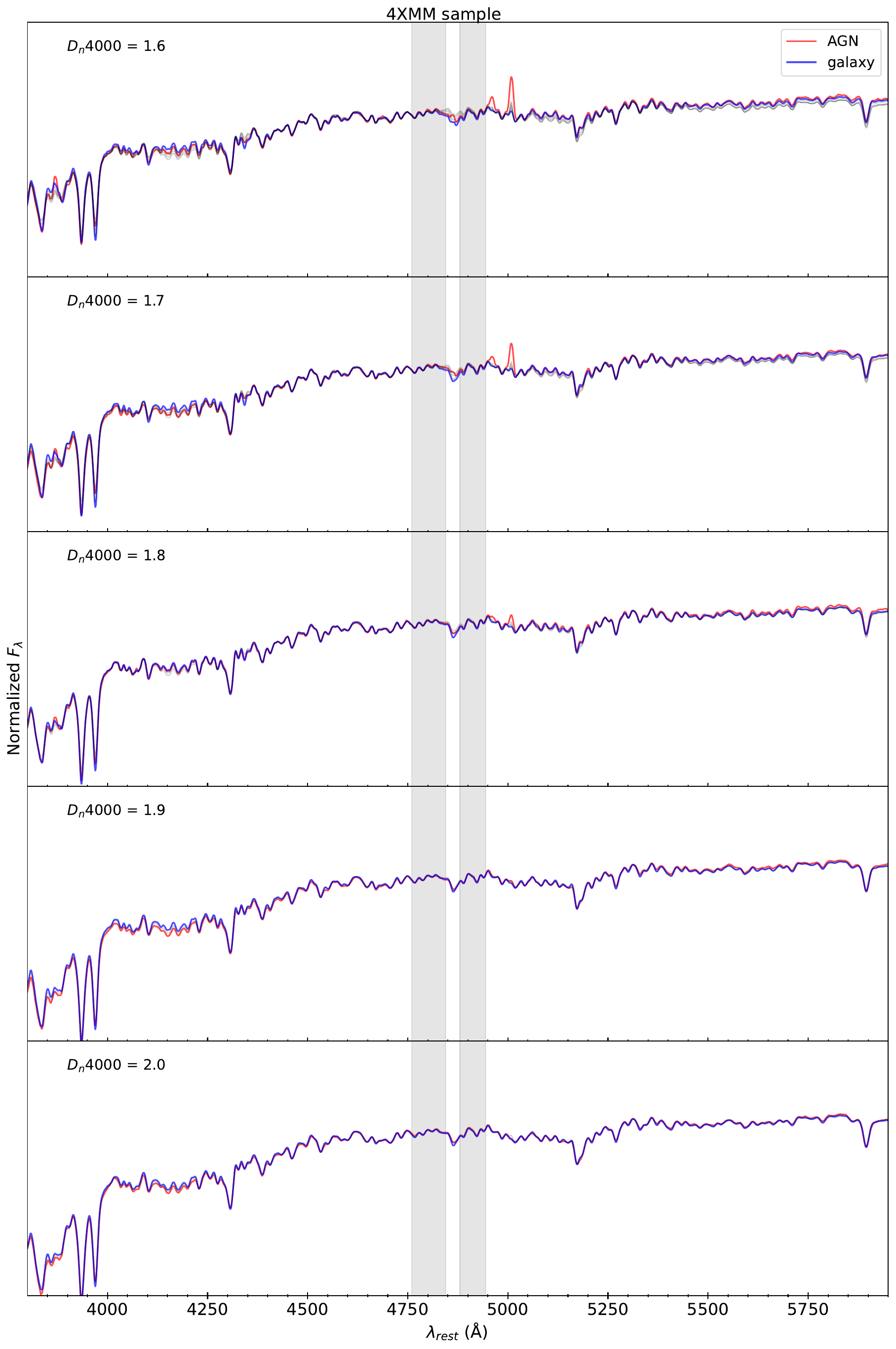}
\caption{Figure~\ref{sdss_dn} continued.
}
\label{sdss_dn_2}
\end{center}
\end{figure*}

\begin{figure*}
\begin{center}
\includegraphics[scale=0.45]{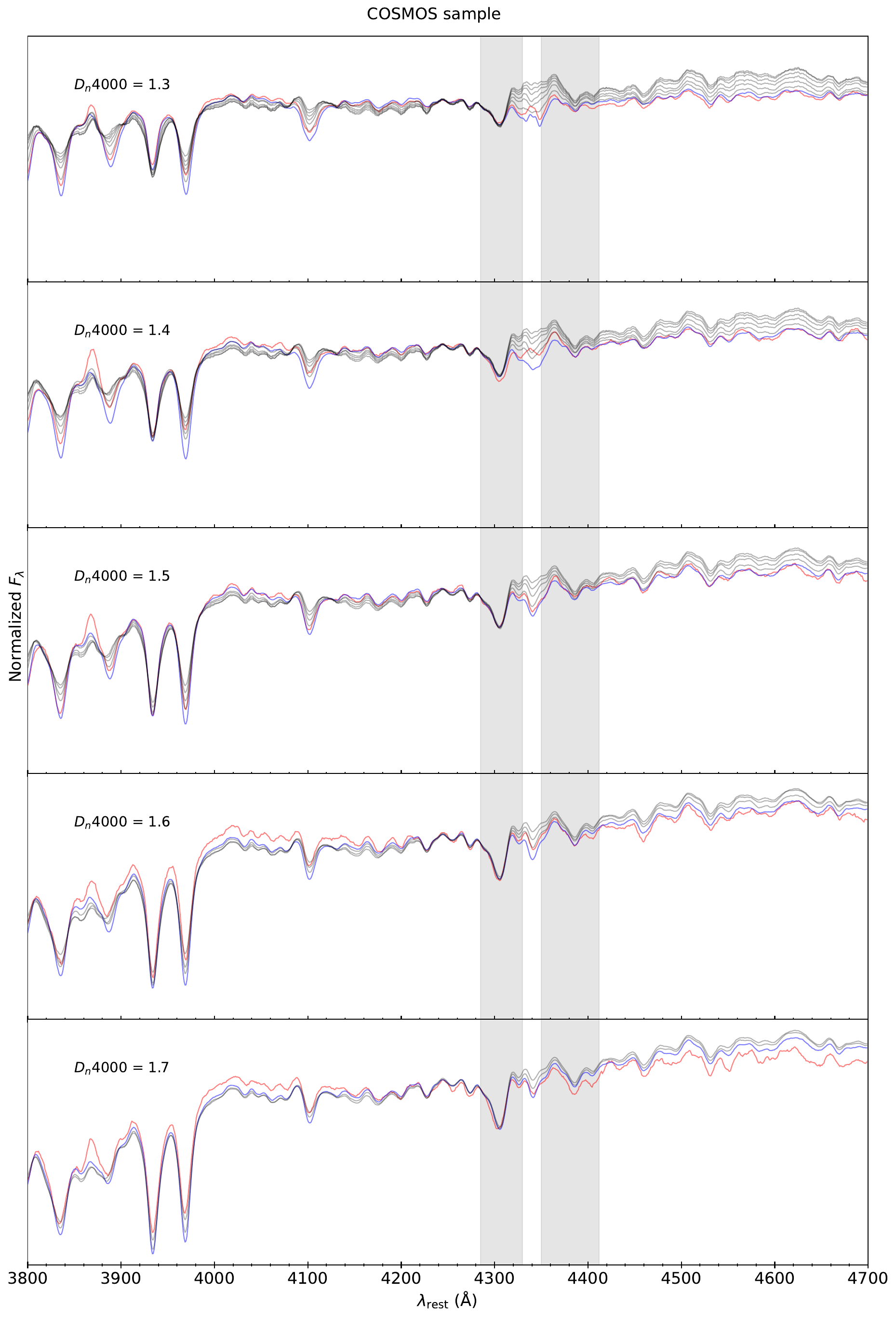}
\caption{Similar to Figure~\ref{sdss_dn}, but for galaxies or AGNs in the COSMOS sample. The synthetic spectra represented by black lines show excess emission around the broad H$\gamma$ wings (indicated by the grey regions) compared to galaxy composite spectra. All the spectra presented are normalized at 4285 \AA.
}
\label{legac_dn}
\end{center}
\end{figure*}

In Figure~\ref{sdssspec}, we present the composite spectra of galaxies/AGNs in the 4XMM sample in different \dn\ bins (\dn\ $=$ 1.1--2.0 with a step of 0.1 and a bin size of 0.1), with the composite quasar spectrum from \citet{VB2001} shown as well.
In Figure~\ref{legacspec}, we present the composite spectra of galaxies/AGNs in the COSMOS sample in different \dn\ bins (\dn\ $=$ 1.3--1.7 with a step of 0.1 and a bin size of 0.1).
Here, AGNs are defined as objects with log \lx/\mstar\ $> 31$. Galaxies are defined as objects not detected in the X-ray.
The composite spectra are created by normalizing each individual spectrum in the subsample at rest-frame 4050 \AA, and taking the median value. The presented composite spectra are smoothed with a boxcar with a width of 10 \AA. We can see that the composite spectra of AGN look generally similar to those of galaxies, and do not exhibit signs of any broad line. 
In Figures~\ref{sdss_dn} and \ref{sdss_dn_2}, we present the 4XMM sample AGN and galaxy composite spectra in different \dn\ bins in different panels; in each panel, we also present spectra constituted by the composite galaxy spectrum from bins with larger \dn\ values and the \citet{VB2001} quasar spectrum, but can mimic the \dn\ value of this panel. We can see that including the quasar emission will lead to a small bump around the H$\beta$ line region, which is not obvious among AGN composite spectra in our sample.
We also perform the same procedures for the AGN and galaxy composite spectra in the COSMOS sample, and the results are presented in Figure~\ref{legac_dn}. Including the quasar emission will make the bump around the H$\gamma$ line region noticeable in the spectra, which is not obvious among our AGN composite spectra.
The lack of apparent differences in the broad emission-line regions of our AGN composite spectra compared to galaxy composite spectra indicates that severe contamination is unlikely.

\begin{figure}
\begin{center}
\includegraphics[scale=0.45]{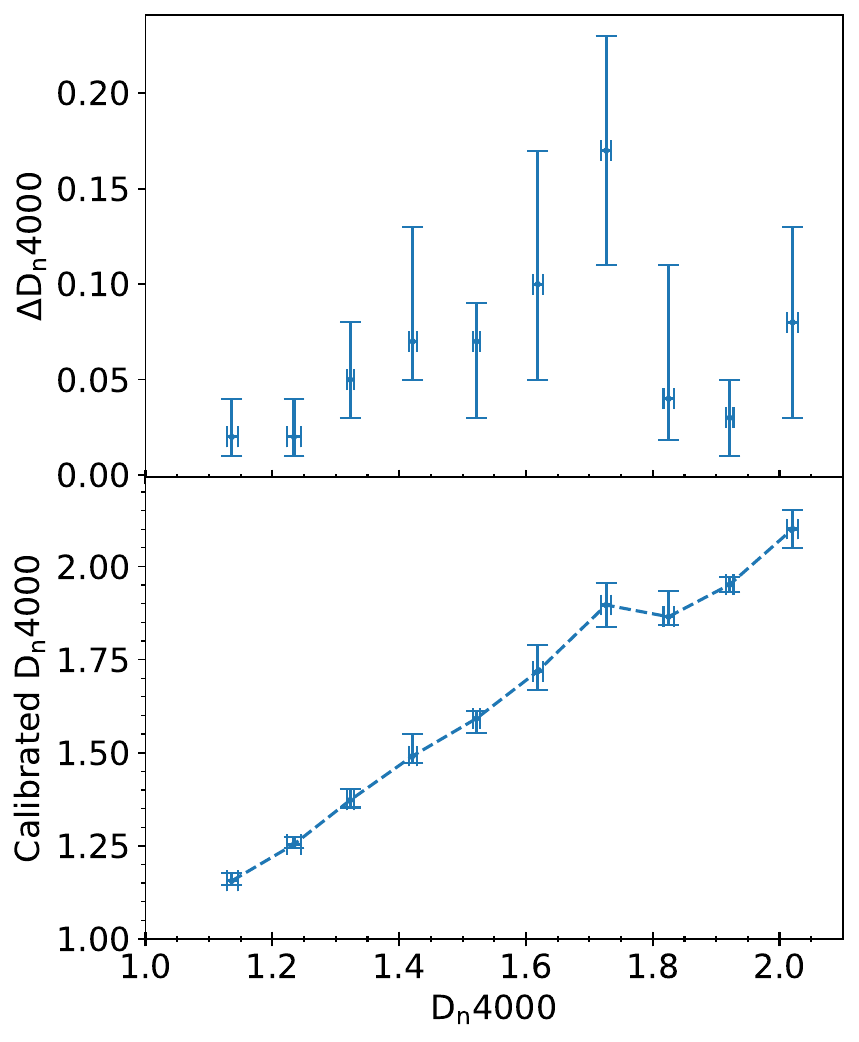}
\caption{\textit{Top:} The underestimation of \dn\ (due to contamination from underlying AGN emission) vs. observed \dn\ of AGNs in the 4XMM sample, with error bars representing the 1$\sigma$ confidence intervals obtained from bootstrapping. \textit{Bottom:} ``Calibrated'' \dn\ values (when accounting for the bias) vs. \dn\ values reported when assuming no AGN component for AGNs in the 4XMM sample. We have verified that our results do not change qualitatively when utilizing these ``calibrated'' \dn\ values for AGNs.}
\label{sdss_dn_bias}
\end{center}
\end{figure}

\begin{figure}
\begin{center}
\includegraphics[scale=0.45]{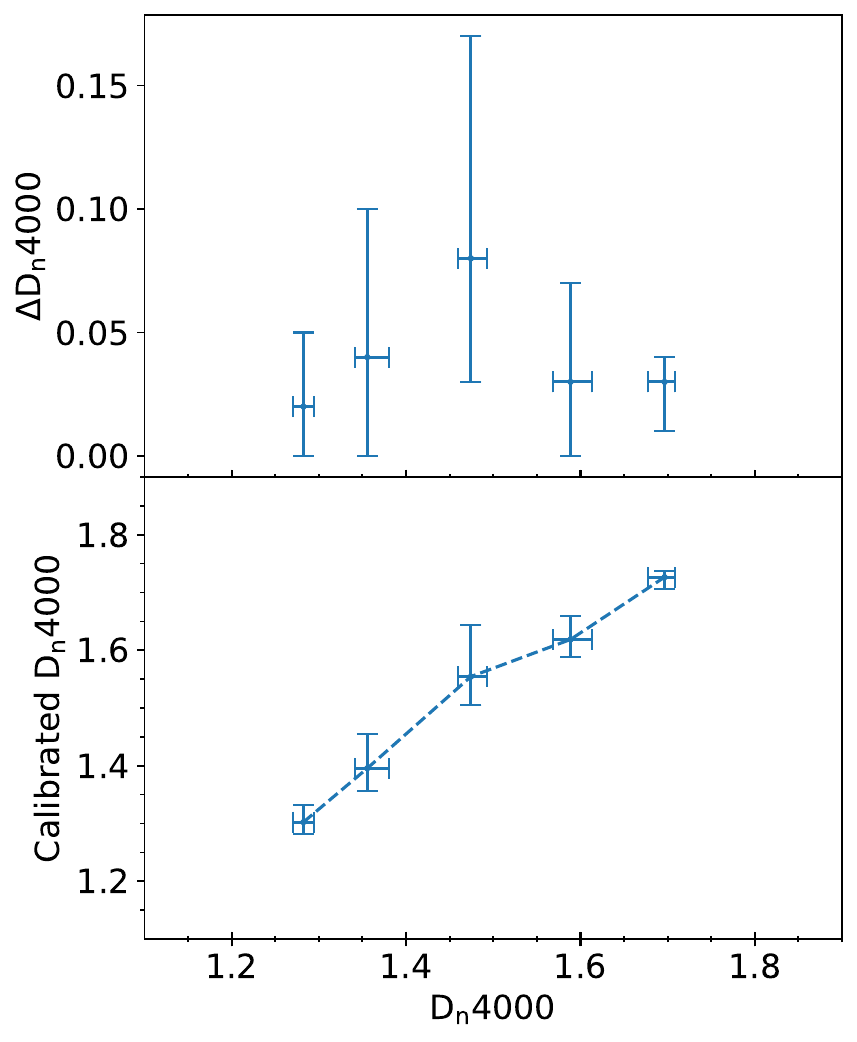}
\caption{Similar to Figure~\ref{sdss_dn_bias}, but for AGNs in the COSMOS sample. 
}
\label{legac_dn_bias}
\end{center}
\end{figure}

To quantify the bias in \dn\ measurements of AGNs related to ``hidden'' AGN emission, we utilize the ratio of the integrated flux in the shaded regions presented in Figures~\ref{sdss_dn}, \ref{sdss_dn_2}, or \ref{legac_dn} between the composite AGN spectrum and galaxy spectrum; these regions characterize the broad H$\beta$ wings for SDSS spectra and the broad H$\gamma$ wings for LEGA-C spectra. We create galaxy composite spectra in the \dn\ grid with a step of 0.01, and mix the galaxy spectra with quasar template to make the \dn\ value equal to that of the AGN composite spectrum in the given \dn\ bin. 
For a given \dn\ bin, when the ratio of the integrated flux in the shaded regions between the synthetic quasar plus galaxy spectrum and the composite galaxy spectrum is close to that between the composite AGN spectrum and the composite galaxy spectrum, we think \dn\ of the galaxy component in the synthetic spectrum represents the true  \dn\ of the AGN composite spectrum, so that the bias can be estimated.
The uncertainty of the bias could be obtained by bootstrapping AGNs in different \dn\ bins, creating different composite AGN spectra, and repeating the above procedures. The results are presented in Figure~\ref{sdss_dn_bias} for the 4XMM sample and Figure~\ref{legac_dn_bias} for the COSMOS sample.
We can see that the bias is generally small ($\lesssim 0.1$--0.2), and ``calibrating'' the measured \dn\ values of AGNs with bias will not change the general \dn\ trend.
We have also verified that our results do not change qualitatively when utilizing ``calibrated'' \dn\ values (i.e., the measured values plus the bias estimated from the above method) for AGNs.

\section{The probability density distribution of a galaxy hosting an AGN as a function of \lx/\mstar\ in the 4XMM sample} \label{a-lxpdf}

\begin{figure*}
\begin{center}
\includegraphics[scale=0.5]{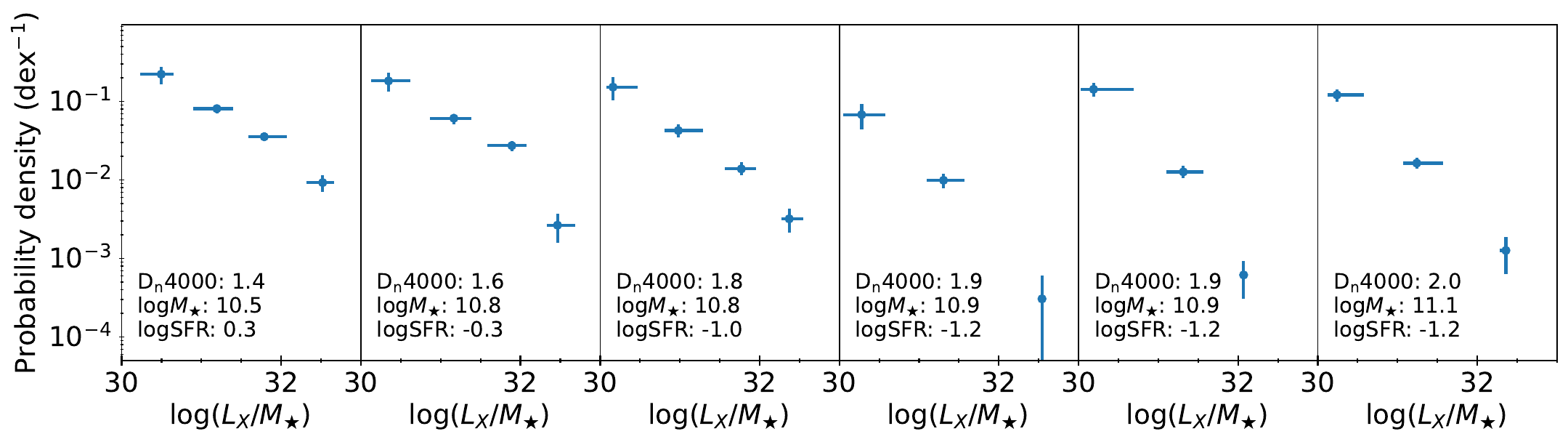}
\caption{Distributions showing the probability of finding an AGN as a function of specific black hole accretion rate. Different panels have different \dn.}
\label{4xmm_full_agnf_sbhar}
\end{center}
\end{figure*}

We note that, in Section~\ref{ss-agn-dn}, the AGN fraction is defined as a single fraction with \lx/\mstar\ greater than a given value, which we take to be $10^{32}$. 
To test whether this arbitrary threshold will affect our results, we further model the probability of finding an AGN as a function of \lx/\mstar\ in the 4XMM sample for all the bins in Figure~\ref{agnf_age}. For the COSMOS sample, this type of analysis is limited by the small number of AGNs detected in each bin.
We could see that this probability density distribution of a galaxy hosting an AGN as a function of \lx/\mstar\ as shown in Figure~\ref{4xmm_full_agnf_sbhar} follows a rough linear relation in the log-log space, similar to what has been found by \citet{Birchall2022}. 
Thus, the \lx/\mstar\ threshold adopted when calculating the AGN fraction would not materially affect the results.

\section{AGN fraction (defined with \lx) as a function of \dn\ when controlling for \mstar, SFR, and $z$} \label{a-lx}

\begin{figure}
\begin{center}
\includegraphics[scale=0.45]{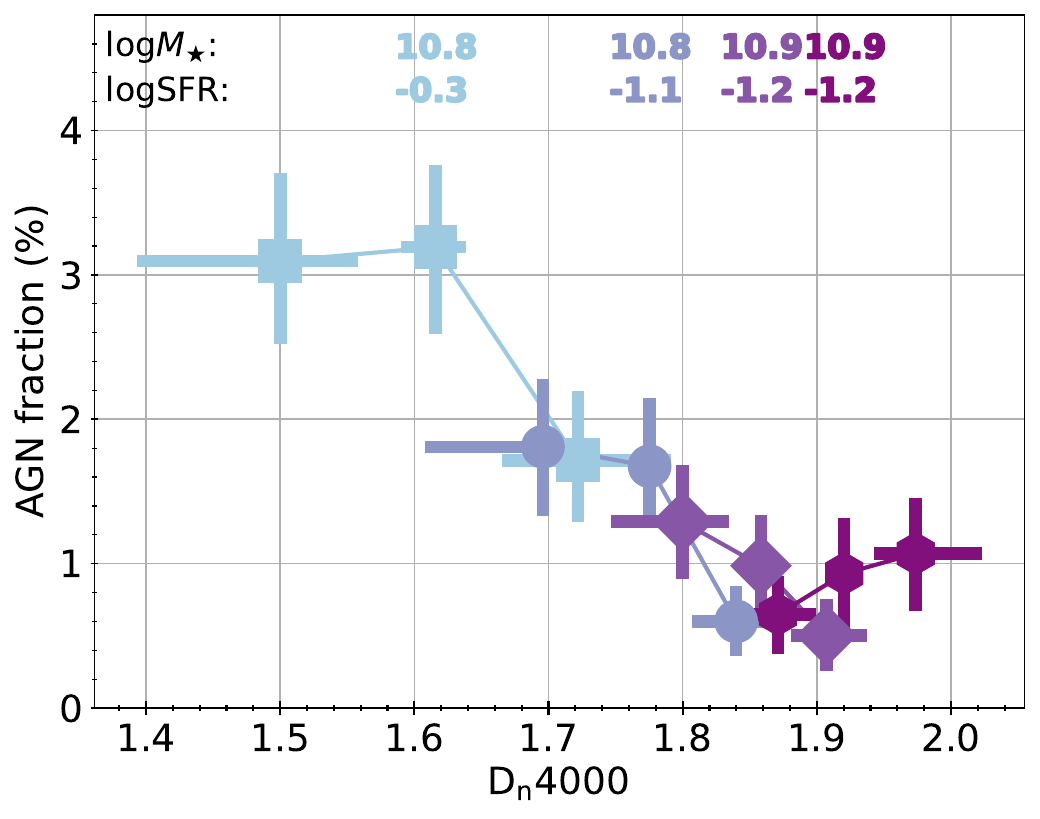}
\caption{AGN fraction (defined by log \lx\ $>$ 42) as a function of \dn\ among galaxies in the 4XMM sample when controlling for \mstar, SFR, and $z$.
Different symbols and colors represent a set of subsamples with similar \mstar, SFR, and $z$ values (as listed on top of the panel with the same color). The horizontal position of each data point represents the median \dn\ of the sources in each sample, with $x$-axis error bars demonstrating the 16th and 84th percentiles of the \dn\ values.
The $y$-axis error bars represent the 1$\sigma$ confidence interval of AGN fraction from bootstrapping. 
}
\label{agnf_lx_age_rw}
\end{center}
\end{figure}

\begin{figure}
\begin{center}
\includegraphics[scale=0.45]{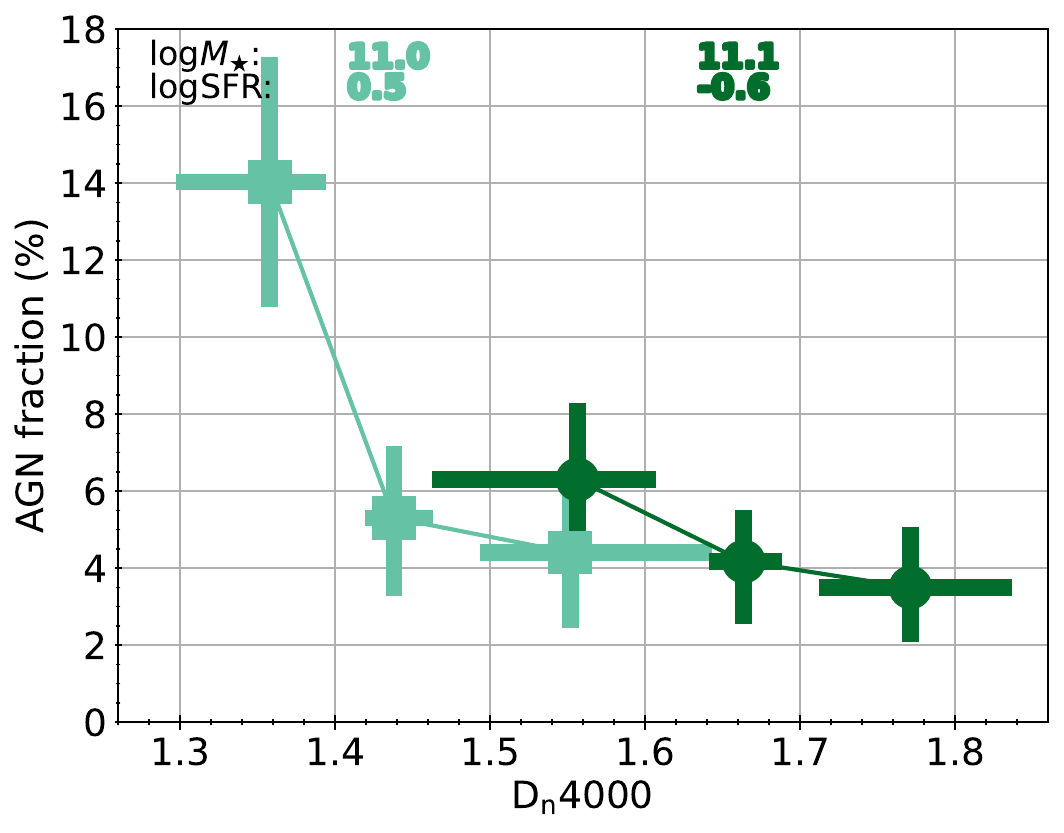}
\caption{Similar to Figure~\ref{agnf_lx_age_rw}, but for AGN fraction as a function of \dn\ among galaxies in the COSMOS sample.
}
\label{legac_agnf_lx_age_rw}
\end{center}
\end{figure}

In Section~\ref{ss-agn-dn-ms}, we examined how the fraction of log \lx/\mstar\ $>$ 32 AGN changes with \dn\ when controlling for \mstar, SFR, and $z$. In Figure~\ref{agnf_lx_age_rw}/\ref{legac_agnf_lx_age_rw}, we present how the fraction of log \lx\ $>$~42 AGN (in the case, AGNs are defined with luminosity rather than with ``specific black hole accretion rate'') changes with \dn\ when controlling for \mstar, SFR, and $z$ for the 4XMM/COSMOS sample. 
We could see that similar to the trends we observed in Figures \ref{agnf_age_rw} and \ref{legac_agnf_age_rw}, the log \lx\ $>$~42 AGN fraction displays a consistent decreasing trend with \dn\ among galaxies in the 4XMM sample (except for the oldest galaxies) and the COSMOS sample.

\section{AGN fraction as a function of \dn\ when controlling for morphology} \label{a-morph}

\begin{figure}
\begin{center}
\includegraphics[scale=0.45]{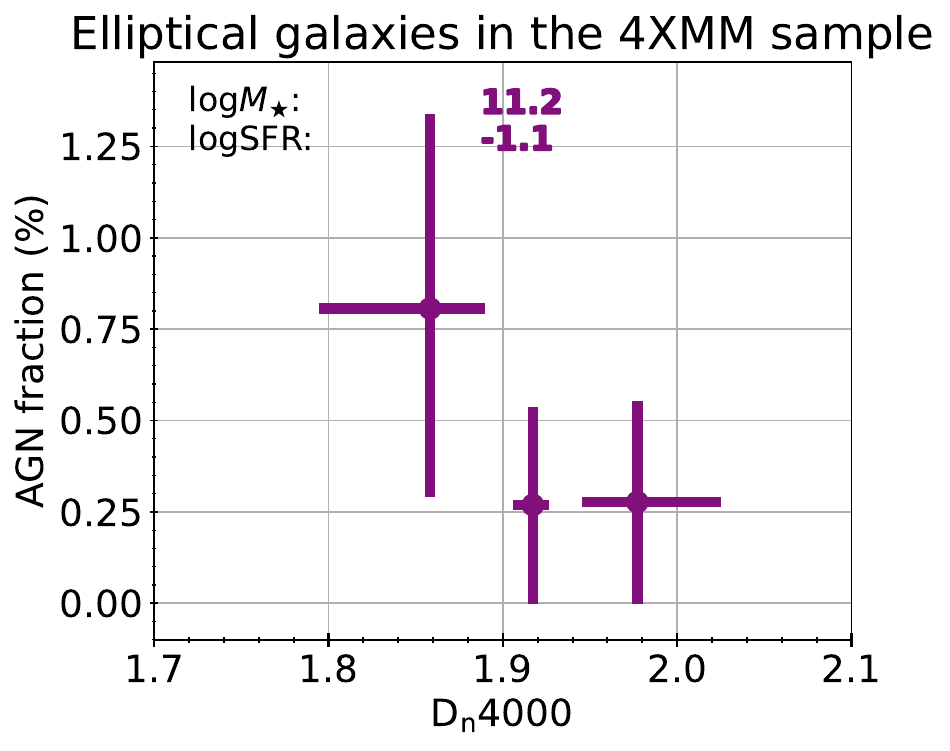}
\includegraphics[scale=0.45]{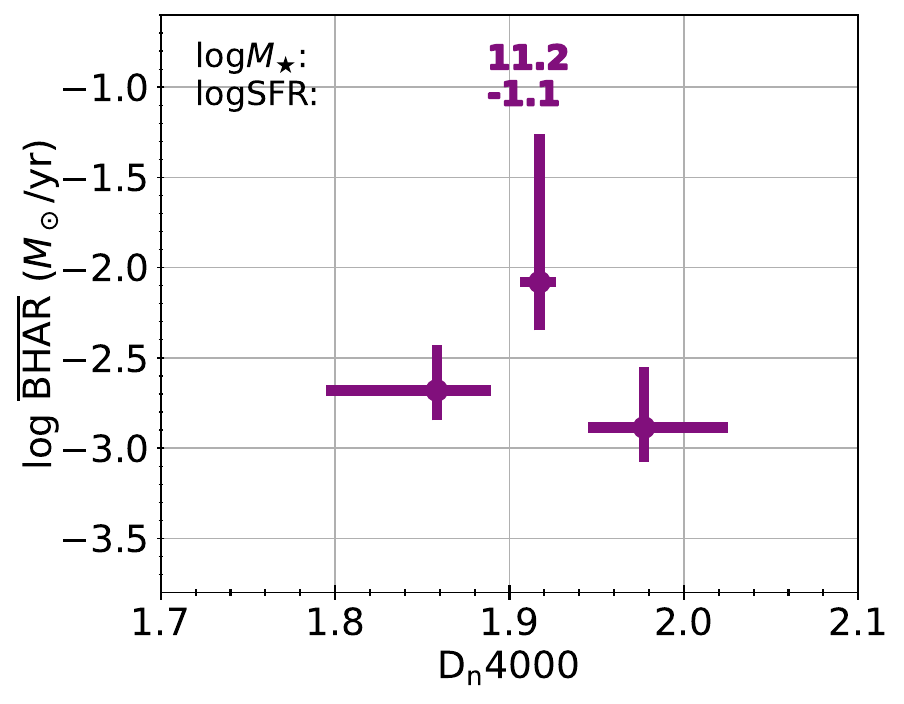}
\caption{\textit{Top:} AGN fraction as a function of \dn\ among elliptical galaxies in the 4XMM sample when controlling for \mstar, SFR, and $z$. All the subsamples share similar \mstar, SFR, and $z$ values. The horizontal position of each data point represents the median \dn\ of the sources in each subsample, with $x$-axis error bars demonstrating the 16th and 84th percentiles of the \dn\ values.
The $y$-axis error bars represent the 1$\sigma$ confidence interval of AGN fraction from bootstrapping. 
\textit{Bottom:} \bhar\ as a function of \dn\ among elliptical galaxies in the 4XMM sample when controlling for \mstar, SFR, and $z$. All the subsamples share similar \mstar, SFR, and $z$ values. The horizontal position of each data point represents the median \dn\ of the sources in each subsample, with $x$-axis error bars demonstrating the 16th and 84th percentiles of the \dn\ values.
The $y$-axis error bars represent the 1$\sigma$ confidence interval of \bhar\ from bootstrapping. 
}
\label{age_rw_e}
\end{center}
\end{figure}

\begin{figure}
\begin{center}
\includegraphics[scale=0.45]{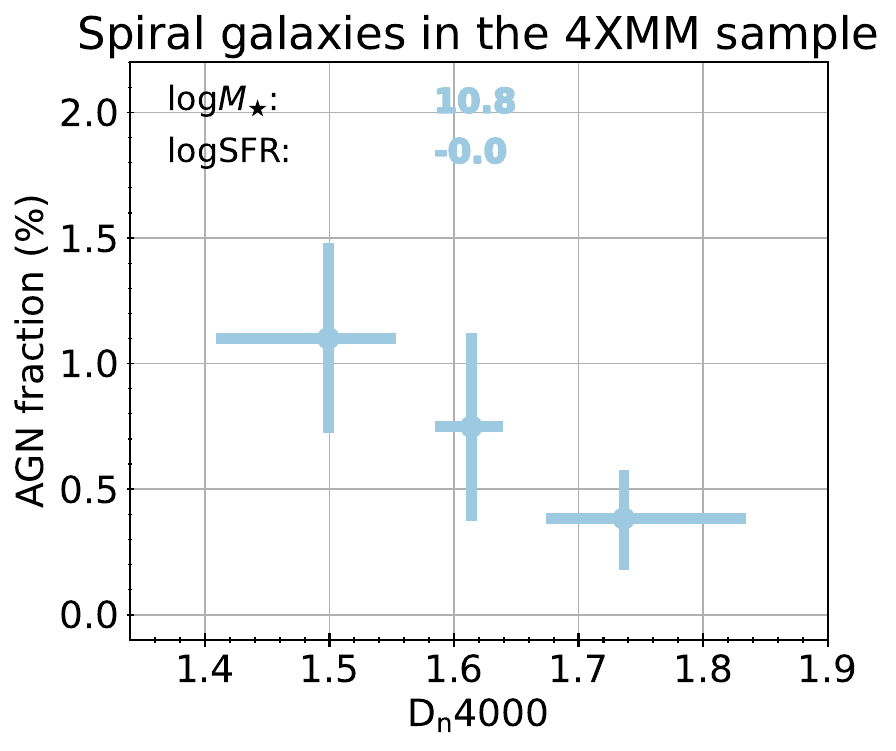}
\includegraphics[scale=0.45]{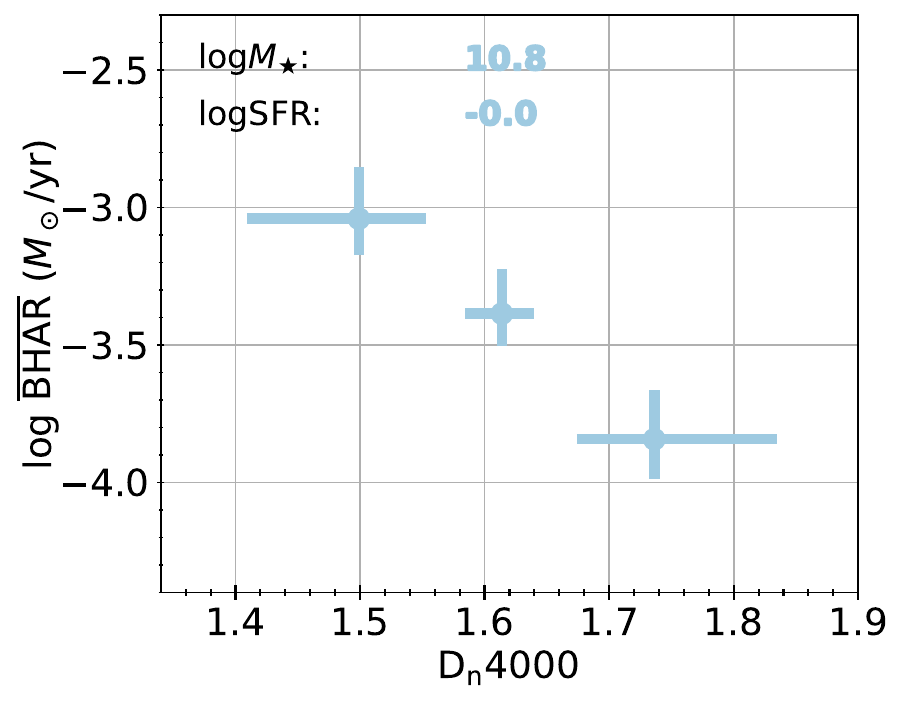}
\caption{\textit{Top:} Similar to the top panel of Figure~\ref{age_rw_e}, but for spiral galaxies in the 4XMM sample.
\textit{Bottom:} Similar to the bottom panel of Figure~\ref{age_rw_e}, but for spiral galaxies in the 4XMM sample.
}
\label{age_rw_s}
\end{center}
\end{figure}

\begin{figure}
\begin{center}
\includegraphics[scale=0.45]{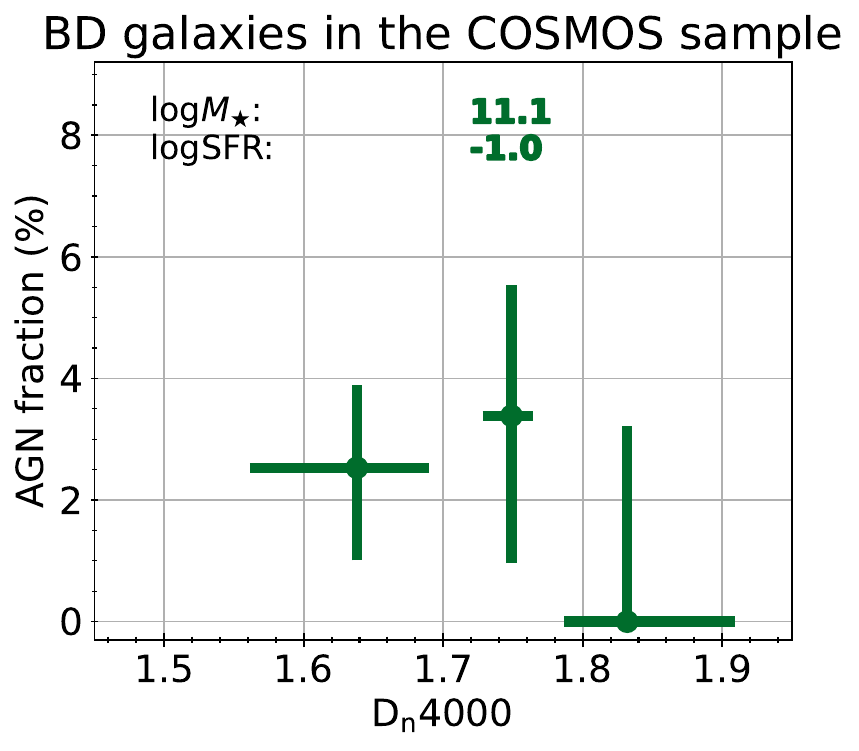}
\includegraphics[scale=0.45]{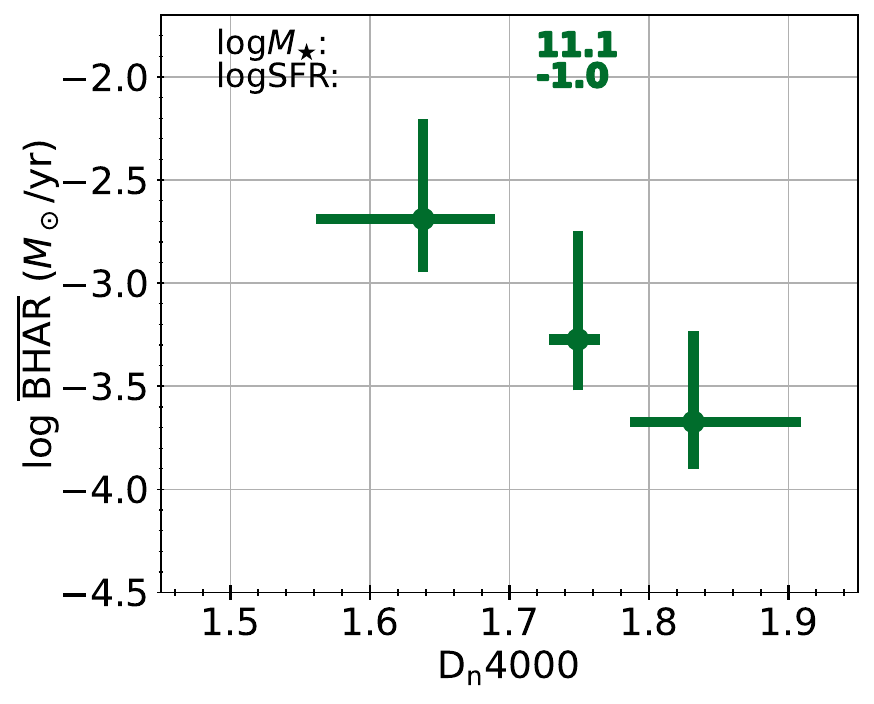}
\caption{\textit{Top:} Similar to the top panel of Figure~\ref{age_rw_e}, but for BD galaxies in the COSMOS sample.
\textit{Bottom:} Similar to the bottom panel of Figure~\ref{age_rw_e}, but for BD galaxies in the COSMOS sample.
}
\label{legac_age_rw_bd}
\end{center}
\end{figure}

\begin{figure}
\begin{center}
\includegraphics[scale=0.45]{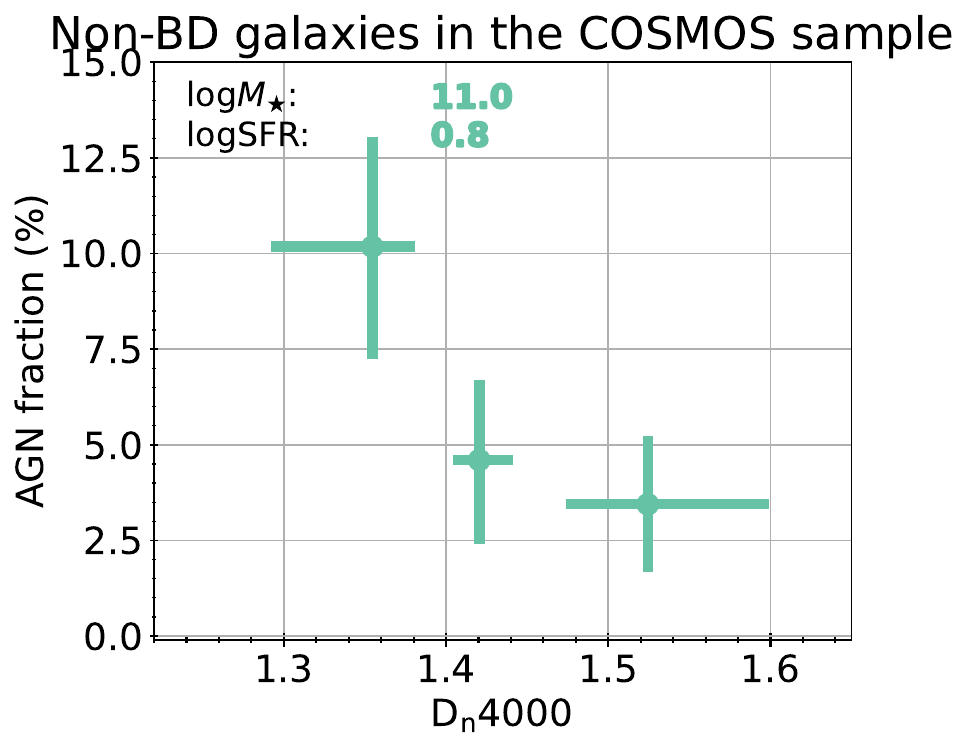}
\includegraphics[scale=0.45]{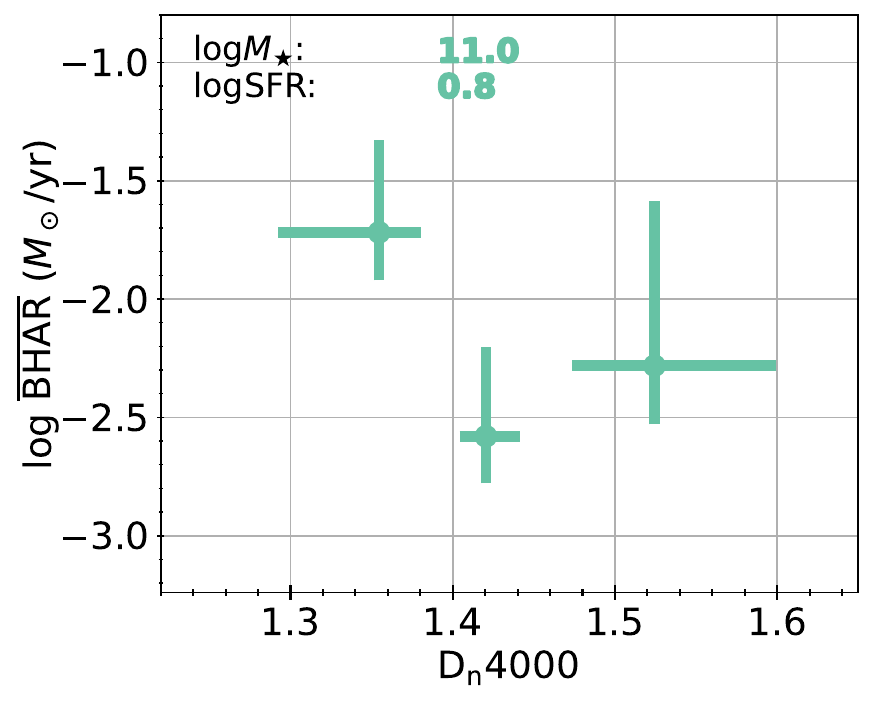}
\caption{\textit{Top:} Similar to the top panel of Figure~\ref{age_rw_e}, but for Non-BD galaxies in the COSMOS sample.
\textit{Bottom:} Similar to the bottom panel of Figure~\ref{age_rw_e}, but for Non-BD galaxies in the COSMOS sample.
}
\label{legac_age_rw_nonbd}
\end{center}
\end{figure}

We also tested whether host-galaxy morphology has an influence on the observed trends with \dn, as the dominant galaxy morphological type changes with the age of galaxies, and it is plausible that different morphological types have different BH fuelling patterns.
For the 4XMM sample, we perform the elliptical/spiral morphological classification with the Galaxy Zoo data \citep{Lintott2008}, utilizing the {\sc CLEAN} criterion developed by \citet{Land2008}.
For elliptical galaxies, we can see that the AGN fraction/\bhar\ does not decrease significantly with \dn\ at \dn\ $\sim$ 1.8--2.0 when controlling for \mstar, SFR, and $z$ in Figure~\ref{age_rw_e}, similar to what we observed in Figure~\ref{agnf_age_rw}.
For spiral galaxies, there is a clear trend that the AGN fraction/\bhar\ drops with \dn\ (see Figure~\ref{age_rw_s}).
For the COSMOS sample, we adopt the bulge-dominated (BD) and non-bulge-dominated (non-BD) morphological classification from \citet{Ni2021} and perform the same analyses for BD galaxies and non-BD galaxies separately.
We note that, among 449 BD galaxies in the COSMOS sample, only 7 of these galaxies have log \lx/\mstar\ $> 32$.
Thus, for the analyses here, we do not adopt the stringent \lx/\mstar\ threshold as we did for the previous analyses. We directly adopt the fraction of objects with log \lx\ $>$ 42 as the AGN fraction (without any sensitivity correction), and the results could be seen in Figure~\ref{legac_age_rw_bd}. How \bhar\ varies with \dn\ when controlling for \mstar, SFR, and $z$ is also presented.
The results for the non-BD galaxies could be seen in Figure~\ref{legac_age_rw_nonbd}.
We can see that a decreasing trend of AGN fraction and \bhar\ is present (though not very significant) among BD galaxies and Non-BD galaxies separately.
These results suggest that the observed variation in AGN activity/BH growth with \dn\ is unlikely due to pure morphological effects.

We also note that host-galaxy structural properties are unlikely to cause the difference in AGN fraction/\bhar\ associated with \dn\ observed in our samples. In \citet{Ni2021}, it has been found that BH growth is closely related to host-galaxy compactness (represented by the projected mass density of the central 1 kpc, \sigmaone) among star-forming galaxies; higher \sigmaone\ values are associated with higher levels of AGN activity/BH growth. As older galaxies tend to be more compact, the higher level of AGN activity/BH growth among younger galaxies is unlikely a result of varying structural properties.

\bsp	
\label{lastpage}
\end{document}